\documentclass{aa}

 \usepackage{graphicx}
 \usepackage{txfonts}
\usepackage[]{hyperref}
\usepackage{comment}
\usepackage{placeins}
\usepackage{flushend}

\newcommand{\hi}{\sc Hi}

\bibpunct{(}{)}{;}{a}{}{,} 
\usepackage[dvipsnames]{xcolor}

\begin{document}

     \title{Atomic gas properties at the positions of supernovae Type Ia, II, and Ib/c
     }


   \author{Bruno \v{S}laus \inst{\ref{inst:uam}}\thanks{e-mail: \emph{bruno.slaus@gmail.com}
                }$^{\mbox{*}}$
            \and   
           Natalia Gotkiewicz
           \inst{\ref{inst:uam}}\thanks{e-mail: \emph{natalia.gotkiewicz@protonmail.com}
                }$^{\mbox{*}}$
           \and
           Micha{\l} J. Micha{\l}owski\inst{\ref{inst:uam}
           }
           \thanks{e-mail: \emph{mj.michalowski@gmail.com}
                }
           \and
           Aleksandra Le\'sniewska\inst{\ref{inst:cop},\ref{inst:uam}}
           \and
           Przemys{\l}aw Nowaczyk\inst{\ref{inst:uam}}
           \and
           Oleh Ryzhov\inst{\ref{inst:uam}}
           \and
           Mart\'in Solar\inst{\ref{inst:uam}}
           \and
           Jakub Nadolny\inst{\ref{inst:uam},\ref{inst:canary}, \ref{inst:tenerife}}
           \and
            Wojciech Dimitrov\inst{\ref{inst:uam}}
           }

  \institute{ Astronomical Observatory Institute, Faculty of Physics and Astronomy, Adam Mickiewicz University, ul. Słoneczna 36, 60-286 Poznań, Poland,  \label{inst:uam}
 \and
  DARK, Niels Bohr Institute, University of Copenhagen, Jagtvej 155A, DK-2200 Copenhagen N, Denmark \label{inst:cop}
  \and
Instituto de Astrof{\'\i}sica de Canarias, E-38205 La Laguna, Tenerife, Spain  \label{inst:canary} 
  \and
Universidad de La Laguna, Dept. Astrof{\'\i}sica, E-38206 La Laguna, Tenerife, Spain \label{inst:tenerife} 
}

   \date{Received ; }

 \abstract
 {Understanding which stars explode as which type of supernovae (SNe) is crucial to measure their contribution to the metal production and feedback halting star formation. Most of the studies of the gas in the environment of SNe are limited by a small sample size ($<10$), hampering any statistical conclusions about their nature.}
 {The goal of this paper is to present the first analysis of atomic gas properties at the positions of a statistically significant sample of SNe in order to constrain their nature.}
 {We selected 133 SNe (29 Ia, 77 II, 27 Ib/c) which have exploded in galaxies with existing atomic gas data. In order to test whether SN positions trace enhancements in the atomic gas distribution, as compared to the atomic gas in the whole extent of the host galaxy, we analyzed the fraction of pixels on the {\hi} map which are fainter than the pixel in which SN is located and the fraction of the {\hi} flux contributed by these pixels. We compare these estimates with both theoretical expectations and simulated models.}
 {All types of SNe deviate from the completely random distribution. From the three types of SNe, Type II showed the largest offset from the {\hi} distribution, preferring even higher concentrations of atomic gas. This type of SNe deviated also from being proportional to the stellar surface density of the host. The results are, however, complicated by the limits of the survey in size, and in the un-even resolution of the {\hi} observations. Furthermore, by direct comparison between the three SN types, we observed that the distributions of these populations are still consistent with each other, i.e. with being drawn from the same underlying distribution. }
 {The obtained results fail to ascertain that Ib/c core-collapse SNe, and possibly also Type II SNe, are connected with the densest concentrations of atomic gas in their hosts, unlike what has been suggested for GRBs and Ic-BL SNe. 
Hence, the birth of progenitors of Type II and Ib/c SNe is still consistent with being connected with the current star formation in their hosts, whereas the progenitors of GRBs and Type Ic-BL SNe require more special conditions to form, for example low metallicity.}


\maketitle
\nolinenumbers

\section{Introduction}
\def\thefootnote{\mbox{*}}\footnotetext{The contribution of the first two authors was comparable}\def\thefootnote{\arabic{footnote}}

Supernovae (SNe) play an important role in  
the evolution of galaxies, affecting chemical content and star formation of their host galaxies via the feedback mechanism. According to their observational characteristics, SNe are classified into different types, namely Type Ia, Type Ib/c and Type II SNe. 
It is currently believed that Type II SNe arise from core collapse of massive stars ($>8\, \mathrm{M_\odot}$; \citealt{Smart2009}), with relatively short life-times.
Type Ib SNe are classified by the lack of hydrogen, while Type Ic by the lack of hydrogen and helium in their spectra (\citealt{Filippenko1997}). Together, type Ib and Ic SNe are referred to as stripped envelope SNe, as it is believed that their outer layers are stripped either by the strong stellar winds of the star itself, or by a companion (\citealt{Smith2014}). Type Ia SNe arise from binary systems with a white dwarf (\citealt{Maoz2014}), and have longer lifetimes from their formation till the explosion.

Understanding which stars explode as which type of SNe is crucial to measure their contribution to the metal production and feedback which affects star formation and the interstellar medium (\citealt{McKee1977}).
However, the progenitors of most types of SNe are  poorly understood and only several Type IIb (possessing a prominent hydrogen line only in the initial phase of the explosion) and II \citep[e.g.][]{maund09,maund14,vandyk13,folatelli15,fraser16,jencson22,smith22,vandyk23} and one Type Ib \citep{eldridge16,folatelli16}
SN progenitors have been identified in pre-explosion images and confirmed by subsequent disappearance, turning out to have initial masses of $8$--$20\ M_\odot $.

The nature of SN progenitors can be statistically inferred from their positions in the host galaxies (\citealt{Anderson2015}). Prior studies showed that the positions of SNe are not randomly distributed but instead depend on the location within the host galaxy \citep[e.g.][]{fruchter06,leloudas11,kuncarayakti18}, however, the details of the interstellar medium (ISM) at these locations are still unclear.

Molecular gas is usually the fuel of star formation. Carbon monoxide (CO) observations, being a good tracer of molecular gas of galaxies, was used to infer the nature of SNe and gamma-ray bursts (GRBs). Such observations for small samples of GRB hosts showed normal amounts of molecular gas compared to star-forming galaxies with similar star formation rates and metallicities \citep{hatsukade14,hatsukade19,hatsukade20co,stanway15,michalowski16,michalowski18co,arabsalmani18,arabsalmani20,deugartepostigo20,deugartepostigo25,nadolny23}.

The amount of molecular gas at the positions of Type Ib/c SNe turned out to be higher than that at the positions of Type II SNe, suggesting a stronger association with ongoing star formation \citep{galbany17}. These observations probed kiloparsec scales, but a similar effect was found with high-resolution observations probing 50-100\,pc scales with a sample of seven Type Ib/c SNe \citep{maykerchen23}. On the other hand, a larger sample of Type Ic SNe studied in the same way, revealed no difference compared to Type II SN locations, which was used to advocate for similar initial masses of progenitors of Type II and Ic SNe, and therefore a binary nature for the latter \citep{solar24}. 

Additionally, the hosts of two super-luminous SNe were observed \citep{arabsalmani19b,hatsukade20b} both suggesting higher gas densities at the locations of the events. A prototypical fast blue optical transient,  AT\,2018cow, was observed by \citet{morokumamatsui19}, arguing for the host galaxy to be a normal star-forming dwarf galaxy, with regards to its gas content. Furthermore, several fast radio bursts \citep[FRBs;][]{bower18,hsu23,chittidi24,yamanaka24} were observed at CO at kiloparsec resolutions, finding gas-rich host galaxies or disturbed kinetic environments (although differing results also exist, such as e.g. \citet{hatsukade22}, arguing for diverse gas properties). 

A large fraction of ISM within galaxies, however, is present in the form of neutral atomic hydrogen ({\hi}). Another approach to infer the nature of progenitors of various transients is, therefore, to analyse properties of atomic gas close to the locations of their explosions. Concentrations of atomic gas have been found close to the positions of three GRBs \citep{michalowski15hi,michalowski16,arabsalmani15b,arabsalmani19,arabsalmani22,deugartepostigo25,thone24} 
and two broad-lined Type Ic (Ic-BL) SNe \citep{michalowski18,michalowski20} 
with only one counterexample with smooth atomic gas distribution (GRB\,111005A; \citealt{lesniewska22}), but this GRB was atypical due to the lack of a SN, low luminosity, rapid radio afterglow decay, and super-solar metallicity \citep{michalowski18grb,tanga18}. This suggests that the birth of the progenitors of GRBs and Type Ic-BL SNe is connected with inflows of gas from the intergalactic medium fueling an episode of enhanced star formation. 
Similarly, the atomic gas bridge between a galaxy NGC\,2770 and its companion was used to explain its unusually high SN rate (4 SNe in 20 years; \citealt{thone09,thone17}), in a way that the revealed interaction is triggering a recent enhancement in the formation of stars \citep{michalowski20b}.
In a similar vein, the atomic gas line profiles of FRB host galaxies were found to be asymmetric, consistent with a recent galaxy merger and supporting the FRB progenitor mechanisms connected with recent star formation \citep{michalowski21frb,glowacki23,glowacki24,leewaddell23,roxburgh26}.
The atomic gas of the host of transient AT\,2018cow has also been characterised, finding absence of gas concentration in the vicinity \citep{michalowski19}, but also asymmetries in the gas density \citep{roychowdhury19}, that could be indicative of interactions with a companion galaxy.

All of these studies are limited by a small sample size, hampering any statistical conclusions about the nature of the transients. The goal of this paper is to present the first analysis of {\hi} atomic gas properties at the positions of a statistically significant sample of over 130 SNe of various types, in order to constrain their nature regarding the gas content within their host galaxies. We achieve this by comparing their positions with theoretical expectations and simulated models.


\section{Data and methods}  \label{sect:section2}

The SN sample was obtained from The Open Supernova Catalog\footnote{\tt github.com/astrocatalogs/supernovae} (\citealt{snespace}) up to July 2021. We divided SNe into three groups: Type Ia, Type II, and Types Ib, Ib/c, Ic together as Type Ib/c.

The selection of the SN sample was based on two criteria: redshift and availability of atomic gas data. 
We selected SNe with redshifts $z<0.1$, because at higher redshifts there are almost no atomic gas detections. We searched for 
the atomic hydrogen 21\,cm {\hi} data for the SN host galaxies in the NASA/IPAC Extragalactic Database\footnote{{\tt ned.ipac.caltech.edu}}. The data come from two observatories: the Westerbork Synthesis Radio Telescope (WSRT) and the NSF's Karl G. Jansky Very Large Array (VLA), or more specifically, from the Westerbork HI Survey of Spiral and Irregular Galaxies (WHISP; \citealt{swaters02}) and The HI Nearby Galaxy Survey (THINGS; \citealt{walter08}) with the VLA. The WHISP survey sample was selected from the Uppsala General Catalogue of Galaxies (\citealt{Nilson1995}), and restricted to galaxies with high {\hi} content that were possible to be sufficiently resolved with the WSRT. Furthermore, the sample was limited to late-type dwarf galaxies, resulting in $73$ new observations, all listed in \citet{swaters02}. The THINGS survey has observed $34$ nearby galaxies, spanning a wide range of physical properties, with the sources limited to low distances ($<15 \ \mathrm{Mpc}$). The survey, furthermore, excluded early type galaxies, as well as edge-on sources. The complete sample is listed in \citet{walter08}. The final sample with SN identification contains $74$ galaxies, $16$ of which were observed with the VLA.

Usually, the data is available in two or three different resolutions. We always selected the best available resolution. Therefore, if both WSRT and VLA data are present, then the latter was used, because they are of higher resolution. Most of the maps have resolutions between 0.1 and 5\,kpc (with the exact values listed in the Table~\ref{tab:table} in the appendix). All the data cover more than the optical extents of the galaxies. Only the pixels where {\hi} emission was detected, were used in the analysis.

In order to test whether SN positions trace enhancements in the atomic gas distribution, we adopted a similar method as used by \citet{fruchter06} for ultraviolet emission (known as normalized cumulative rank; \citealt{James2006}). We selected the pixel at the SN position from the host galaxy {\hi} image, and then proceeded to calculate the fraction of pixels that are fainter than the pixel in which a given SN exploded ($f_p$). A cumulative distribution plot of these fractions provided us with insight into the positional distribution of SN events.

In order to limit the influence of numerous faint pixels which can wash out the signal traced by this metric, we also analyzed the fraction of flux (instead of pixel number) contributed by pixels fainter than the pixel in which a given SN exploded ($f_f$). This is simply the sum of the {\hi} fluxes of these faint pixels divided by the sum of fluxes of all pixels for a given galaxy.

In order to assess the differences of the distributions of the $f_p$ and $f_f$ values of various samples, we used the Kolmogorov-Smirnov (KS) test, examining all three SN types separately. This method uses the cumulative rank functions in order to test how likely it is for an observed sample to come from a specified distribution, or, if applied between two observed samples, how likely it is for these samples to have arisen from the same underlying distribution.

In addition to the real observed samples of SN events, we also used simulated samples as a comparison. These simulations were created to follow different positional distributions within host galaxies, and are described in detail in Sect.~\ref{Sect:SimulationsREsults}.

\section{Results}
\label{sect:section3}

The sample contains 74 galaxies and 133 SNe, including 29 Ia, 77 II, 12 Ib, 2 Ib/c, 13 Ic (27 Ib/c in total) SNe. All SN hosts, their redshifts, resolutions of the {\hi} data, and $f_p$ and $f_f$ values are shown in Table~\ref{tab:table}. Mean values of $f_p$ and $f_f$ for each SN type are shown in Table~\ref{tab:mean}. As seen from the table, the mean values of $f_p$ are all above $0.8$, while $f_f$ are slightly larger than $0.5$.

\begin{table*}
\caption{Mean values of $f_p$ and $f_f$ with standard deviation errors for all three SNe groups.}
\small
\center
\begin{tabular}{cccc}
\hline \hline
& Ia & II & Ib/c  \\ \hline
$\overline{f_p}$ & 0.81 $\pm$ 0.24 & 0.85 $\pm$ 0.21 & 0.84 $\pm$ 0.20  \\ 
$\overline{f_f}$ & 0.54 $\pm$ 0.31 &  0.59 $\pm$ 0.28 & 0.55 $\pm$ 0.29\\
\hline
\label{tab:mean}
\end{tabular}
\end{table*}

\begin{table*}
\caption{KS test p-values for $f_p$ (left) and $f_f$ (right) between different types of SN. The diagonal shows the one-sided KS test with respect to the uniform distribution for $f_p$ or $f_f$, respectively.}
\small
\center
\begin{tabular}{cccc}
\hline \hline
$f_p$ \vline & Ia & II & Ib/c \\ \hline
Ia & $2.7\times10^{-9}$   & 0.54  & 0.40  \\ 
II &  & $2.3\times10^{-26}$   & 0.51 \\
Ib/c &  &   & $4.9\times10^{-11}$   \\ 
\hline
\label{tab:ksfp}
\end{tabular}
\quad
\quad
\quad
\begin{tabular}{cccc}
\hline \hline
$f_f$ \vline & Ia & II & Ib/c \\ \hline
Ia & 0.51  & 0.63 & 0.61  \\ 
II &  & 0.0040  & 0.55 \\
Ib/c &  &   & 0.13  \\ 
\hline
\label{tab:ksff}
\end{tabular}
\end{table*}

\begin{figure*}
\centering
\begin{tabular}{cc}
\includegraphics[width=0.5\textwidth]{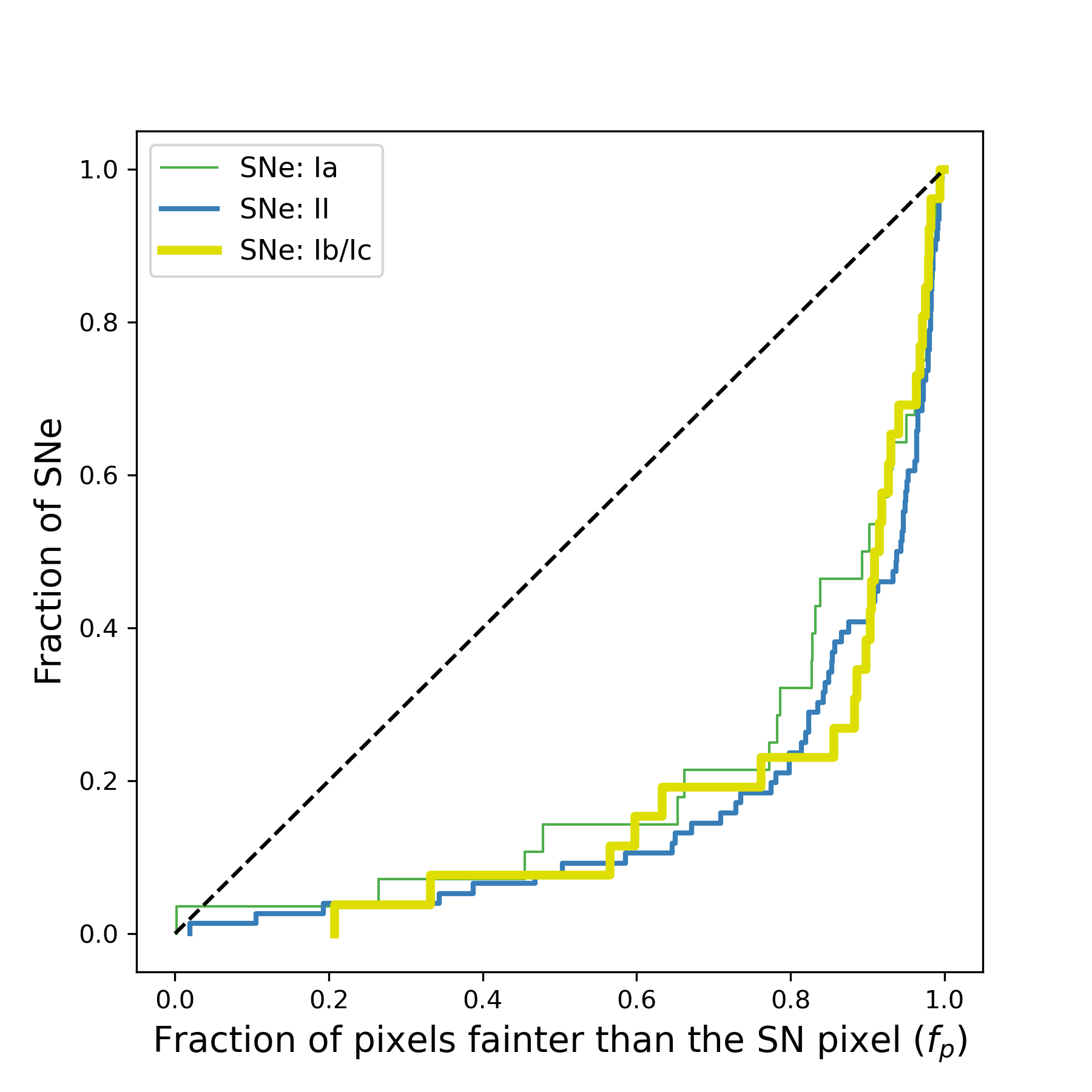} &
\includegraphics[width=0.5\textwidth]{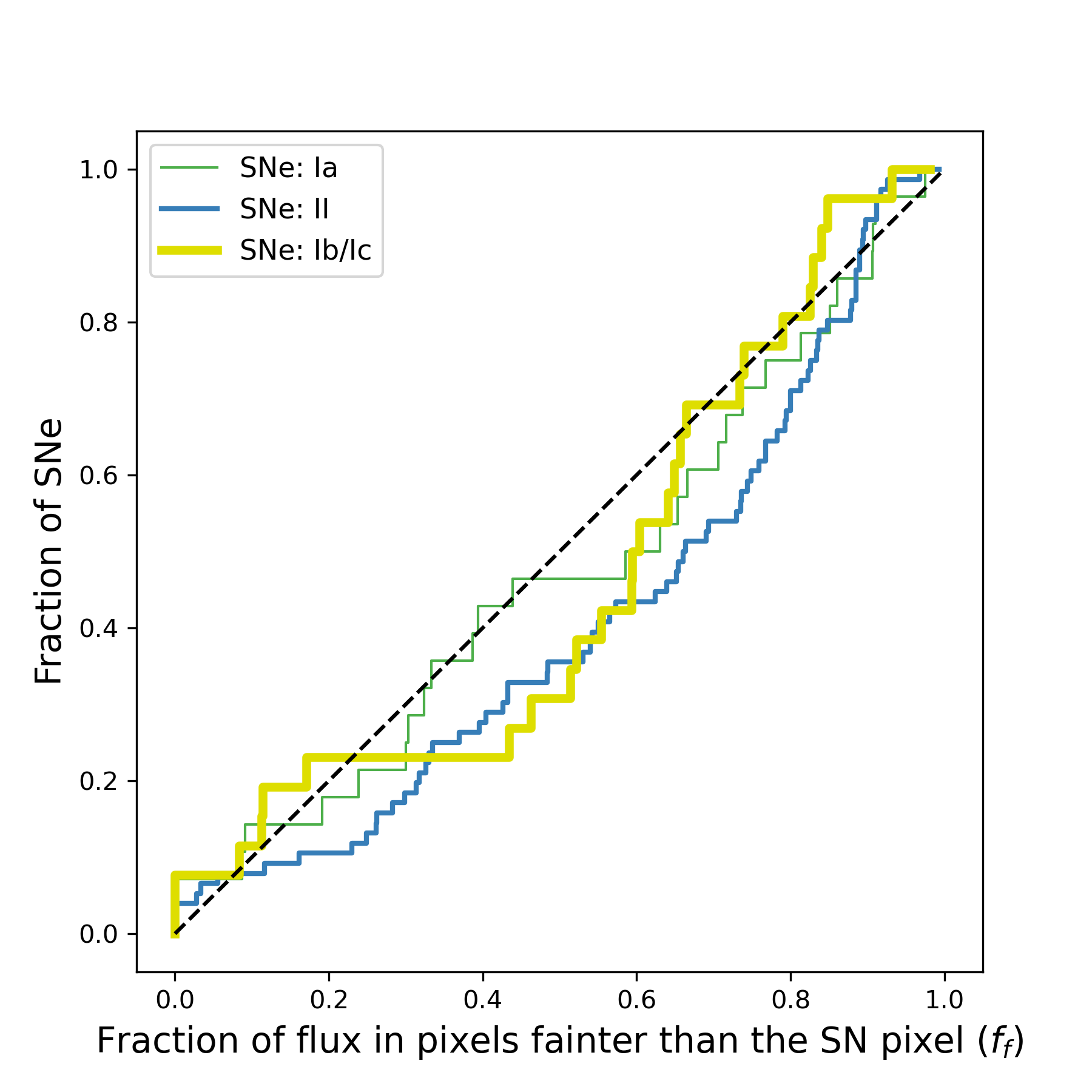}\\
\end{tabular}
\caption{Cumulative distributions of fractions of pixels fainter than the SN pixel, $f_p$ (left) and the fractions of total flux, $f_f$ (right), for all types of SNe. The dashed lines denote the $1-1$ line. For the left plot ($f_p$) this corresponds to the situation in which SNe follow a completely random distribution, while for the plot on the right ($f_f$), it means they follow directly the atomic gas distribution.}
\label{fig:cumulhist}
\end{figure*}

Fig.~\ref{fig:cumulhist} shows the cumulative distributions of $f_p$ and $f_f$ values for each SN type. Tab. \ref{tab:ksfp} presents the results of the KS tests, via their p-values. The diagonal in both tables shows the probability for the one-sided KS test, in which the distribution of a given SN type is compared with the $1-1$ model. For the plot showing the $f_p$ values this line corresponds to SNe being positioned randomly within the host galaxy, while for the plot showing the $f_f$ values it corresponds to SNe tracing the {\hi} distribution of galaxies directly. The non-diagonal entries in the tables show the probabilities of the two-sided KS tests, assessing whether two given SN types could be drawn from the same parent population (i.e. the same probability distribution).

The cumulative distribution plot of $f_p$ values (Fig.~\ref{fig:cumulhist}, left) shows clearly that all three types of SNe deviate strongly from the $1-1$ line, indicating they do not follow a random distribution within the host galaxies. This is further strengthened by the very small values of the p-value of the KS test presented in Tab.~\ref{tab:ksfp}. The differences between the different types of SNe, however, are far less pronounced. The KS test is consistent with all three types of SNe coming from the same distribution.

The cumulative distribution plot of $f_f$ values (Fig.~\ref{fig:cumulhist}, right) traces how well the SN types follow the {\hi} distribution of the host galaxies. Out of all three types of SNe, Type II seems to deviate the most from the $1-1$ line. This would suggest that Type II SNe prefer positions with an even greater abundance of {\hi} gas. This is also seen from the value of the KS test for this population ($p = 0.0040$, see Tab. \ref{tab:ksff}). The same cannot be said for the other two types of SNe. However, a direct comparison between different types of SN events is still consistent with all three of them being drawn from the same underlying distribution.

\subsection{Comparison with simulations}
\label{Sect:SimulationsREsults}

To further explore the locations of the SN events, we compared the observed SNe sample with simulations (following \citealt{Chen2023}, \citealt{Chen2024}). For this purpose, we created three different simulated samples of SNe, following different positional distributions within the {\hi} maps. We then compared these samples with real observations, repeating the methodology from previous sections, using the same {\hi} maps.

Firstly, in order to further test the validity of the methods presented in the last subsection, we simulated two samples of SNe, one following a completely random positional distribution, and the second following the {\hi} emission of the parent galaxies. The first of these samples reproduces the $1-1$ line in the $f_p$ plot, while the second distribution, selected to follow the {\hi} gas emission, traces the $1-1$ line in the $f_f$ plot of Fig \ref{fig:cumulhist}. As expected, the comparison between these simulated samples and the observed SNe data, produced results that were qualitatively identical to the ones presented in Tab. \ref{tab:ksff}. The exact values are listed in Tab. \ref{tab:ksfpffsim}.

\begin{figure*}
\centering
\begin{tabular}{cc}
\includegraphics[width=0.5\textwidth]{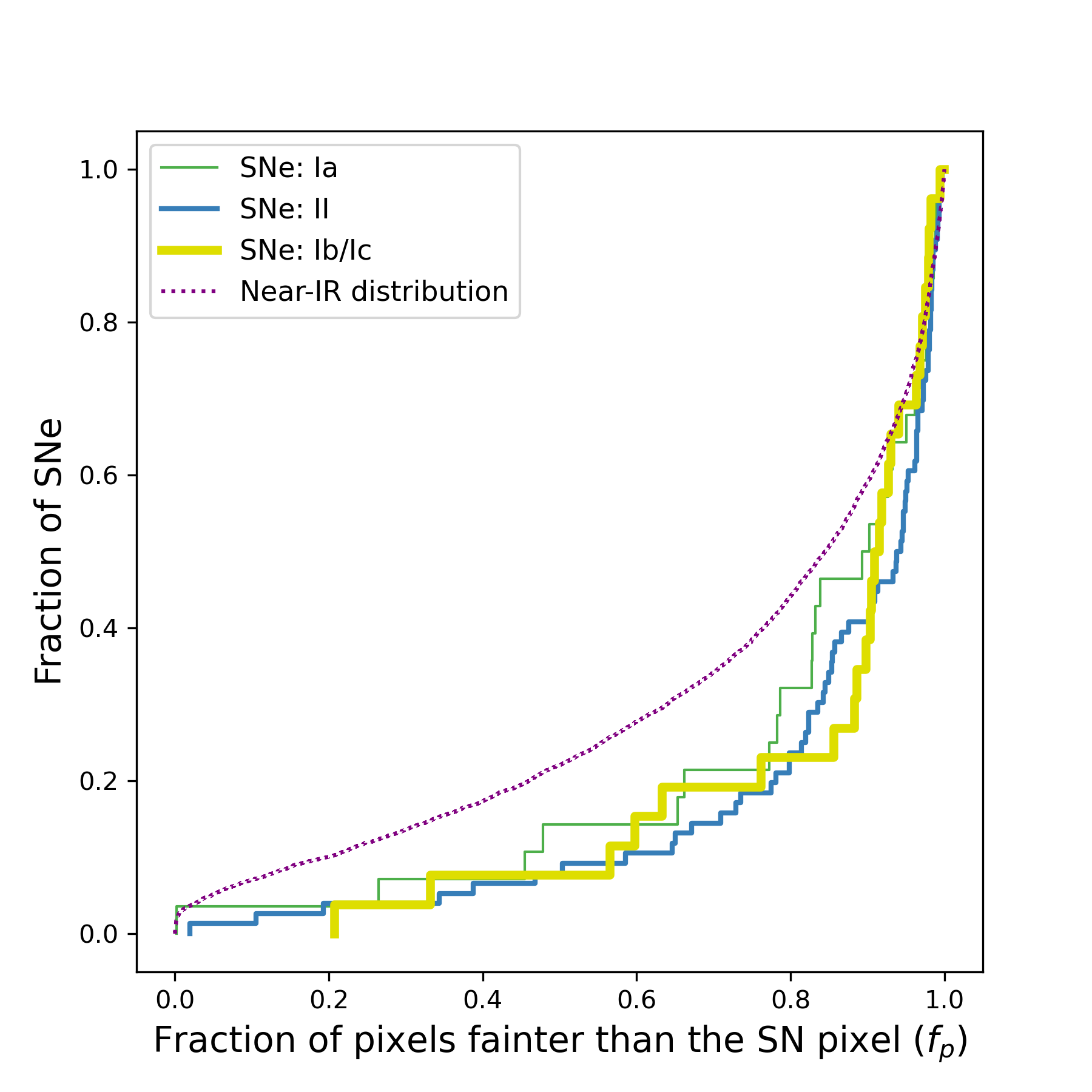} &
\includegraphics[width=0.5\textwidth]{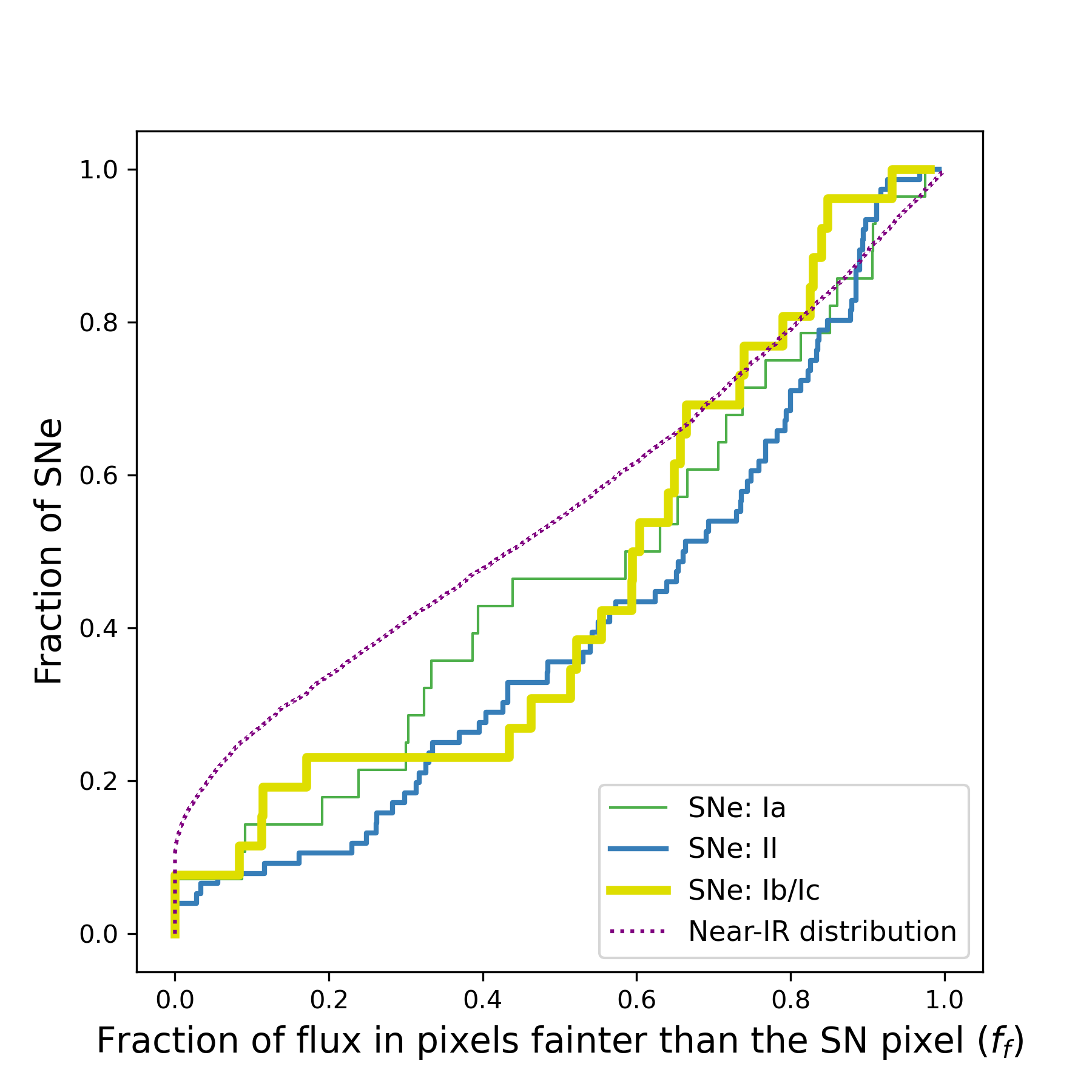}\\
\end{tabular}
\caption{Cumulative distributions of $f_p$ (left) and $f_f$ (right) values for observed (solid lines) SNe and the simulation following the Near-IR emission (dashed line), as denoted in the legend. }
\label{fig:cumulhist_sim}
\end{figure*}

Secondly, we simulated a sample of SNe that follows the stellar mass of the host galaxies. For this purpose, we used Near-Infrared (Near-IR) maps from the Infrared Array Camera (IRAC; \citealt{irac}) of the Spitzer telescope, observed at $3.6 \ \mathrm{\mu m}$, as it is a good tracer of the stellar mass distribution in galaxies. The data comes from the Spitzer Survey of Stellar Structure in Galaxies\footnote{DOI: 10.26131/IRSA425} (S$^4$G) and is described in detail in \citet{Sheth2010} (see also \citealt{Munoz2013}, \citealt{Querejeta2015}, \citealt{Watkins2022}). A clear advantage of this data set is the existence of masks for foreground and background objects for each observed galaxy (S$^4$G Pipeline $2$, \citealt{Sheth2010}), leaving us with a better estimate of the stellar distributions in host galaxies, required for our simulations. Although the dataset does not encompass the complete set of host galaxies used in this work (covering $47$ of the $74$ host galaxies), we decided to prioritize the purity of the data, deeming the number of IRAC-detected galaxies large enough to create a relevant simulated sample. A comparison between the observed and simulated data, using the same {\hi} maps, is presented in Fig.~\ref{fig:cumulhist_sim}, with the KS p-values listed in Tab. \ref{tab:ksfpffsim}. The figure shows both the $f_p$ and $f_f$ cumulative distribution plots. The strongest deviation from the Near-IR distributed simulations was observed for Type II SN. On the other hand, Type Ia SN seem to be consistent with the Near-IR simulated samples. The results for Type Ib/c SN are less conclusive. It is noteworthy that both the Ib/c and the Ia SN samples are smaller than the sample of Type II SN, which influences the results of the KS tests.

\begin{figure*}
\centering
\begin{tabular}{cc}
\includegraphics[width=0.9\textwidth]{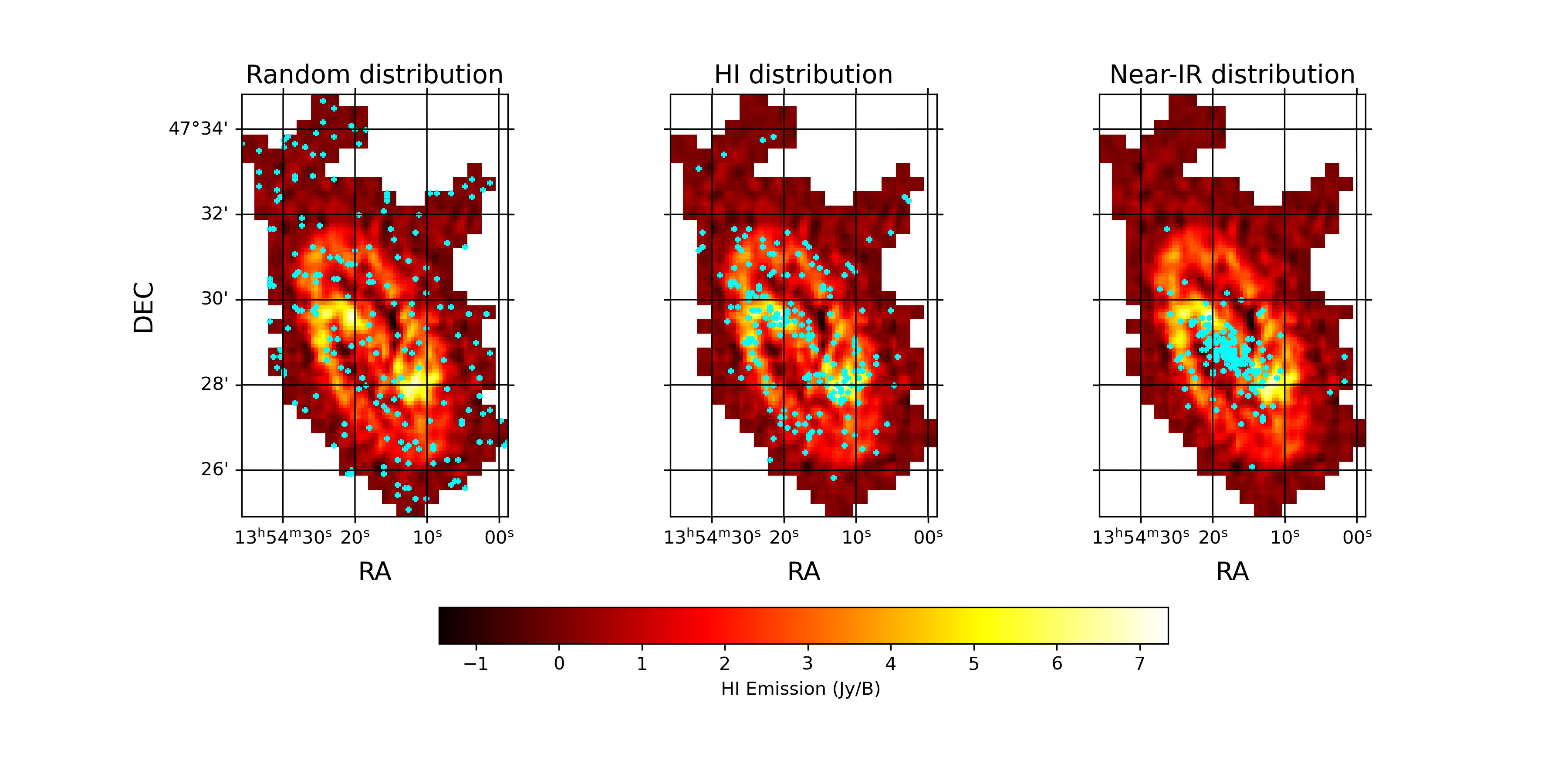}
\end{tabular}
\caption{Positions of simulated SNe (blue crosses) shown over maps of {\hi} emission. The figure corresponds to HI emission of a host galaxy NGC5377. The left part of the figure shows the purely random distribution of SN positions. The middle part of the figure shows the scenario where the SN positions follow the distribution of {\hi} emission. The right part of the figure shows the scenario where the positions follow the stellar distribution of the host galaxy, traced by the Near-IR IRAC $3.6 \ \mathrm{\mu m}$ emission. See Section \ref{Sect:SimulationsREsults} for more details. }
\label{fig:example_sim_maps}
\end{figure*}

The number of simulated SNe was set to $200$ for all of the above-described distributions. An example of the simulated sample for a single host galaxy is shown in Fig.~\ref{fig:example_sim_maps}, where the positions of the simulated SNe are shown over a map of the {\hi} emission.

Lastly, in order to test the validity of the KS test for our purposes, we re-estimated the $f_f$ comparison between different types of SNe, as well as the comparison of our observations with the simulations, using the Anderson-Darling test. The results remained the same (with slightly more agreement between the Near-IR sample and Type Ib/c SNe; see Tab. \ref{tab:ADsim} and Tab. \ref{tab:AD} in the appendix).

\begin{table*}
\caption{KS test $p$-values for $f_p$ and $f_f$ between each SN type and the simulations.}
\small
\center
\begin{tabular}{cccc}
\hline \hline
$f_p$ \vline & Ia & II & Ib/c \\ \hline
Random & $2.2\times10^{-9}$  & $2.5\times10^{-26}$  & $3.9\times10^{-11}$   \\ 
HI & 0.85  & 0.007   & 0.09  \\
Near-IR & 0.12  &  0.0002  & 0.0096   \\ 
\hline
\end{tabular}
\quad
\quad
\quad
\begin{tabular}{cccc}
\hline \hline
$f_f$ \vline & Ia & II & Ib/c \\ \hline
Random & $8.6\times10^{-10}$  & $8.6\times10^{-30}$  & $4.1\times10^{-11}$   \\ 
HI & 0.54  & 0.005   & 0.14  \\
Near-IR & 0.13  &  $2.7\times10^{-5}$  & 0.018   \\ 
\hline
\end{tabular}
\label{tab:ksfpffsim}
\end{table*}

\section{Discussion}

If the positional distribution of SN events, for any type of SN, were completely random, then the cumulative distribution of the $f_p$ values would trace the $1-1$ line and the mean $f_p$ would be around $0.5$. On the other hand, if the frequency of SN explosions of a given type is proportional to the amount of atomic gas at a given location, then the $f_f$ distributions of the pixel {\hi} fluxes would follow the $1-1$ line (Fig.~\ref{fig:cumulhist}). The corresponding mean values of the fractional $f_f$ metric would then be $0.5$.

Our results show that, for all SN types, the mean and median values of $f_p$ (fraction of pixels fainter than the SN pixel) are consistently higher than 0.5 (Table~\ref{tab:mean}) and the one-sided KS tests reveal the distribution to be inconsistent with the $1-1$ line (Table ~\ref{tab:ksfp} and Fig.~\ref{fig:cumulhist}). This can be interpreted as all SN types preferring brighter {\hi} pixels. Physically, this could be because {\hi} disks of galaxies are usually several times larger than the optical disks \citep[e.g.][]{wang13}. Most SNe explodes inside the optical disk \citep{maza76}, where {\hi} density is higher, so the distribution of {\hi} values must be skewed towards high values. A consistent trend is seen from the $f_f$ cumulative distribution (Fig.~\ref{fig:cumulhist}), since for all the SN types the $f_f$ distributions are closer to the $1-1$ line than the $f_p$ values. We can therefore deduce that the positional distribution of SN events follows {\hi} more directly.

Furthermore, Type II SNe seem to deviate from the {\hi} distribution towards positions with even greater {\hi} content. This could imply they follow a different distribution, one that prioritizes {\hi} gas of galaxies more strongly than for the other two types of SNe. The deviation from the $1-1$ line was significant at a $4\sigma$ level, as seen from the KS test. This would imply that SN Type II do prefer environments rich in atomic gas, in a similar way as GRBs prefer UV-bright pixels \citep{fruchter06}. This would, in turn, be consistent with atomic gas being the fuel for the formation of their progenitors, either indirectly (through the transition through molecular gas), or directly \citep{fumagalli08,bigiel10,glover12,krumholz12,hu16,elmegreen16,elmegreen18}. 

On the other hand, this deviation for Type II SN was not obtained from a direct comparison between the three types of SNe, using the KS test. For this diagnostic, we can treat the distribution of Type Ia SNe as the reference point. This is because the evolution of the progenitors of Type Ia SNe is long enough to assume that their position should not be physically connected with the current distribution of gas (\citealt{Maoz2014}, \citealt{Anderson2015_2}). Any enhancement of the $f_p$ or $f_f$ values for another SN type can be interpreted as the preference of the SN progenitors to be born in {\hi}-rich environments. Since we have not found any significant deviation of the distributions of atomic gas of Type II or Ib/c from that of Type Ia SNe, it follows that the observed samples are consistent with being drawn from the same underlying distribution. This would imply conversely that Type II SNe are not connected with the densest concentrations of atomic gas, unlike what has been suggested for GRBs and Type Ic-BL SNe \citep{michalowski15hi,michalowski16,michalowski18,michalowski20,arabsalmani15b,arabsalmani19}. Hence, the birth of progenitors of Type II or Ib/c SNe would be connected with the current star formation in their hosts, instead of to increased star-formation activity due to an inflow of metal-poor gas, suggested for GRBs and Type Ic-BL SNe. For Type Ib/c SN, this is also supported by the one-sided KS test, finding no significant deviation for the $f_f$ values from the $1-1$ line ($p = 0.13$, see Tab. \ref{tab:ksff}). The direct comparisons between the different types of SN are, however, influenced by the sample size of our observations. Both the Ia and the Ib/c samples are much smaller than the Type II sample. A larger sample of SNe could provide further insight into the underlying distributions of three different types of SNe.

\subsection{Insight from simulated samples}

The observed trends, discussed above, were further explored by comparing observations with the simulated samples. Firstly, using simulated SN samples which followed a random positional distribution and a positional distribution following {\hi} content of the galaxy, we reproduced all the trends derived via comparisons with the theoretical $1-1$ lines.

Furthermore, we observed a deviation of Type II SNe from the Near-IR simulated samples. This deviation implies that the Type II SNe do not follow the stellar mass distribution of the host galaxy. It can also be seen from Fig.~\ref{fig:cumulhist_sim} and Tab.~\ref{tab:ksfpffsim} that this difference is even stronger than the deviation from a population following directly the {\hi} gas, implying a larger positional difference. Physically, it would mean that Type II SNe occur in areas with a high abundance of atomic gas, instead of following the stellar content of the host galaxy. This could place our study among similar research that suggested such a trend for GRBs (\citealt{michalowski15hi,michalowski16,michalowski18,michalowski20,arabsalmani15b,arabsalmani19}.)

The results for Type Ib/c SNe are less significant. The KS $p$-values in Tab.~\ref{tab:ksfpffsim} could imply a positional difference, but the results remain uncertain. This could be due to the smaller size of the sample available for this type of SNe.

Type Ia SN do not seem to deviate from the Near-IR distributed simulations, as seen from the obtained $p$-values. Again, a larger sample of these objects would greatly improve these results, as the KS test is affected by the size of the sample. The trend observed for these objects is also consistent with a physical picture of their formation (arising from a long-lived binary system; \citealt{Maoz2014}). In other words, as already stated, owing to the longer evolution of the progenitors of Type Ia SNe, it is reasonable to assume that their position is not correlated with gas over-densities of the host galaxies (e.g. \citealt{Anderson2015_2}).

\subsection{Limits of the survey}
\label{sect:limits}
There are several caveats in our analysis. Firstly, the sample size is limited by the availability of {\hi} observations of SN hosts. If the difference between the locations of SNe of different types is minor, it may become statistically insignificant with our sample size. The situation will improve with on-going and future wide, relatively deep, and high-resolution {\hi} surveys, such as the {\hi} survey with APERture Tile In Focus (APERTIF) on the Westerbork Synthesis Radio Telescope (WSRT; e.g., \citealt{hess21,morganti21}); the Widefield ASKAP L-band Legacy All-sky Blind surveY (WALLABY; \citealt{koribalski20}); MeerKAT International GigaHertz Tiered Extragalactic Exploration survey (MIGHTEE; \citealt{maddox21}),  MeerKAT {\hi} Observations
of Nearby Galactic Objects: Observing Southern Emitters (MHONGOOSE; \citealt{deblok20}), and finally with the Square Kilometre Array. We note also that the {\hi} maps come from two different surveys, the WISP and THINGS, including different samples of galaxies (with the WHISP survey concentrating on dwarf galaxies), as described in Sect. \ref{sect:section2}.  

Secondly, the currently available data we are using may be of too poor resolution. The typical resolution corresponds to several hundred pc, reaching a few kpc for the most distant galaxies (Table~\ref{tab:table} in Appendix \ref{app:A}). This is the resolution at which atomic gas concentrations were detected in the studies on hosts of GRBs and SN Type Ic-BL mentioned above, but the concentrations may also be present at lower scales. This will also be addressed with future surveys, delivering data at arcsec or sub-arcsec resolutions. 

Furthermore, the uneven resolution of the {\hi} maps (varying from galaxy to galaxy) means we trace the correlation between SN events and their environment at different spatial scales. In order to further explore this, we decided to re-estimate the $f_f$ values for Type II SNe, using only high-resolution data. We concentrated on Type II SNe since they alone exhibited a significant deviation from the $1-1$ line, and since only for them the sample was big enough for further subdivisions. By using only {\hi} maps with linear resolution $< 1 \mathrm{kpc}$, the deviation from the $1-1$ line disappears (with a p-value of the one-sided test of $p = 0.57$).

In order to test the effect of resolution on our results, we artificially decreased the resolution of the high resolution maps, and re-estimated the $f_f$ values for Type II SNe. For this test we used only the maps with the highest resolution of $< 0.5\, \mathrm{kpc}$. The resolution was worsened by convolving the images with an appropriate Gaussian, setting the final linear resolution of all the maps first to $1\, \mathrm{kpc}$, and then to $2\, \mathrm{kpc}$. The KS p-values between the $1-1$ line and the Type II subsample dropped from $p=0.61$ for high resolution maps, to $p = 0.012$ for $1\, \mathrm{kpc}$ maps, and $p = 9.11 \cdot 10^{-5}$ for the $2\, \mathrm{kpc}$ maps, indicating a larger deviation from the $1-1$ line as the resolution worsens. It follows therefore, that the deviation of Type II SNe, reported in Sect. \ref{sect:section3} was an effect present only at larger spatial scales, not observable on {\hi} maps with the highest resolution. At this point, however, we do not have a large enough sample of high resolution data to explore this further. We, furthermore, did not wish to artificially worsen the high resolution maps, opting instead to always use the best possible resolution available. The uneven resolutions are therefore left as a caveat of this study. Lastly, no direct correlation was found between the linear resolution and the $f_f$ values, with the Pearson and Spearman coefficients being $p = 0.007$ and $p = 0.001$, respectively.

\section{Conclusions}

We have analyzed the atomic gas data for hosts of 133 SNe (29 Ia, 77 II, 27 Ib/c). We have found that all three types deviate strongly from the completely random positional distribution. Type II SN, furthermore, seemed to deviate also from a direct {\hi} distribution, preferring even higher abundances of atomic gas, although this result was complicated by the uneven resolution of the host galaxy images. We also observed, by comparison with simulations, that Type II SNe deviate from the stellar distribution of the host galaxies. Although less significant, a similar trend of deviating from the stellar content of the host galaxy could also be present for Type Ib/c SNe. 

On the other hand, a direct comparison between the three types of SNe was consistent with them being drawn from the same underlying sample. We observed no statistically significant deviation between these samples, which is affected by small sample sizes for Type Ia and Ib/c SNe. Furthermore, as discussed in Sect. \ref{sect:limits}, the deviation of Type II SN towards even more atomic gas, was induced by the inclusion of low resolution maps, and was not reproducible by using only the subsample with high resolution {\hi} maps. A further increase in the sample size of SNe and better resolution of {\hi} maps will further improve this results.

This is the first statistically significant survey of HI gas abundance at the positions of detected SNe. We fail to find a clear connection of Ib/c core-collapse SNe with the densest concentrations of atomic gas in their hosts, as has been claimed for GRBs and Type Ic-BL SNe. The results for Type II SN are less clear. While the KS test provides us with reasons to believe this population of SNe prefers over-densities of neutral hydrogen even greater than following the {\hi} distribution directly, the conclusions are complicated by the limits of the survey, namely the uneven resolution and the limited sample size. Hence, the progenitors of regular Type Ib/c SNe, and possibly also Type II SNe, are still consistent with being connected to the current level of star formation, whereas the progenitors of GRBs and  Type Ic-BL SNe require more special conditions to form, for example low metallicity.

\begin{acknowledgements}
M.J.M.~and A.L.~acknowledge the support of the National Science Centre, Poland through the SONATA BIS grant 2018/30/E/ST9/00208. 
M.J.M.~acknowledges the support of
the Polish National Agency for Academic Exchange Bekker grant BPN/BEK/2022/1/00110.
 This research was funded in whole or in part by National Science Centre, Poland through the grants 2023/49/B/ST9/00066 and 2024/53/N/ST9/00350 and . Supported by the Foundation for Polish Science (FNP).

The National Radio Astronomy Observatory is a facility of the National Science Foundation operated under cooperative agreement by Associated Universities, Inc. 
The Westerbork Synthesis Radio Telescope is operated
by the ASTRON (Netherlands Institute for Radio Astronomy)
with support from the Netherlands Foundation for Scientific
Research (NWO).
This work made use of THINGS, `The HI Nearby Galaxy Survey' \citep{walter08}.
This research has made use of 
the NASA/IPAC Extragalactic Database (NED) which is operated by the Jet Propulsion Laboratory, California Institute of Technology, under contract with the National Aeronautics and Space Administration;
SAOImage DS9, developed by Smithsonian Astrophysical Observatory \citep{ds9};
Edward Wright cosmology calculator \citep{wrightcalc};
and NASA's Astrophysics Data System Bibliographic Services. This research was funded in whole or in part by National Science Centre, Poland through the PRELUDIUM grant 2024/53/N/ST9/00350. Supported by the Foundation for Polish Science (FNP).
\end{acknowledgements}

\bibliographystyle{aa}  
\bibliography{bib}

\begin{appendix}
\section{Long Figures and Tables}
\label{app:A}

Moment 0 {\hi} maps are shown in Fig.~\ref{fig:mapIa}, for each host galaxy. Tab. \ref{tab:table} shows the redshifts, resolutions, and the $f_p$ and $f_f$ statistics for individual SN. The linear resolutions were determined via distances from the HyperLeda database\footnote{{\tt http://atlas.obs-hp.fr/hyperleda/}} and the  NASA/IPAC Extragalactic Database.

\begin{figure}[!b]
\centering
\begin{tabular}{llll}
\includegraphics[width=0.25\textwidth]{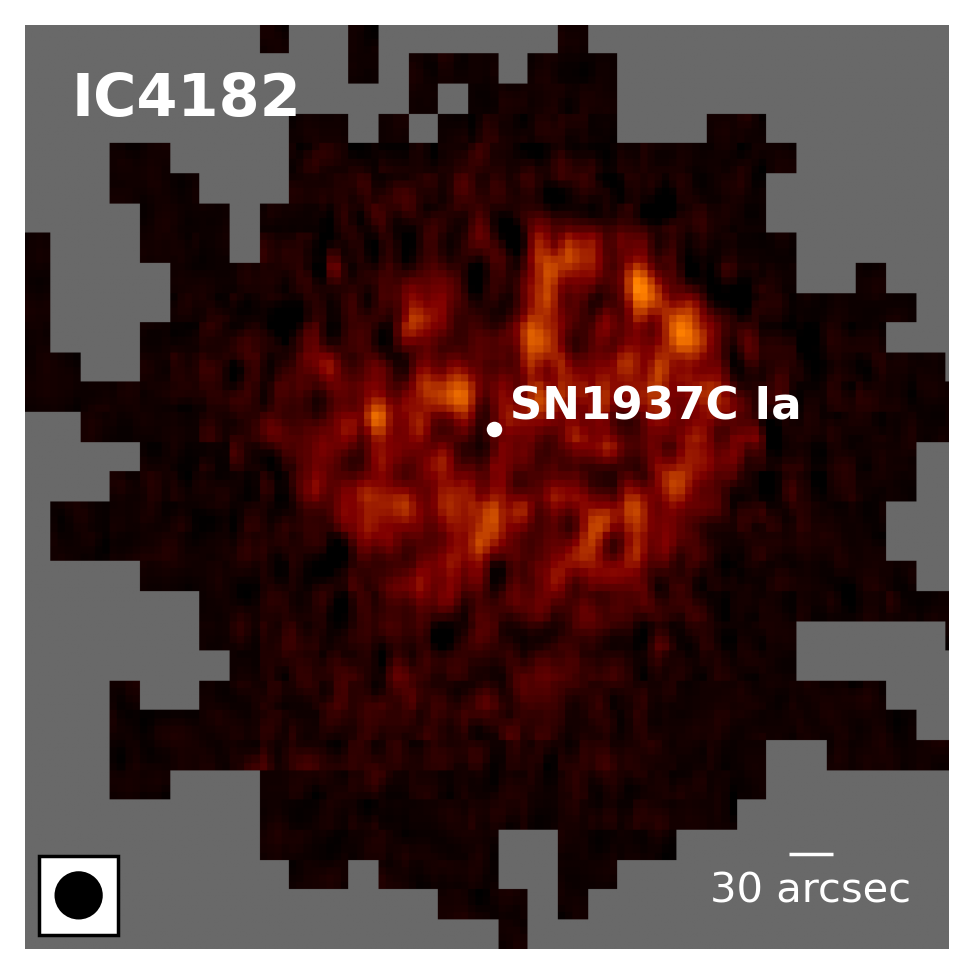}
\includegraphics[width=0.25\textwidth]{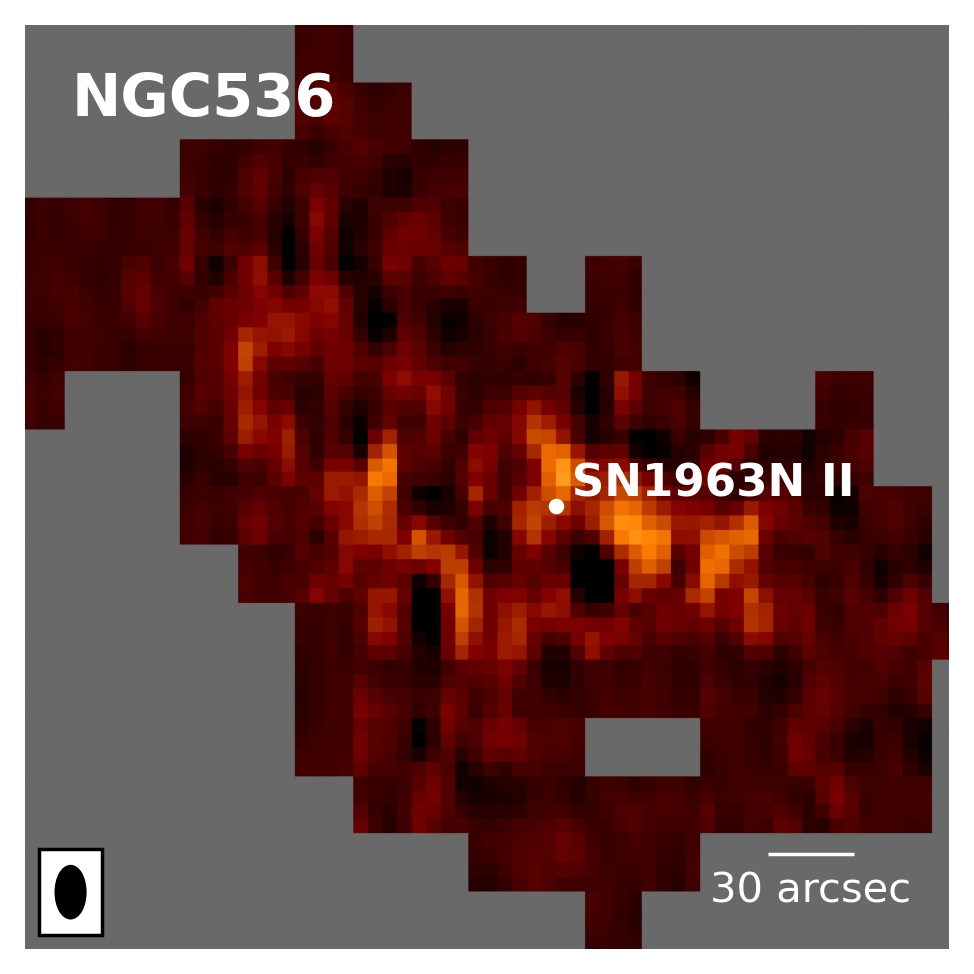}
\includegraphics[width=0.25\textwidth]{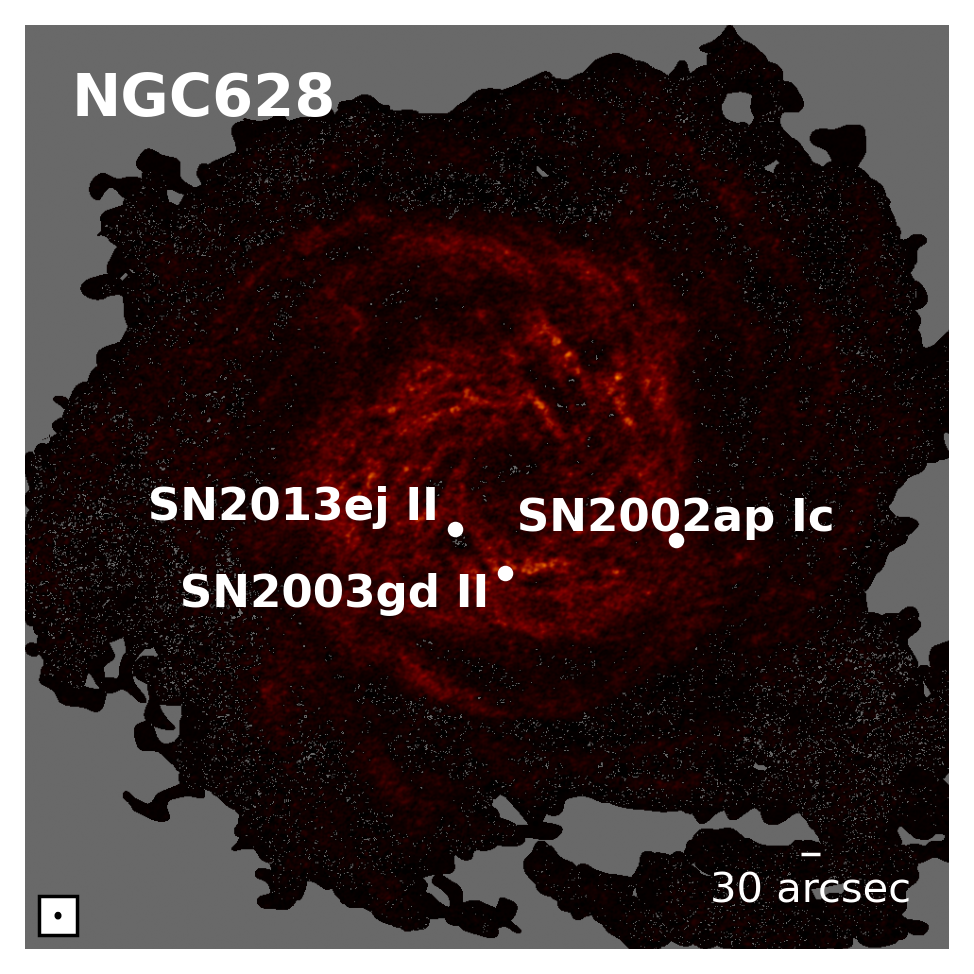}
\includegraphics[width=0.25\textwidth]{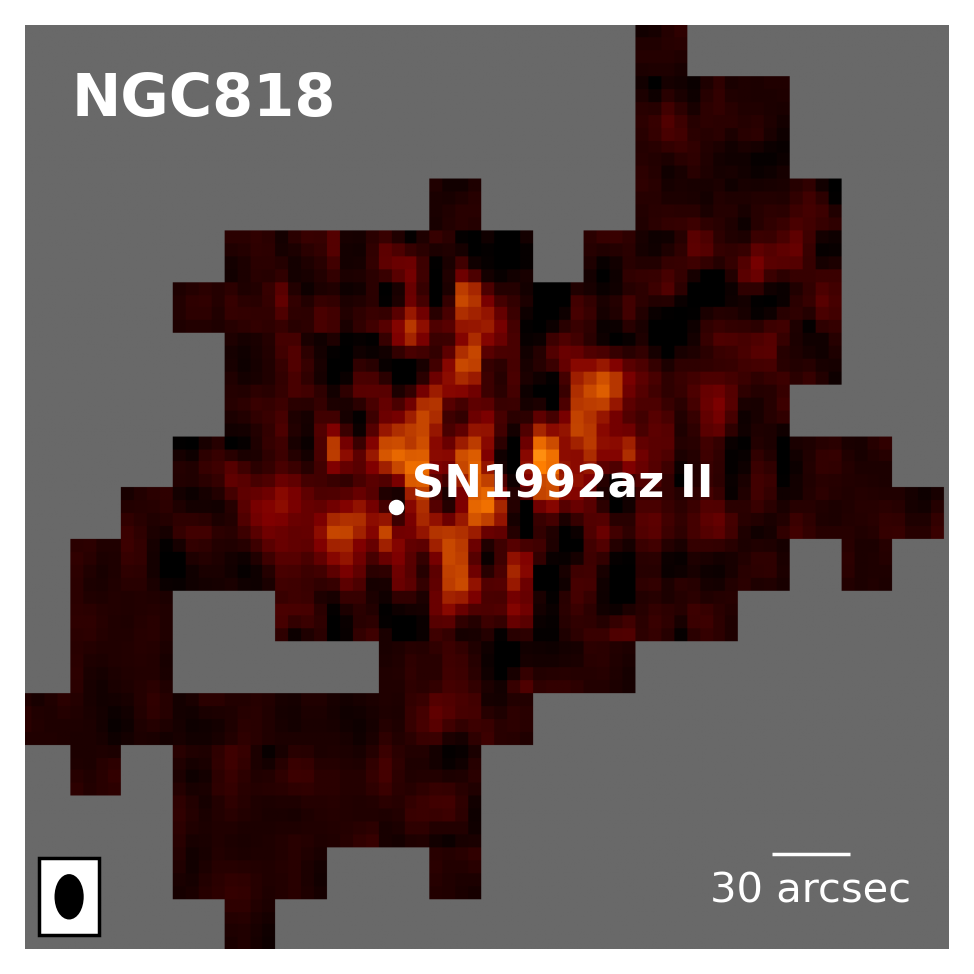}\\
\includegraphics[width=0.25\textwidth]{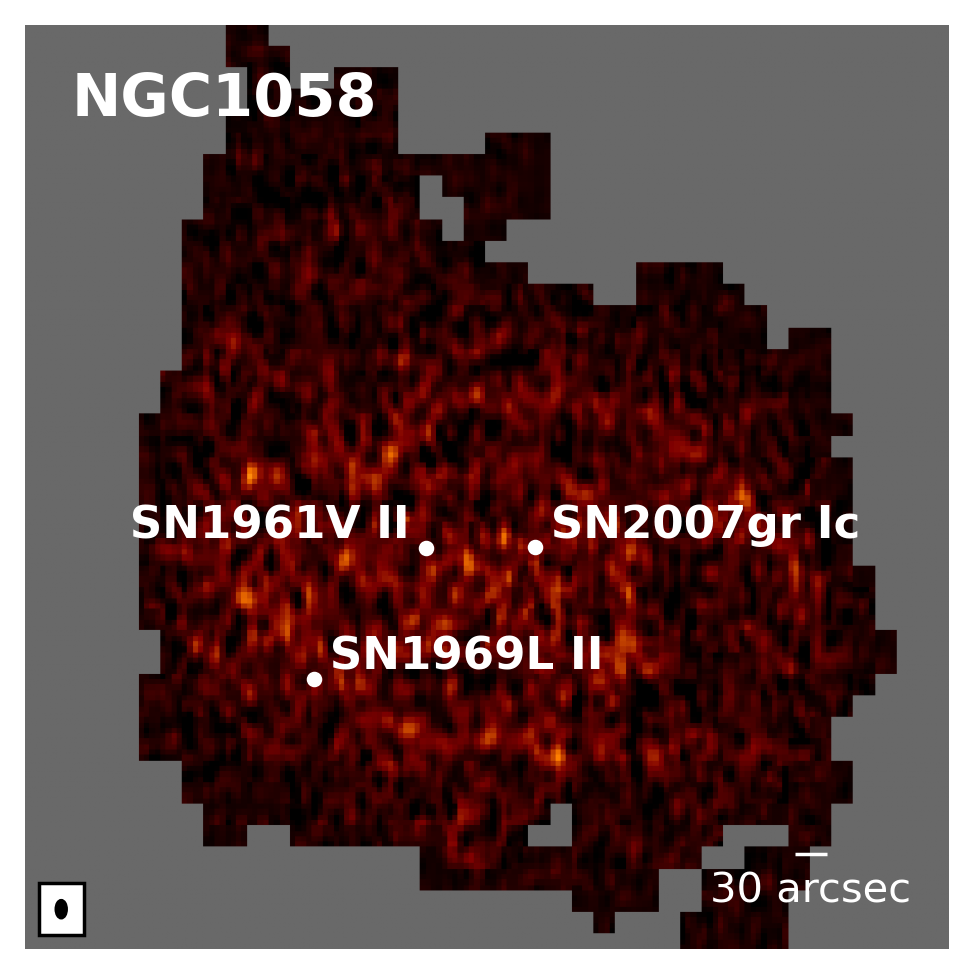}
\includegraphics[width=0.25\textwidth]{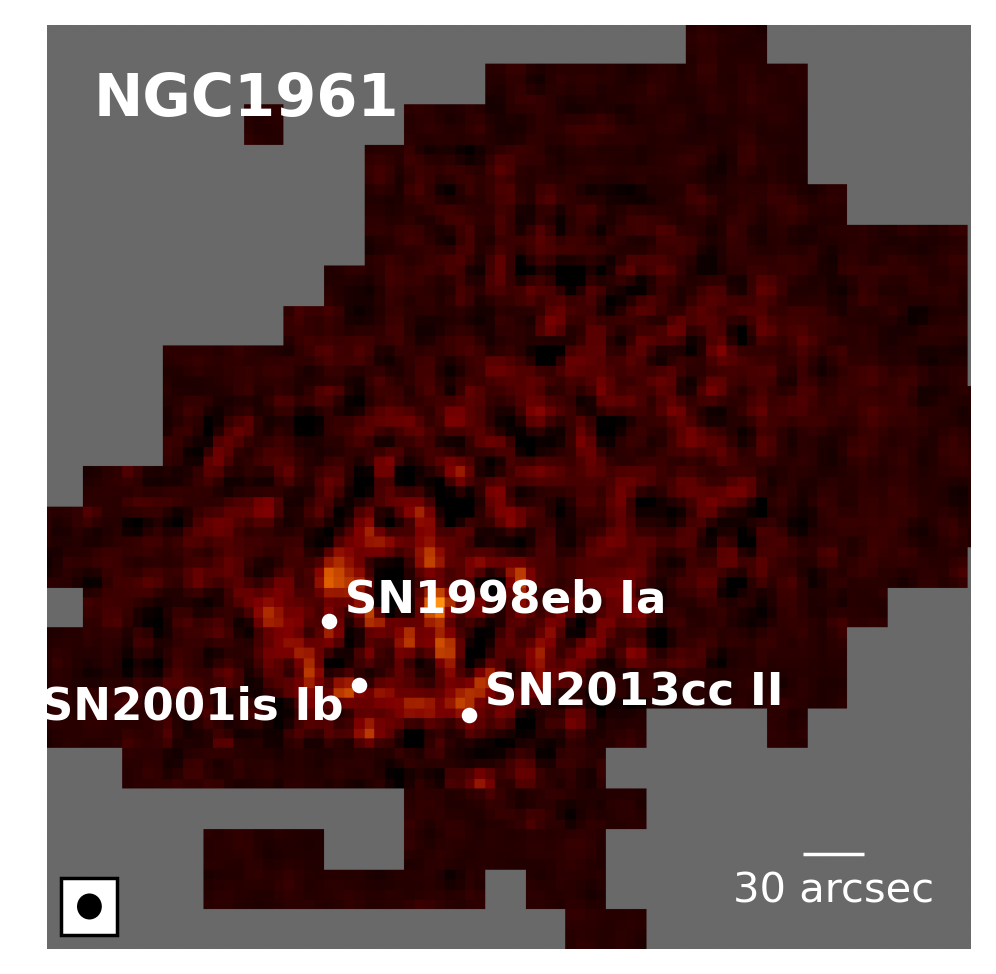}
\includegraphics[width=0.25\textwidth]{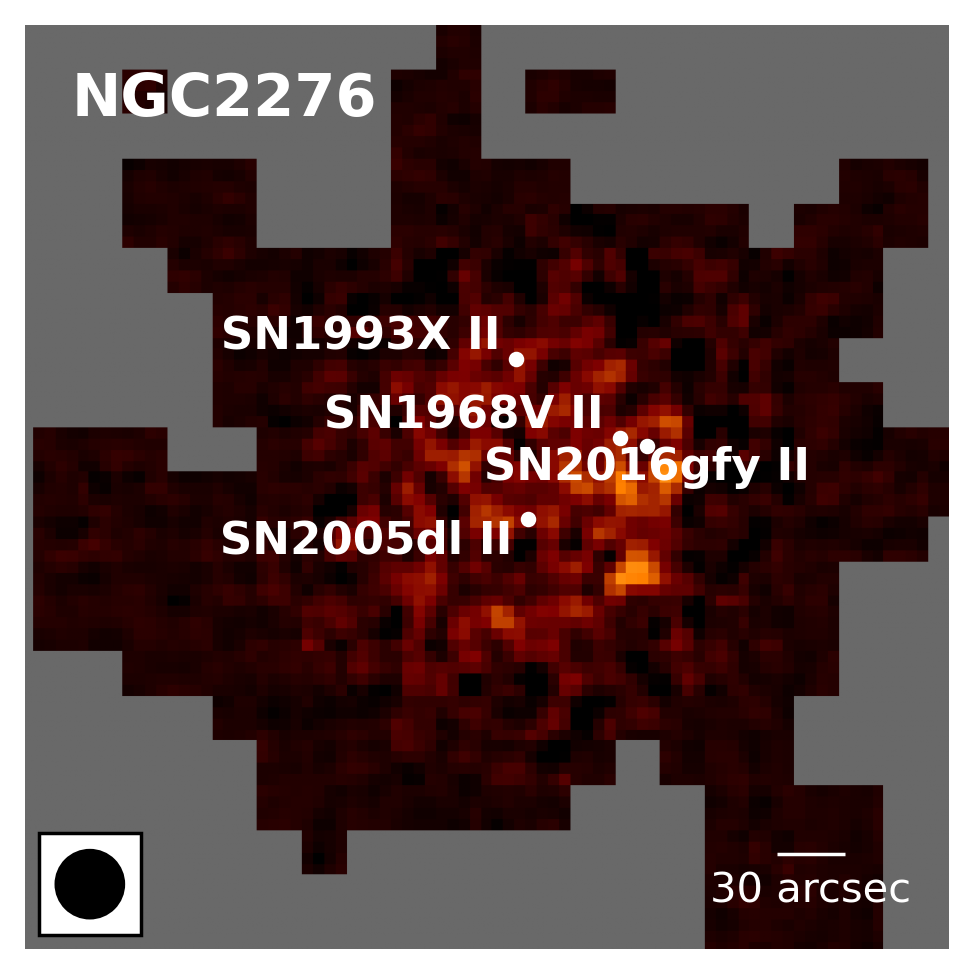}
\includegraphics[width=0.25\textwidth]{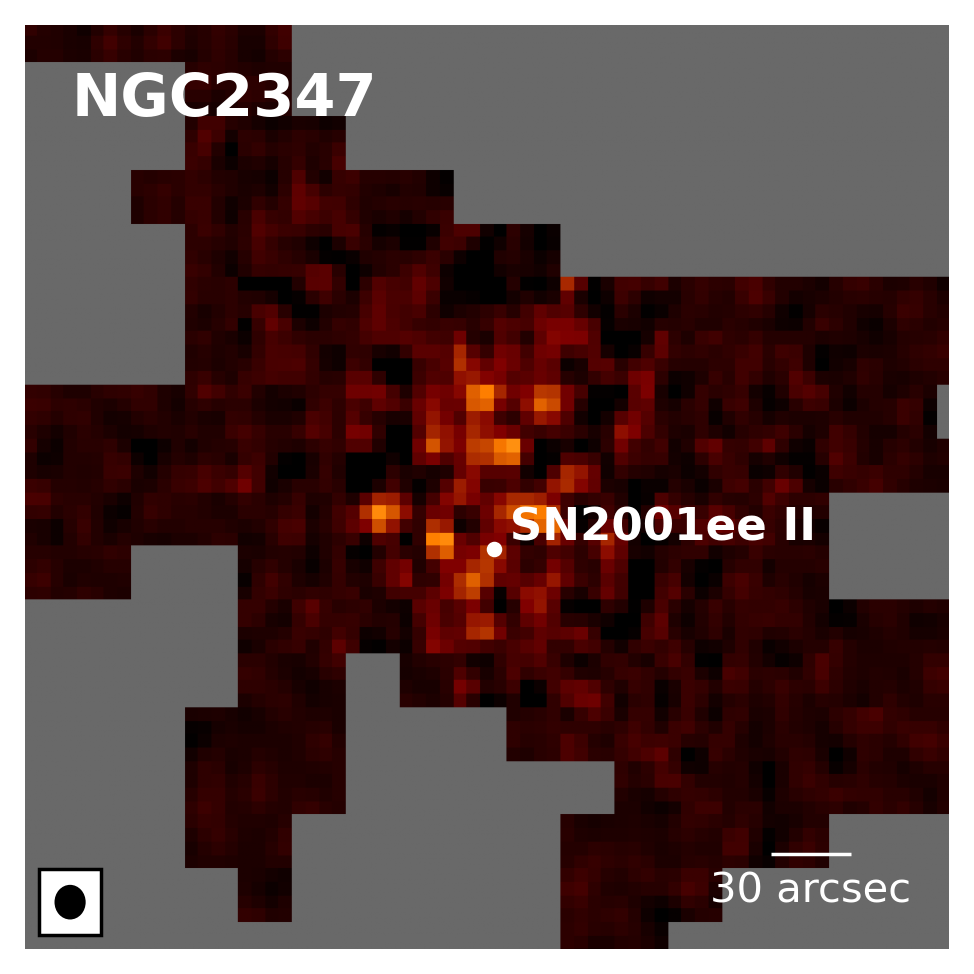}\\
\includegraphics[width=0.25\textwidth]{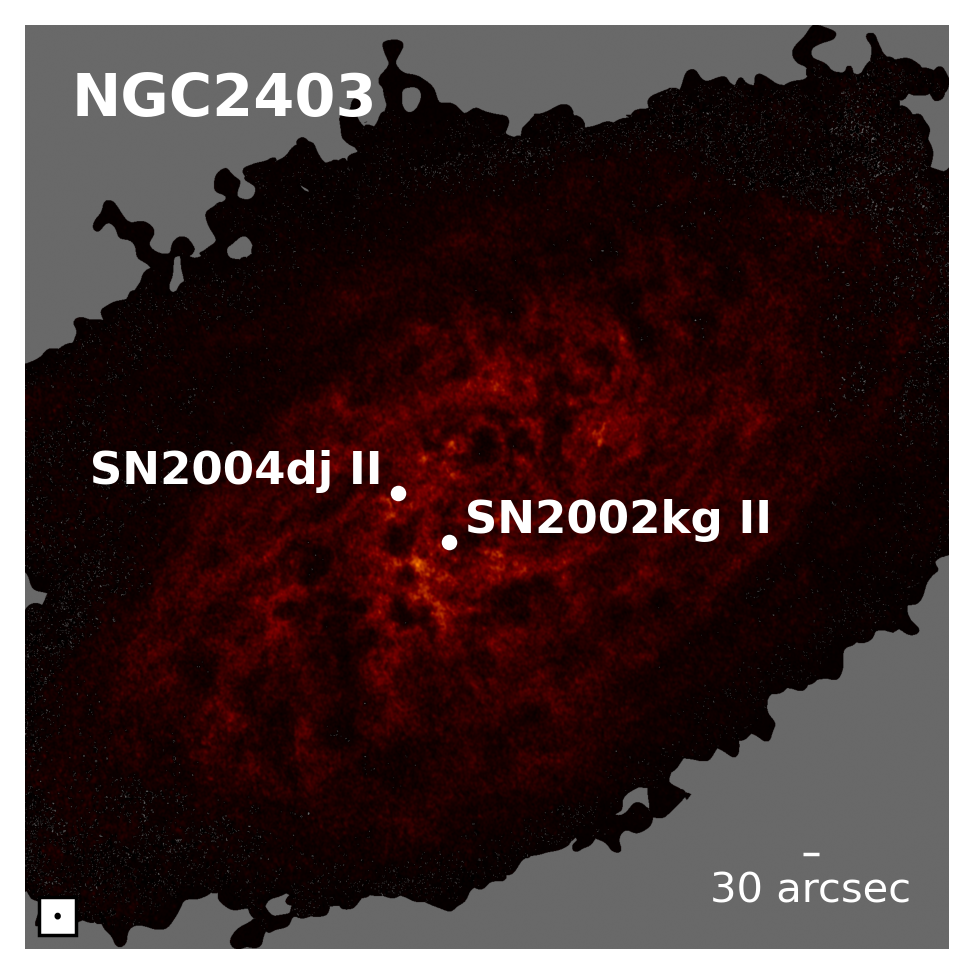}
\includegraphics[width=0.25\textwidth]{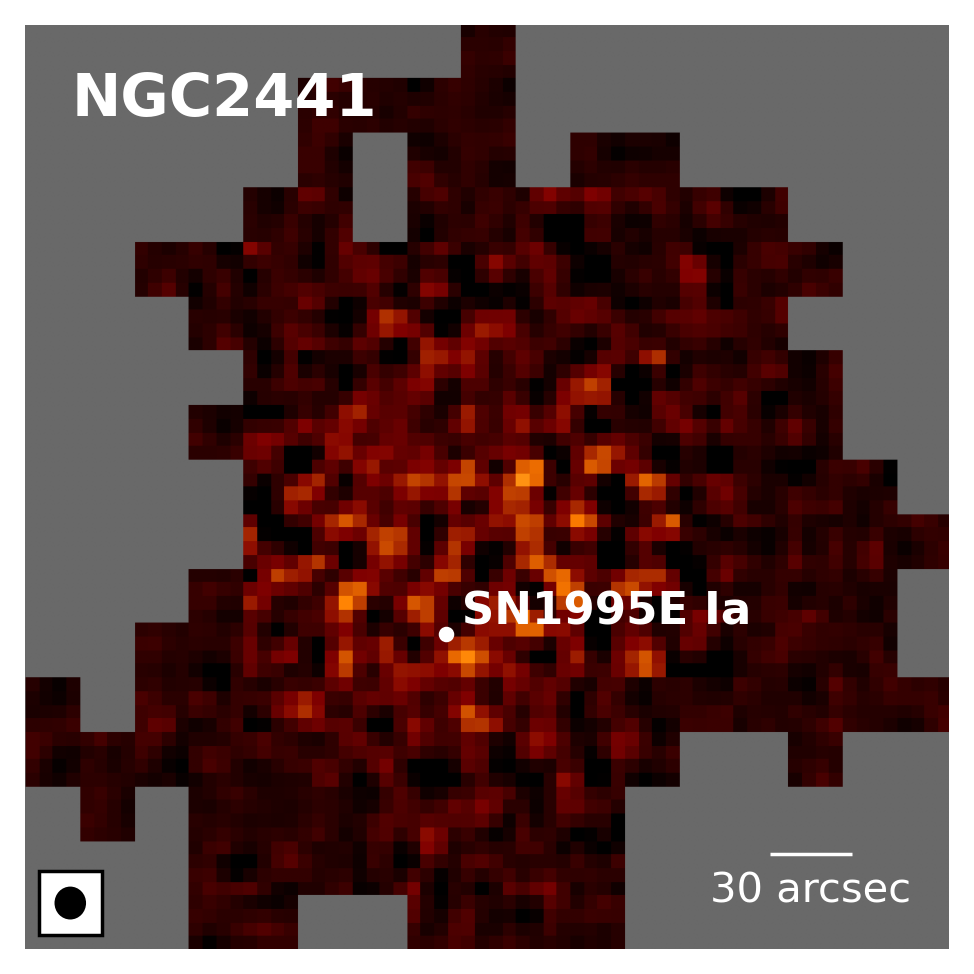}
\includegraphics[width=0.25\textwidth]{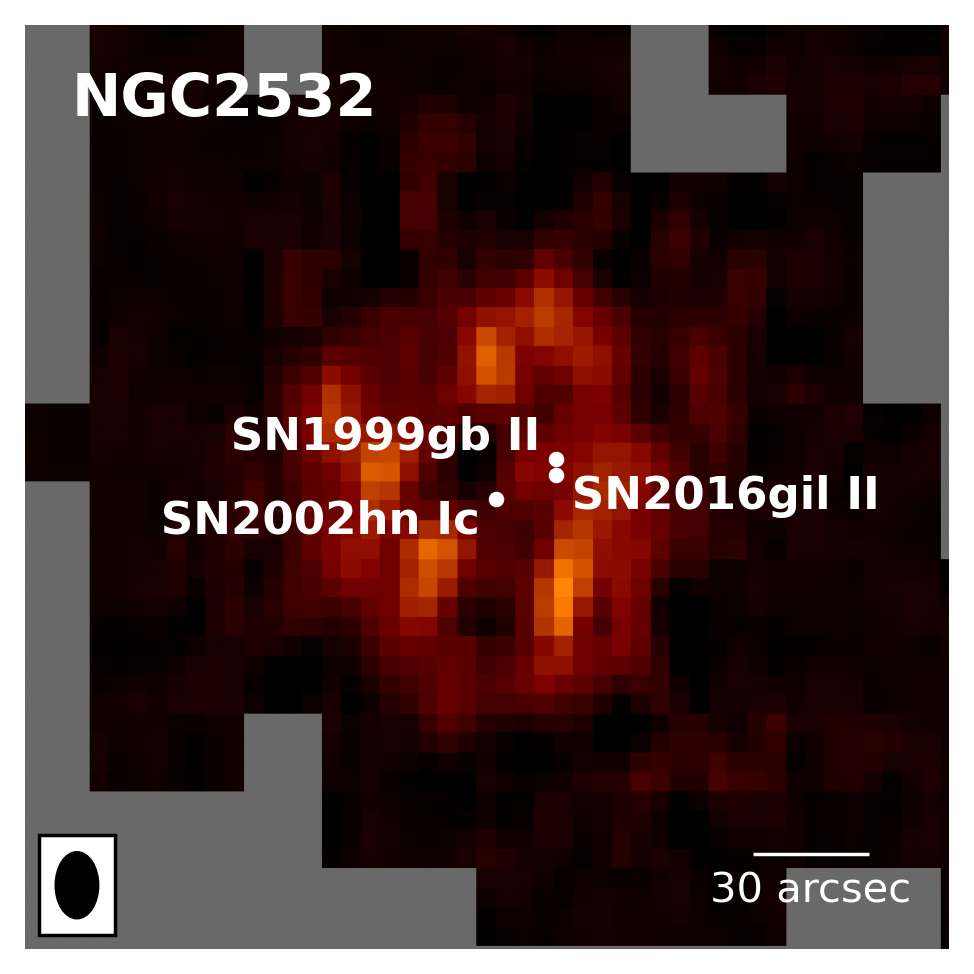}
\includegraphics[width=0.25\textwidth]{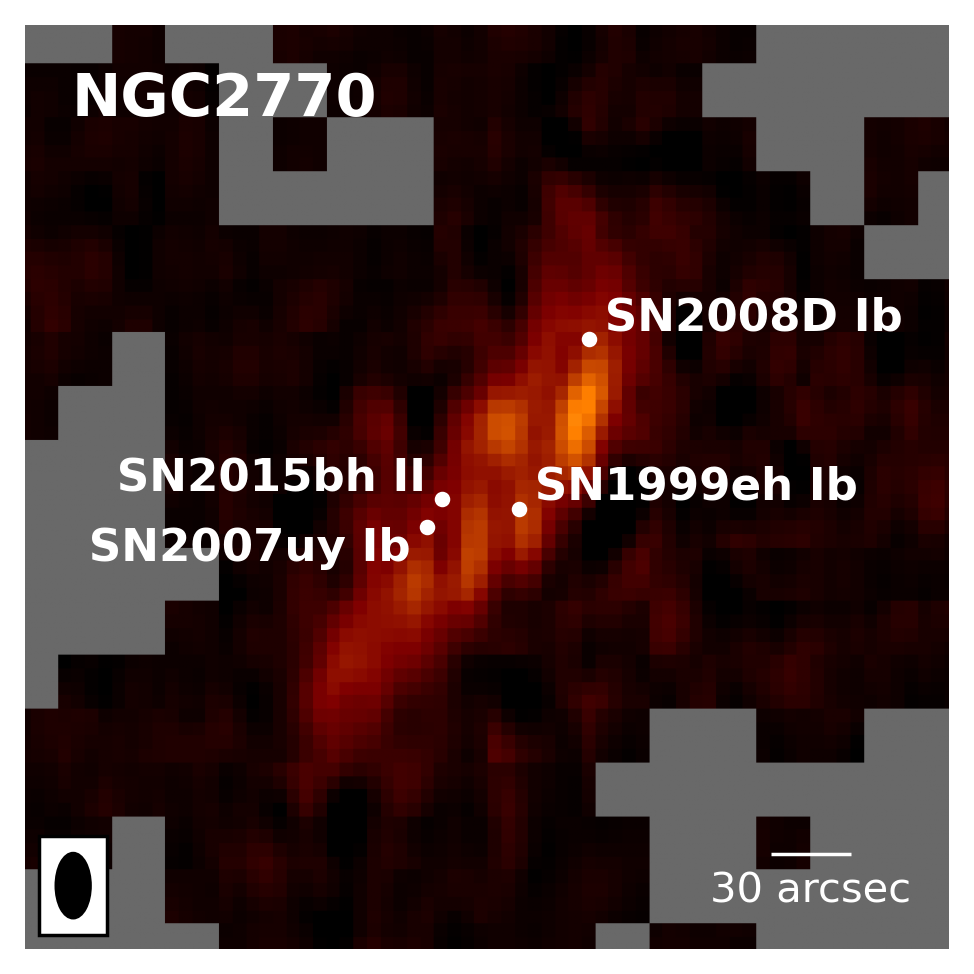}\\
\includegraphics[width=0.25\textwidth]{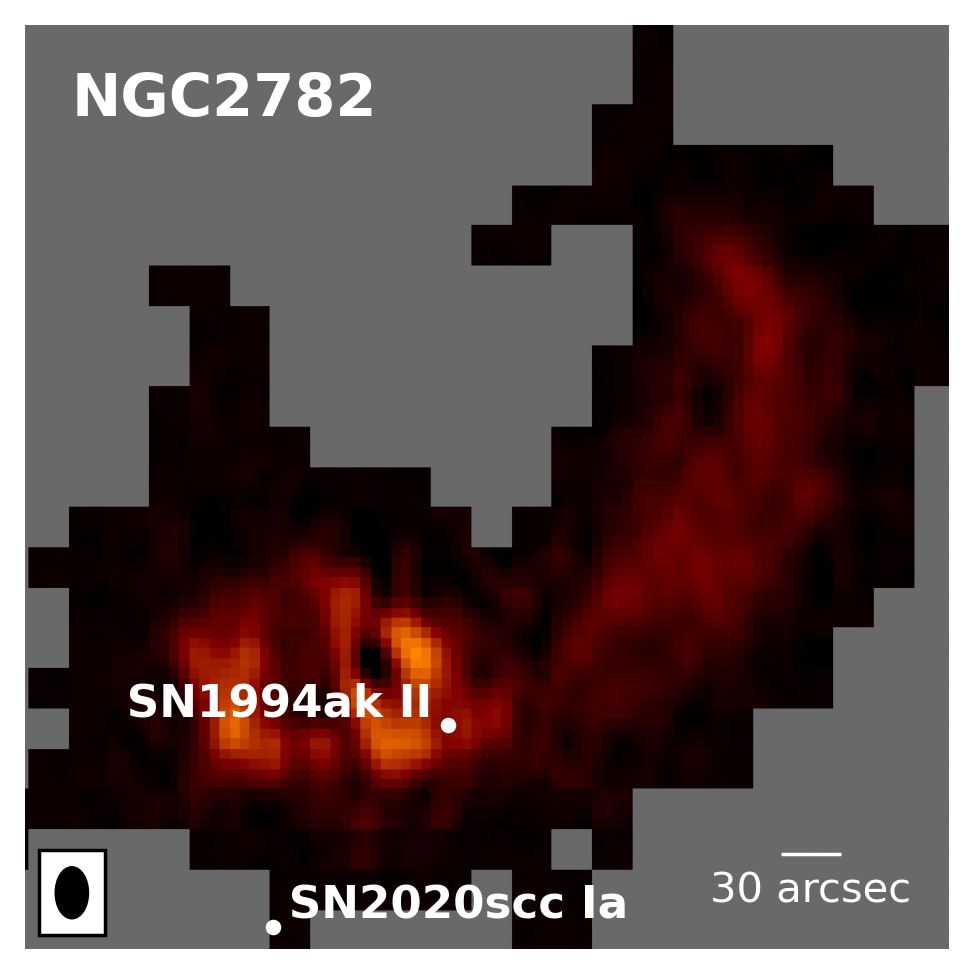}
\includegraphics[width=0.25\textwidth]{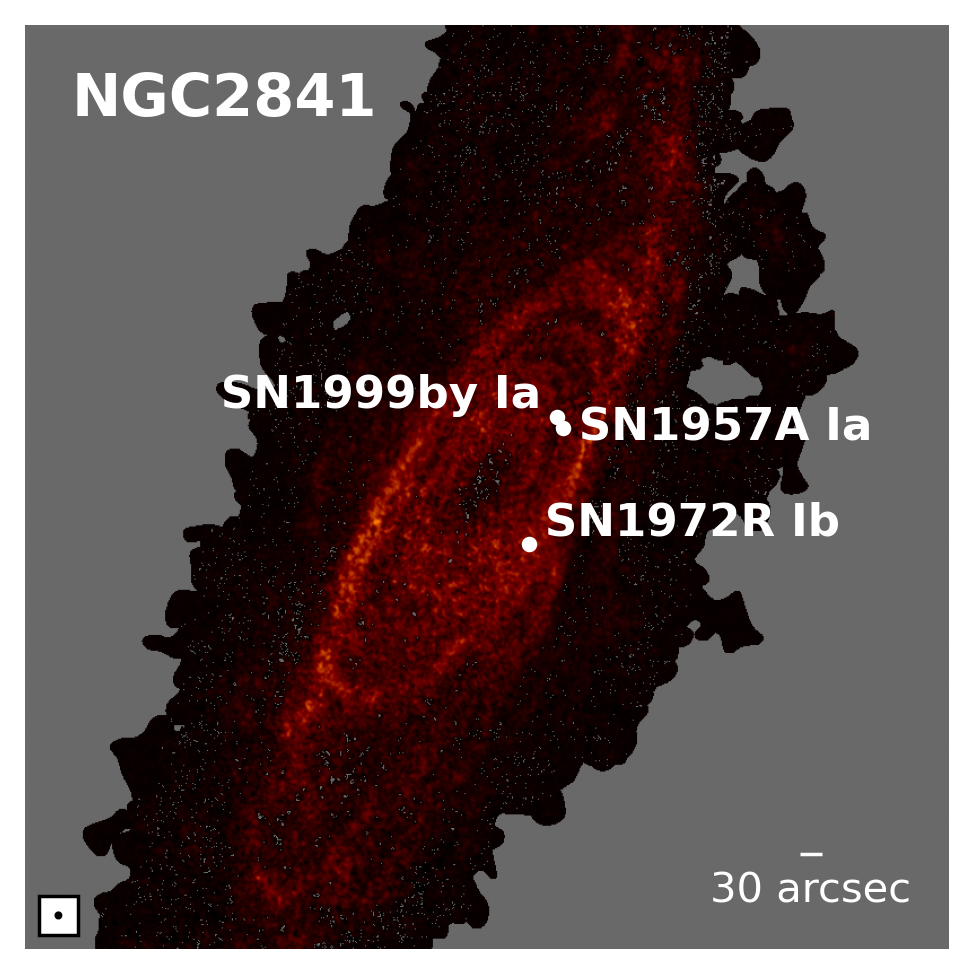}
\includegraphics[width=0.25\textwidth]{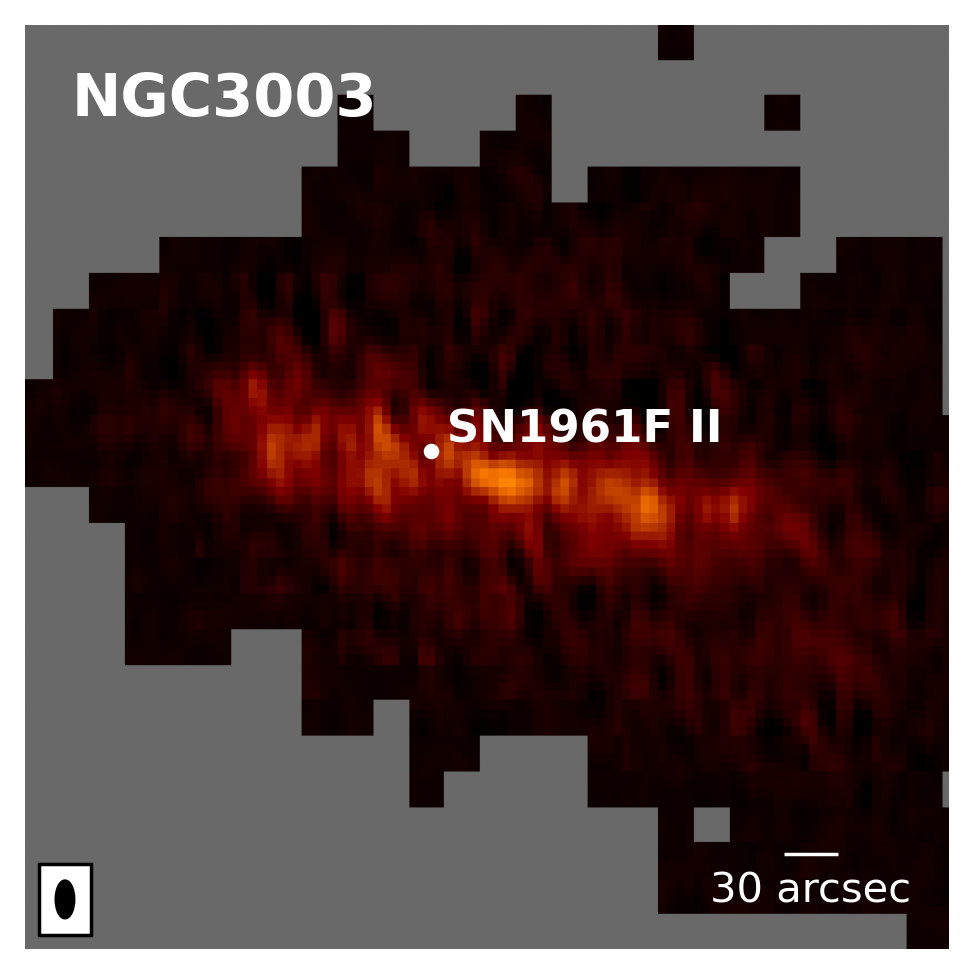}
\includegraphics[width=0.25\textwidth]{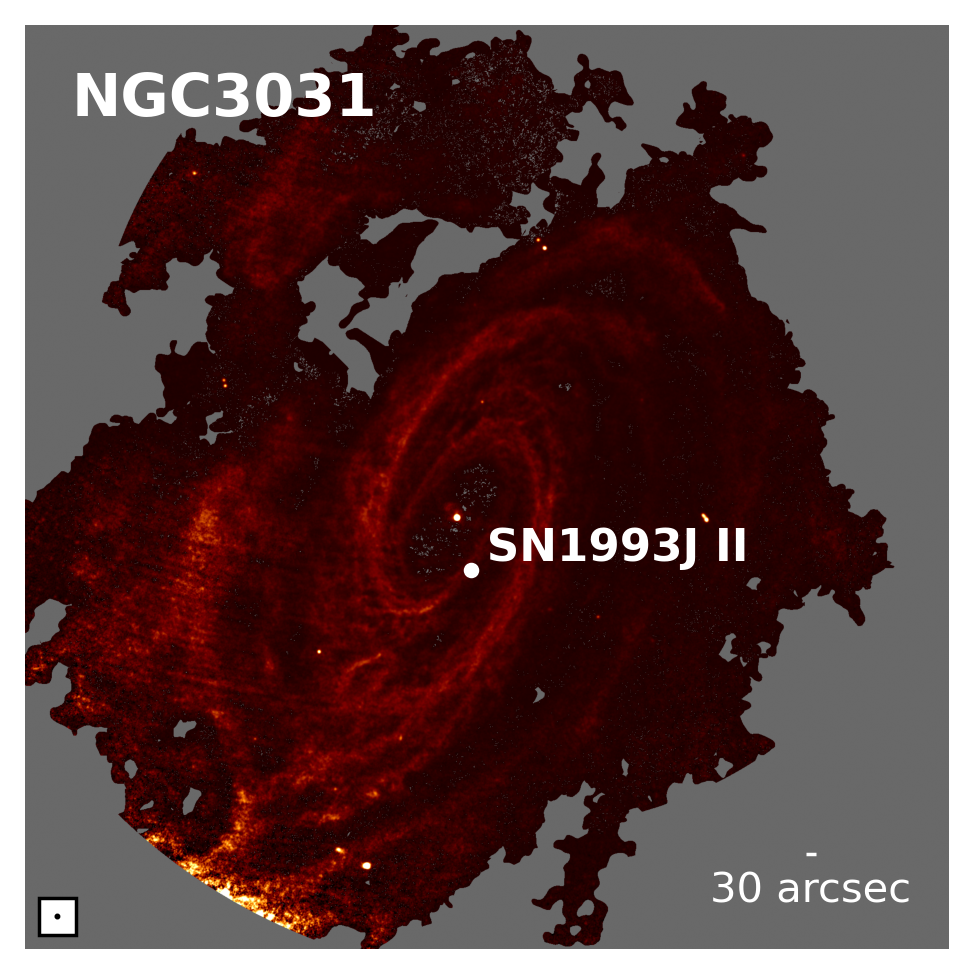}\\
\end{tabular}
\caption{{\hi} moment0 maps of galaxies that hosted SNe, as denoted in the upper left corner of each figure. The SN positions are marked as white circles. The physical scale is shown with the bar and the beam size with the ellipse in the lower left corners. The gray area was not part of the used maps, as it lacked {\hi} detections.}
\label{fig:mapIa}
\end{figure}

\addtocounter{figure}{-1}
\begin{figure*}
\centering
\begin{tabular}{llll}
\includegraphics[width=0.25\textwidth]{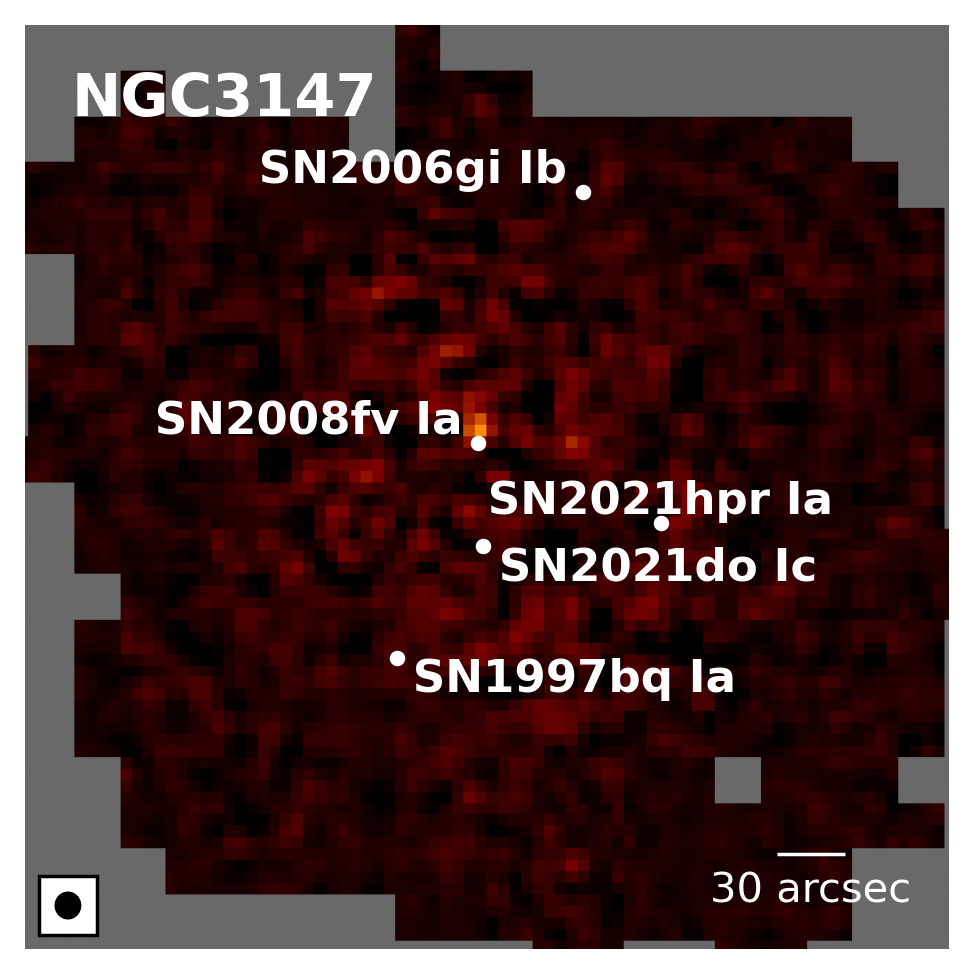}
\includegraphics[width=0.25\textwidth]{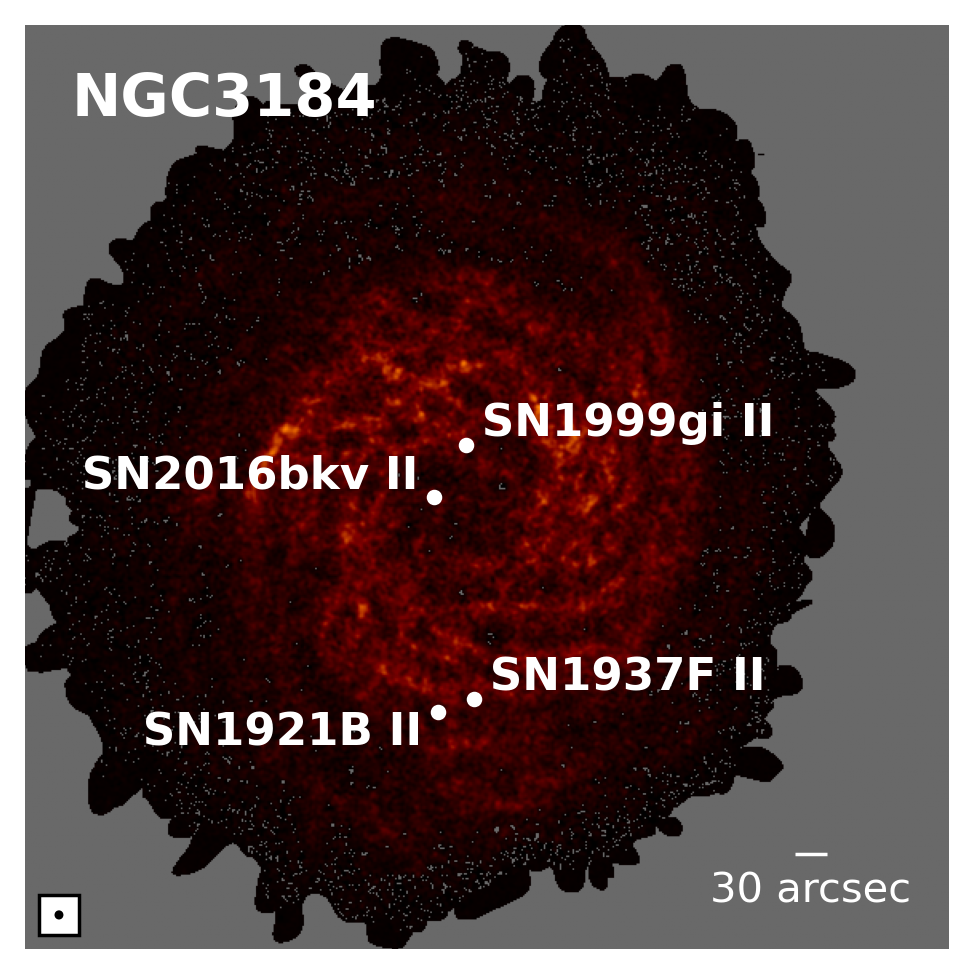}
\includegraphics[width=0.25\textwidth]{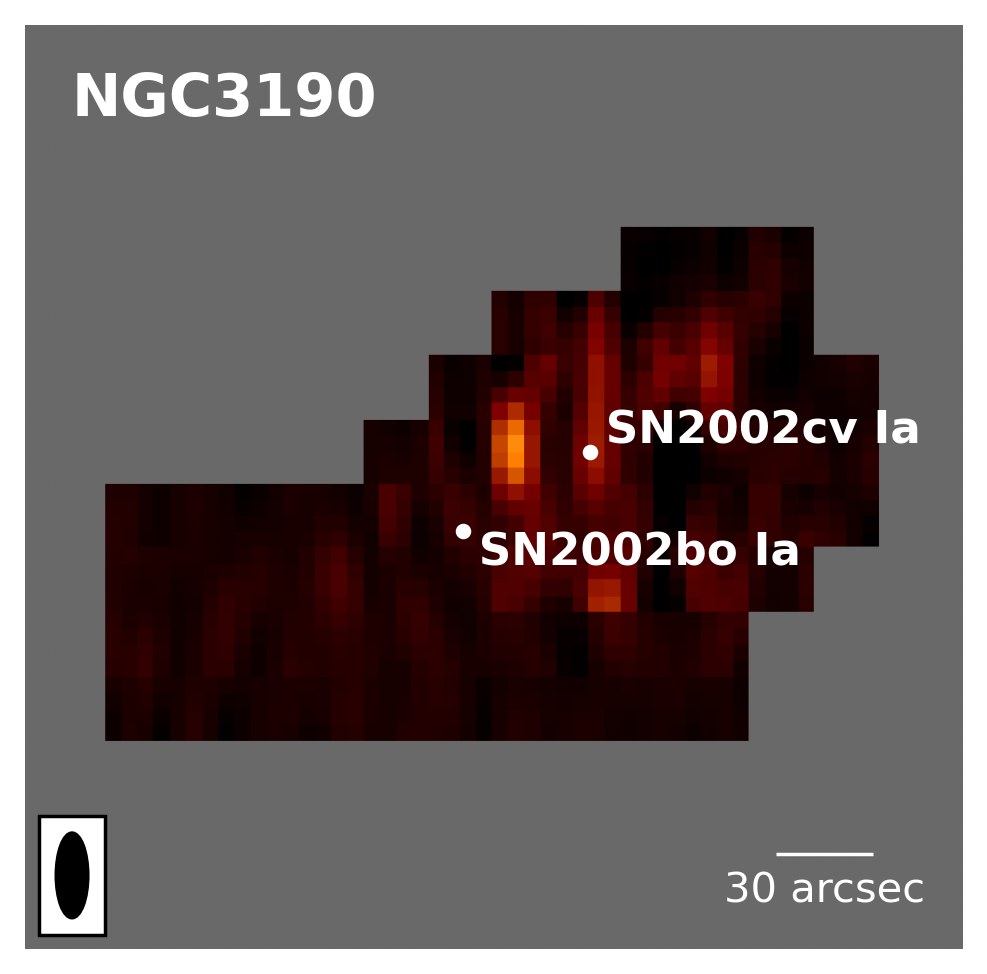}
\includegraphics[width=0.25\textwidth]{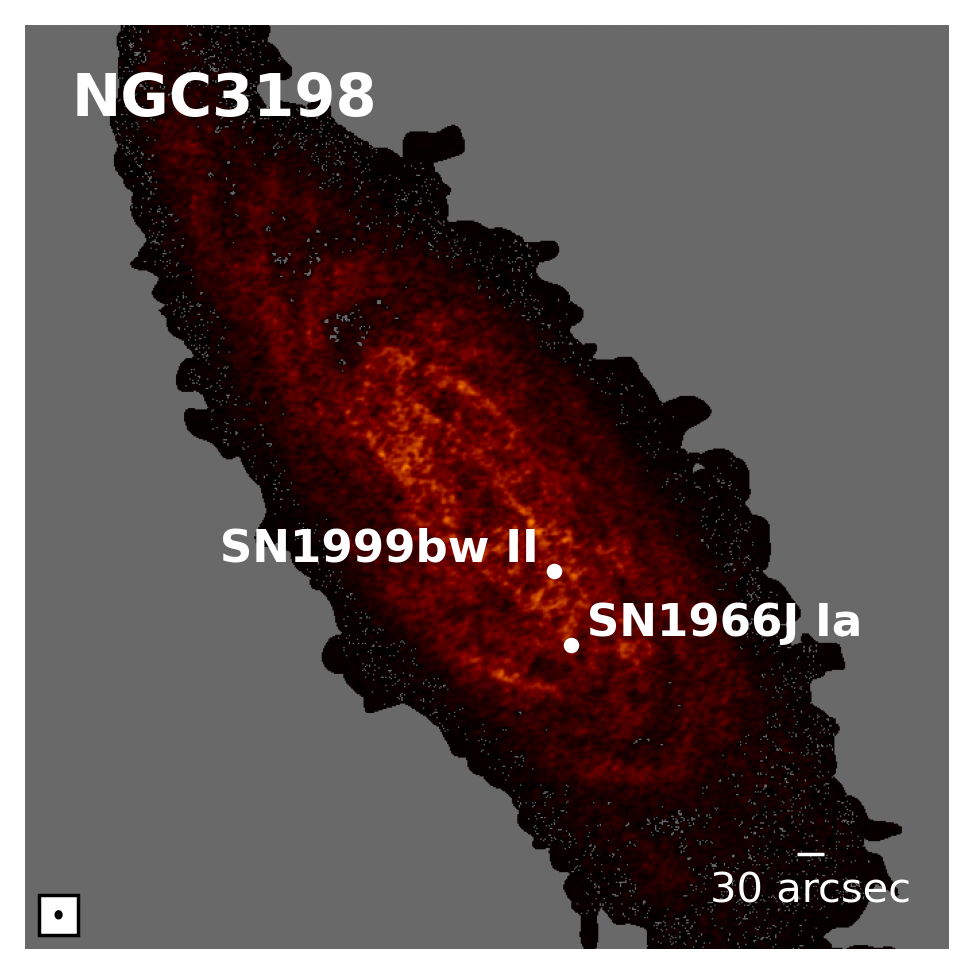}\\
\includegraphics[width=0.25\textwidth]{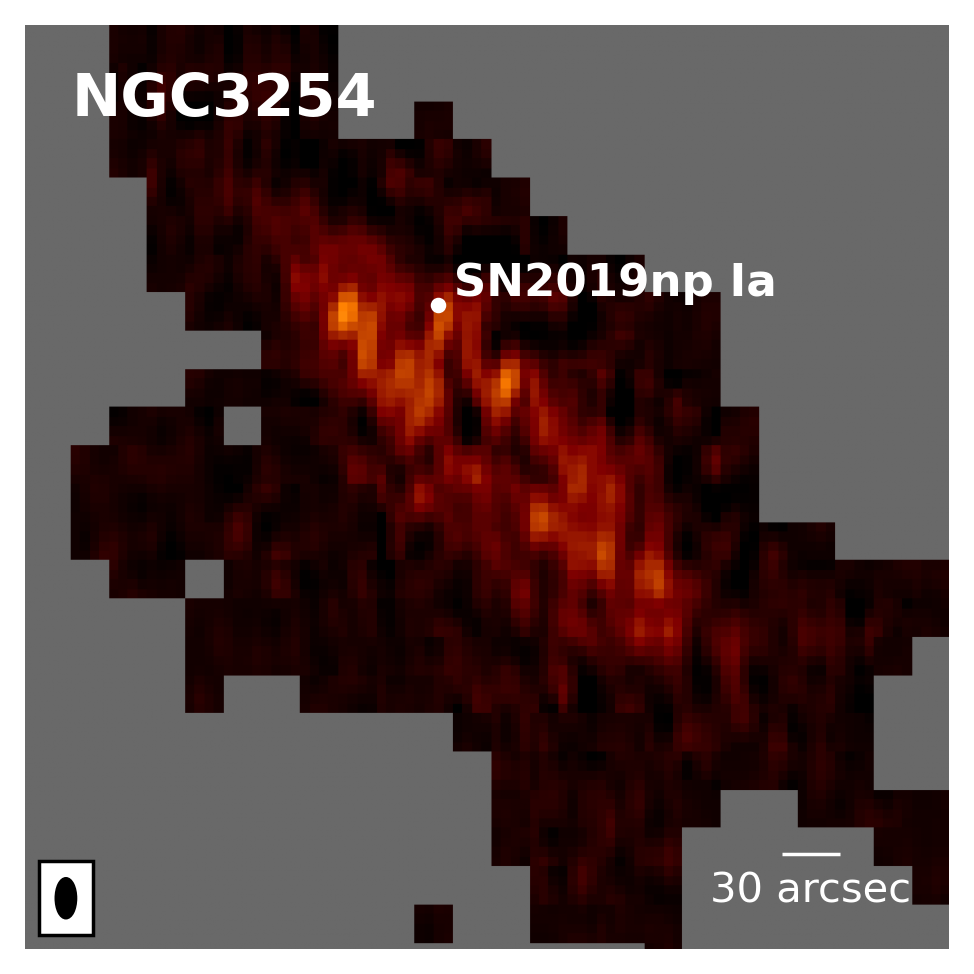}
\includegraphics[width=0.25\textwidth]{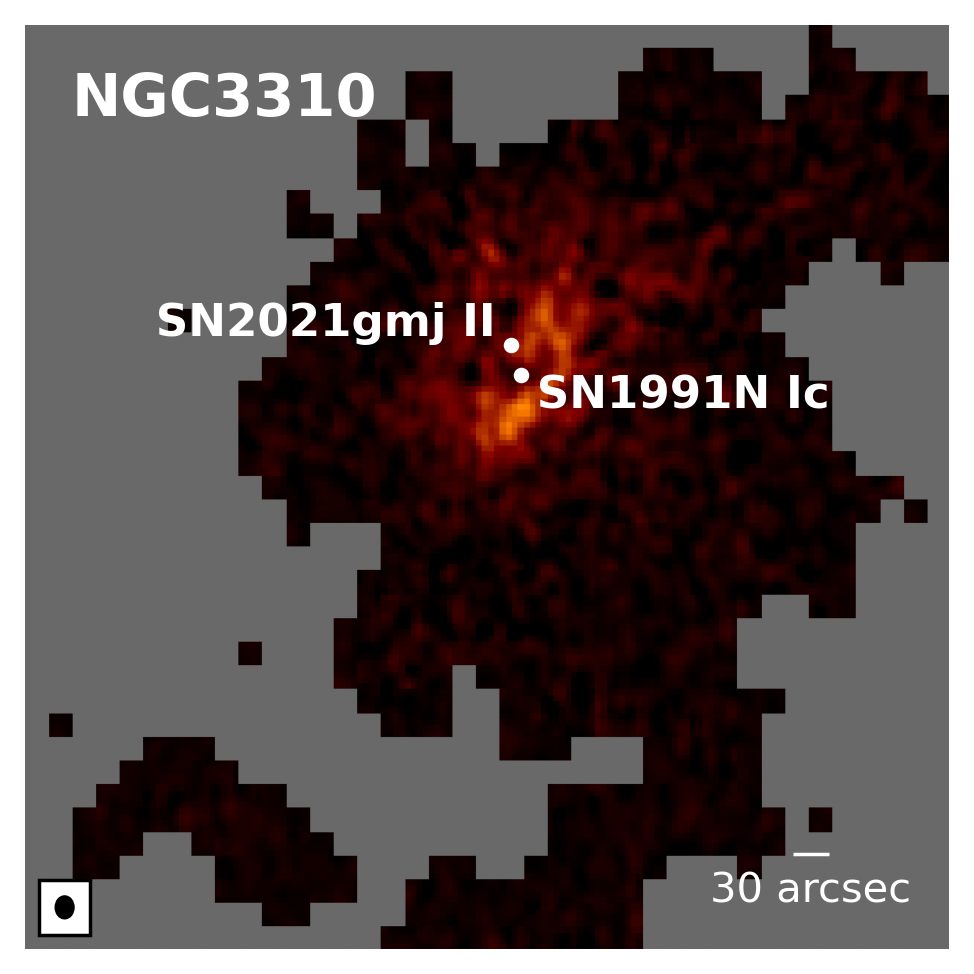}
\includegraphics[width=0.25\textwidth]{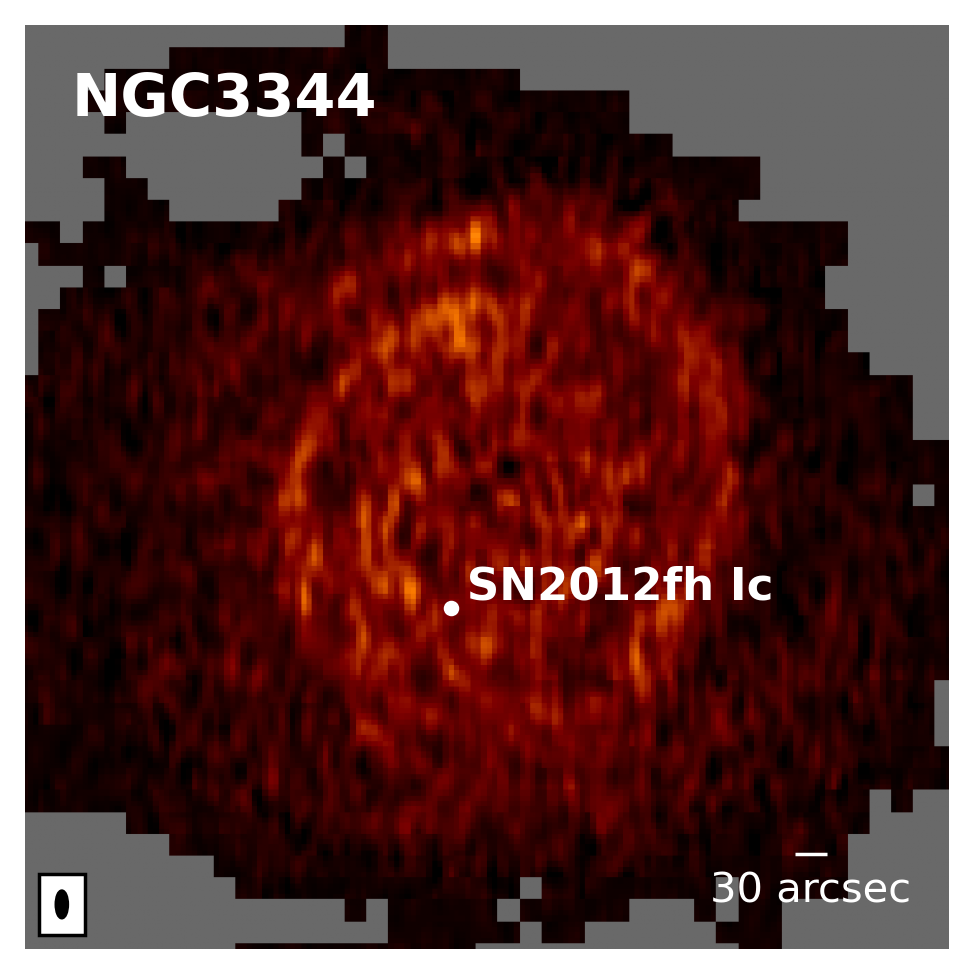}
\includegraphics[width=0.25\textwidth]{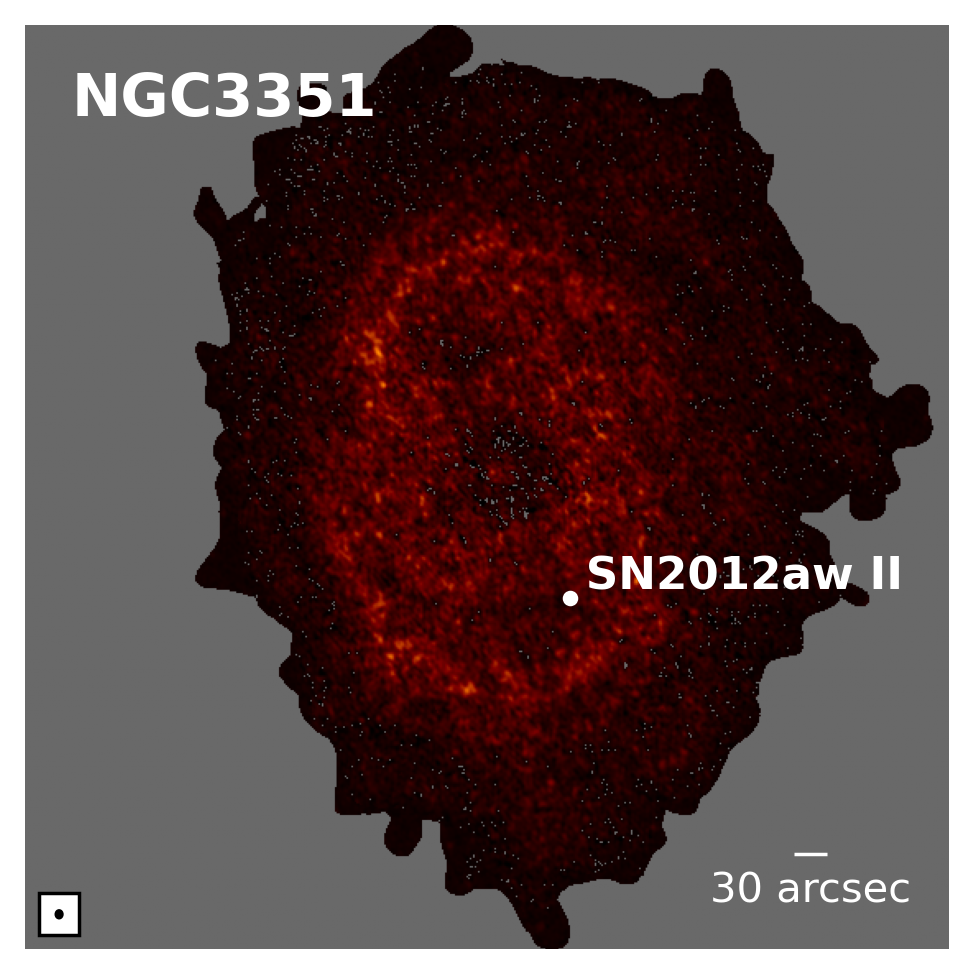}\\
\includegraphics[width=0.25\textwidth]{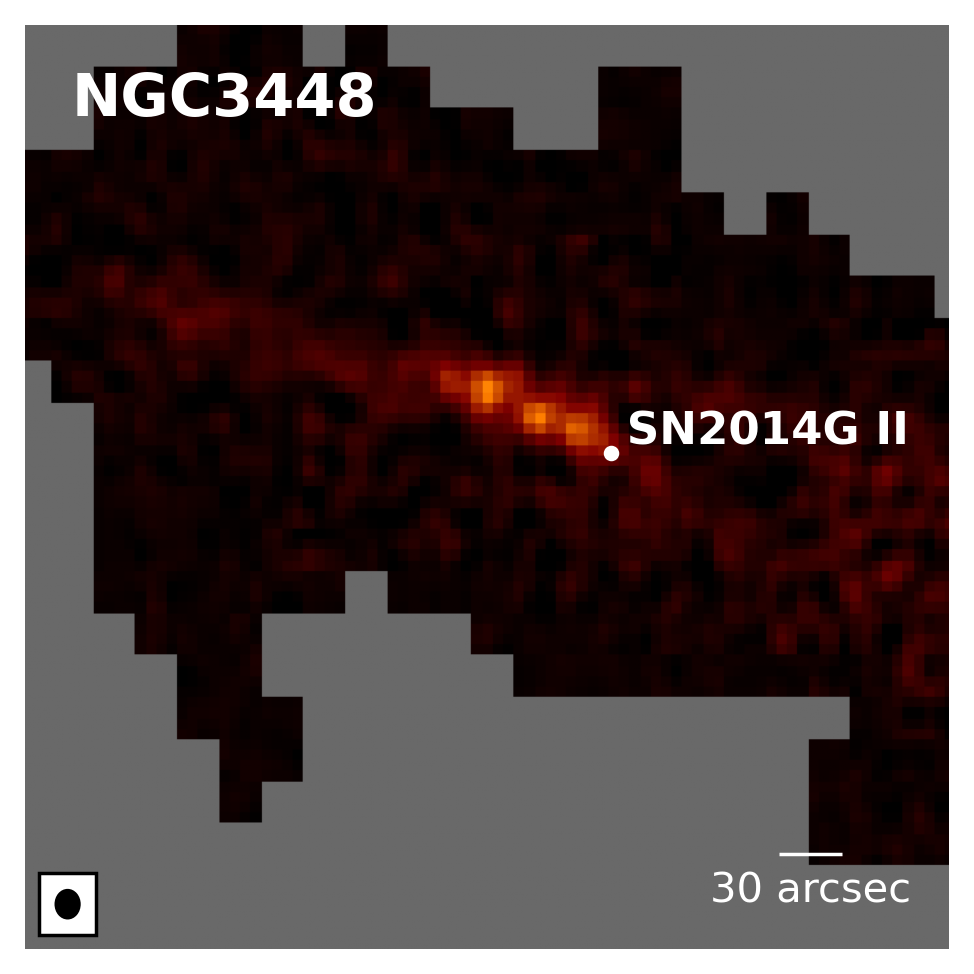}
\includegraphics[width=0.25\textwidth]{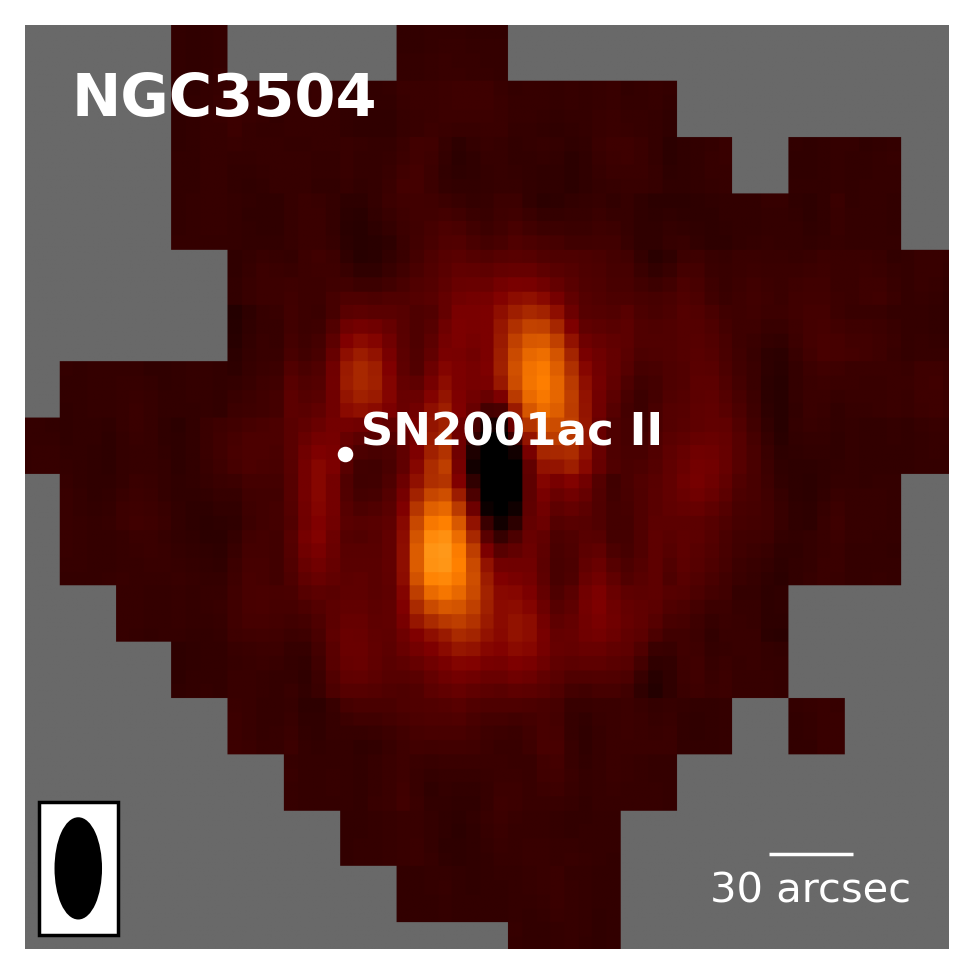}
\includegraphics[width=0.25\textwidth]{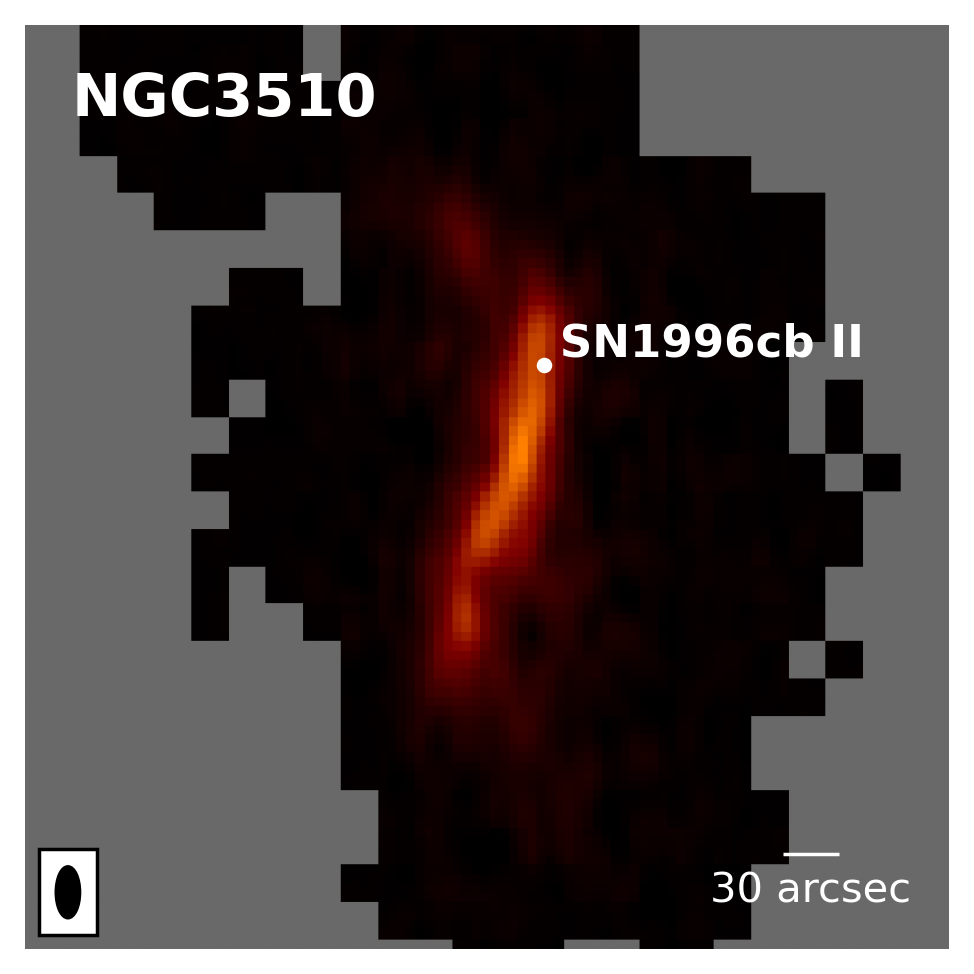}
\includegraphics[width=0.25\textwidth]{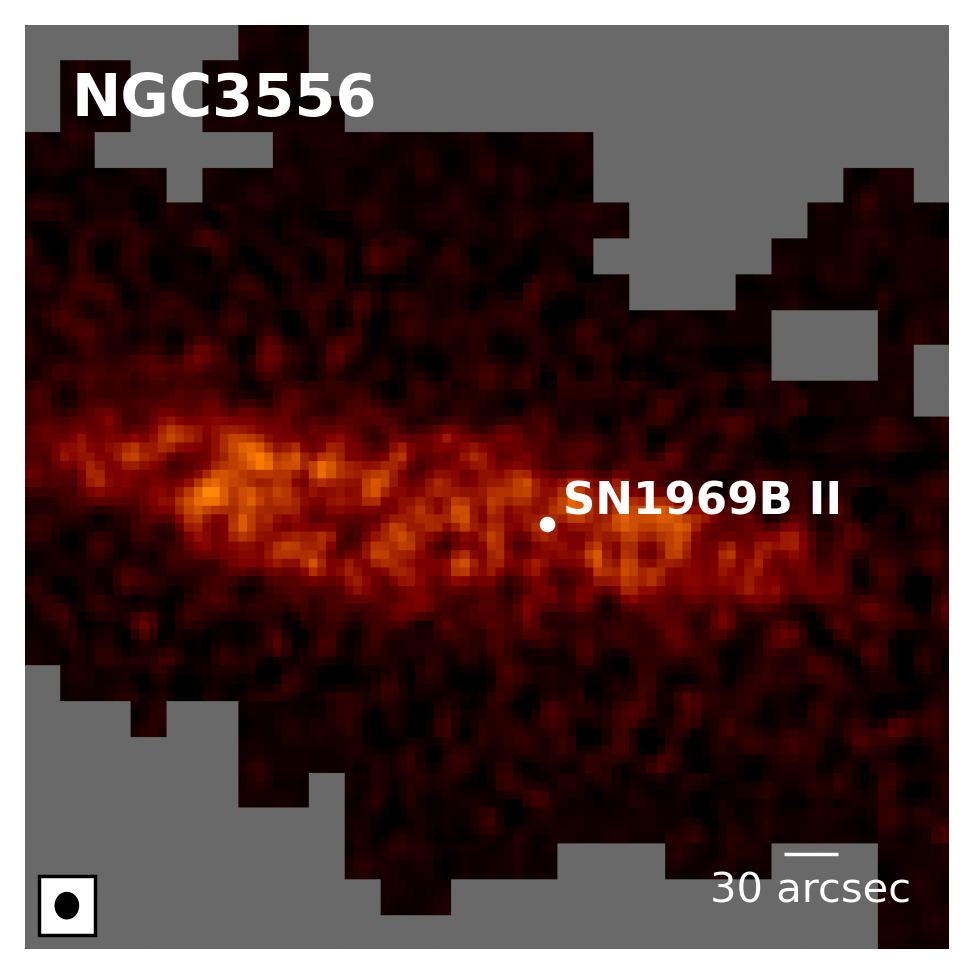}\\
\includegraphics[width=0.25\textwidth]{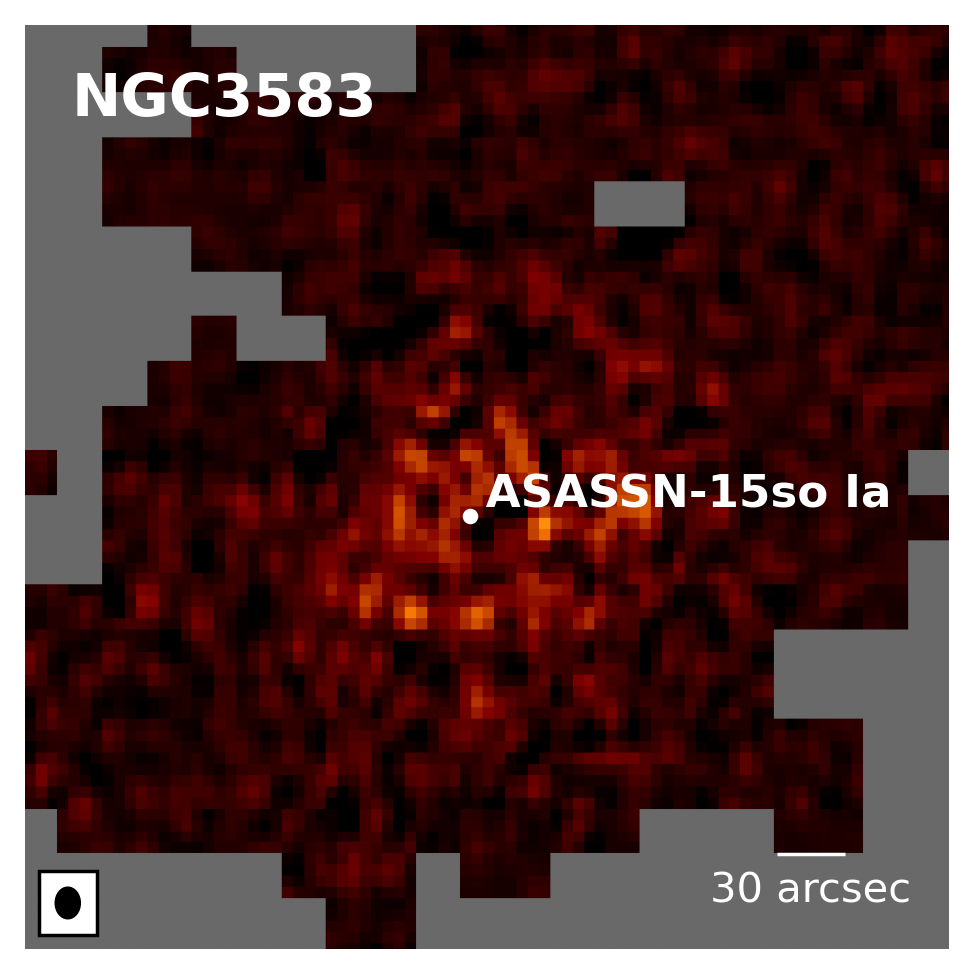}
\includegraphics[width=0.25\textwidth]{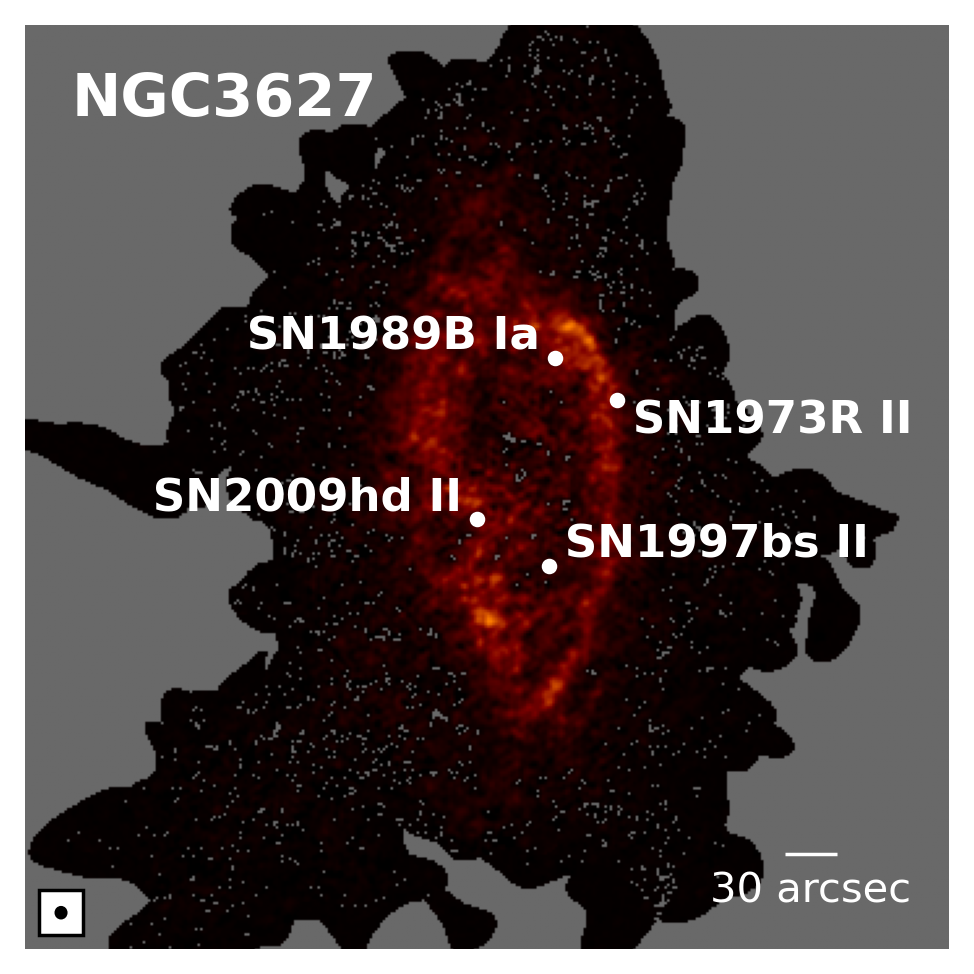}
\includegraphics[width=0.25\textwidth]{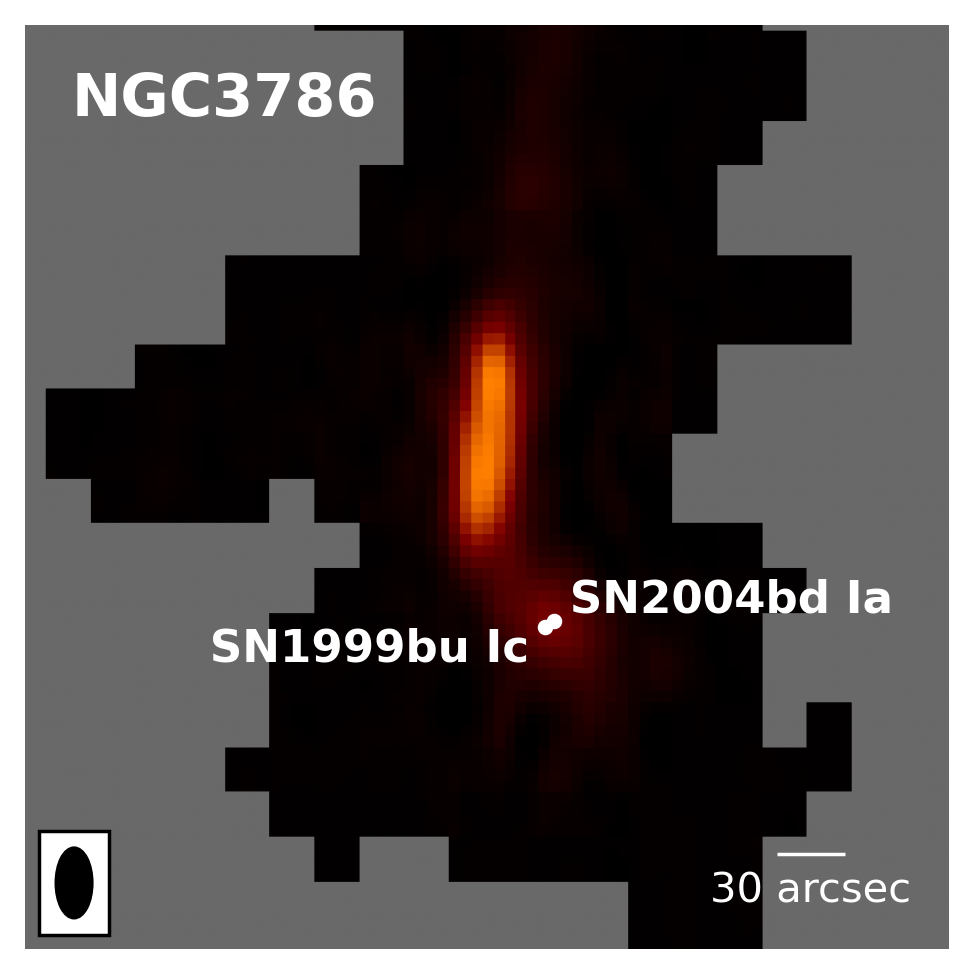}
\includegraphics[width=0.25\textwidth]{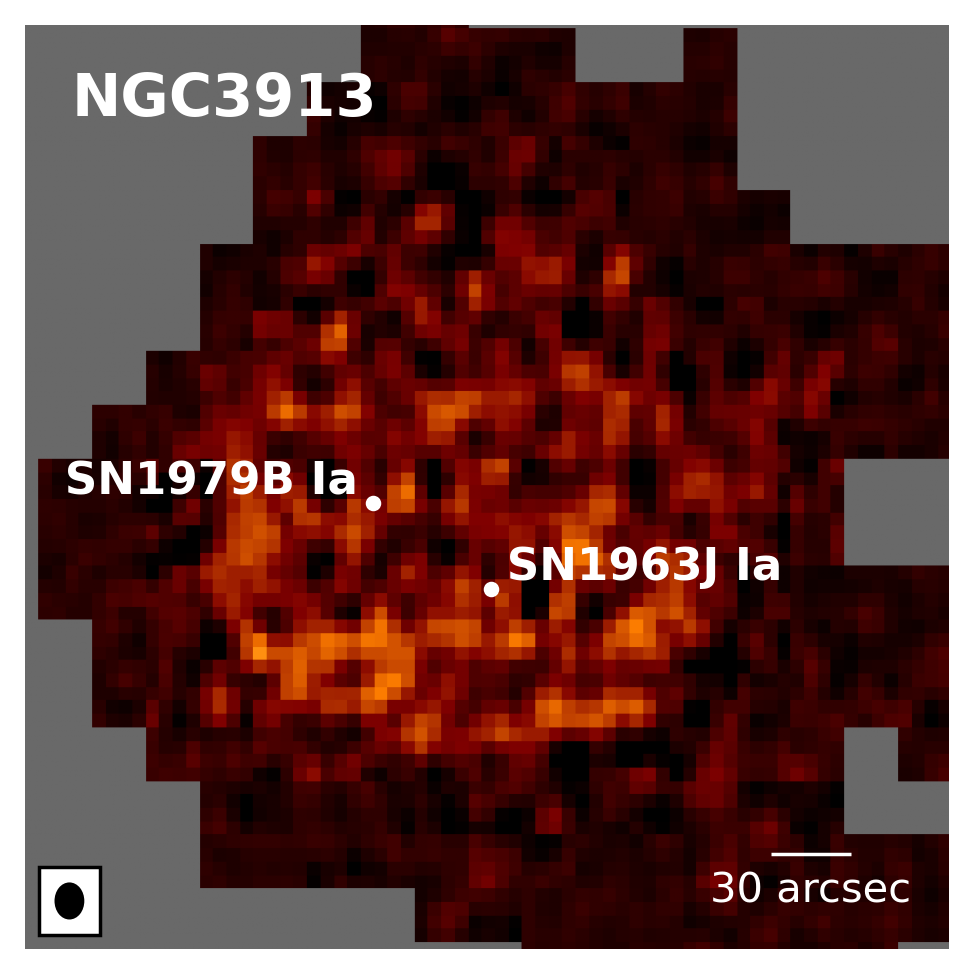}\\
\includegraphics[width=0.25\textwidth]{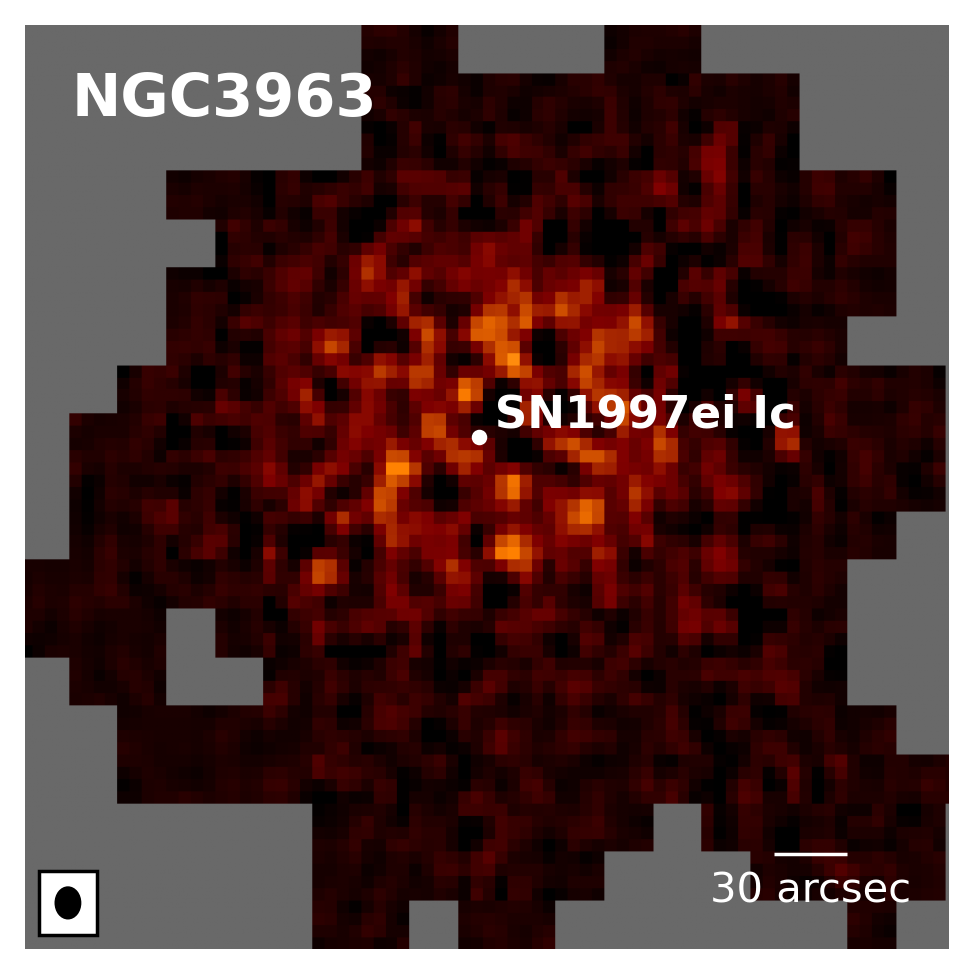}
\includegraphics[width=0.25\textwidth]{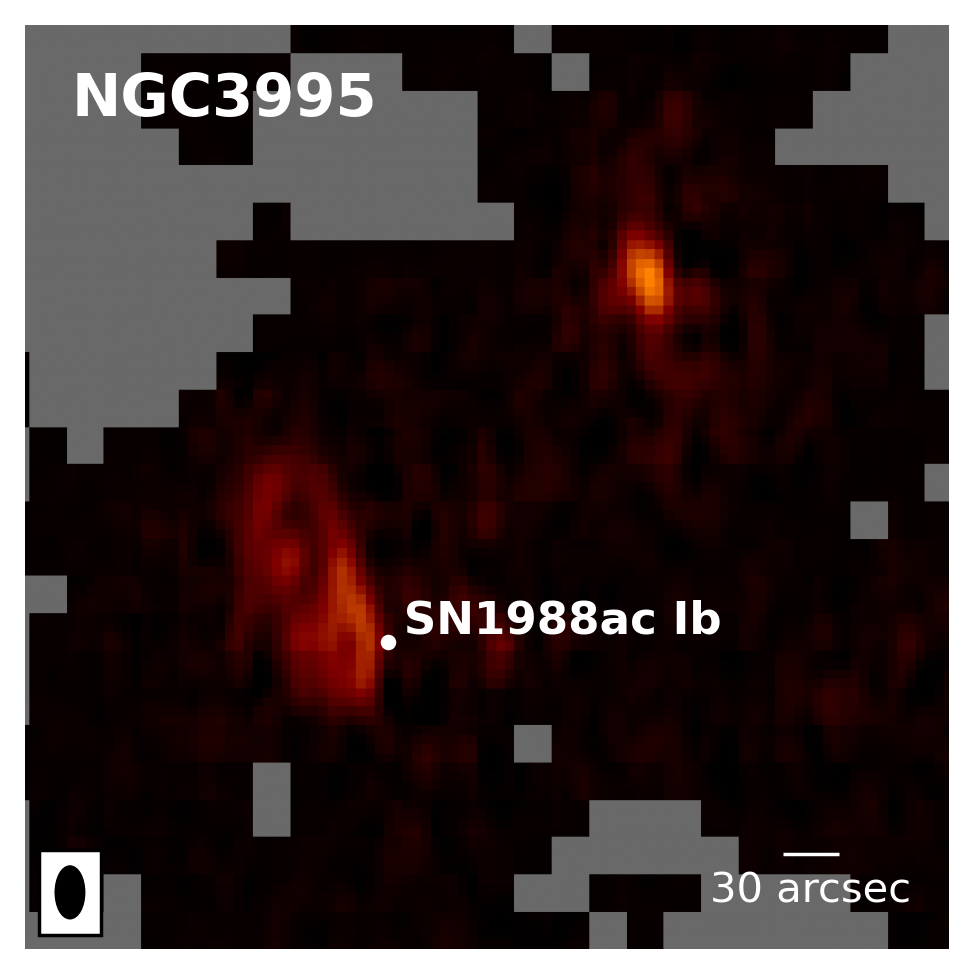}
\includegraphics[width=0.25\textwidth]{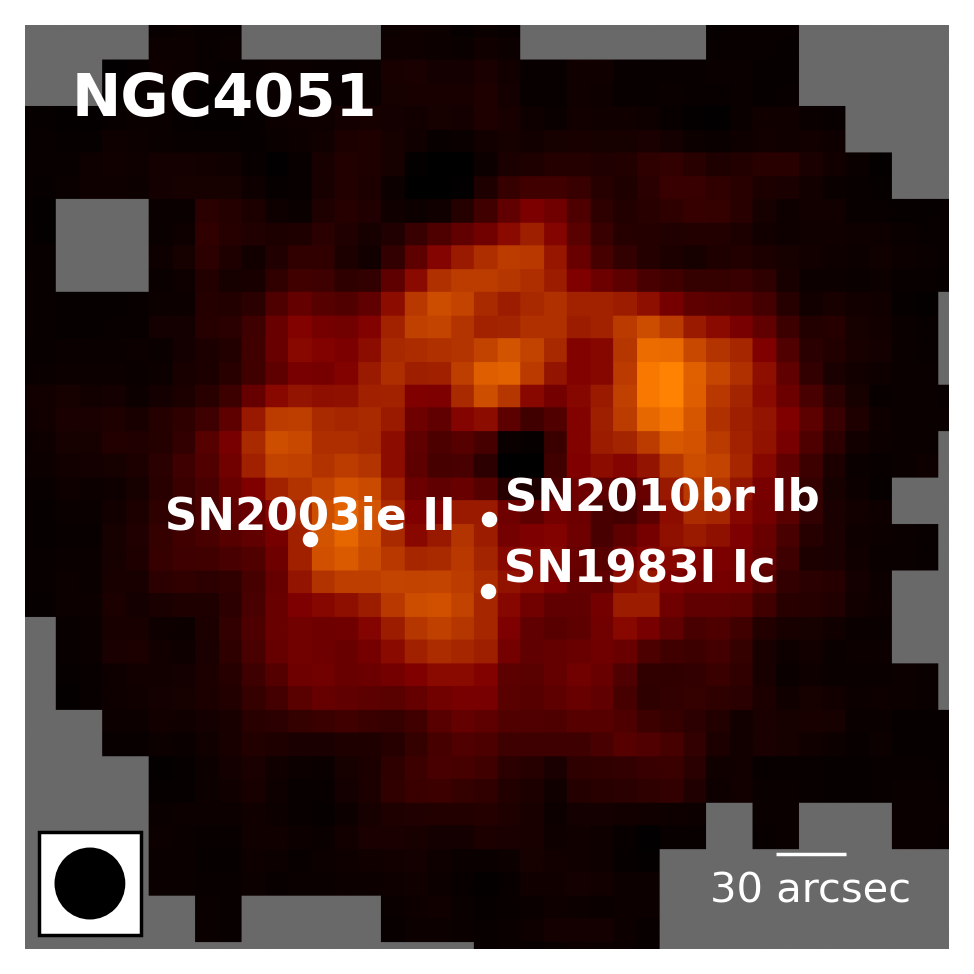}
\includegraphics[width=0.25\textwidth]{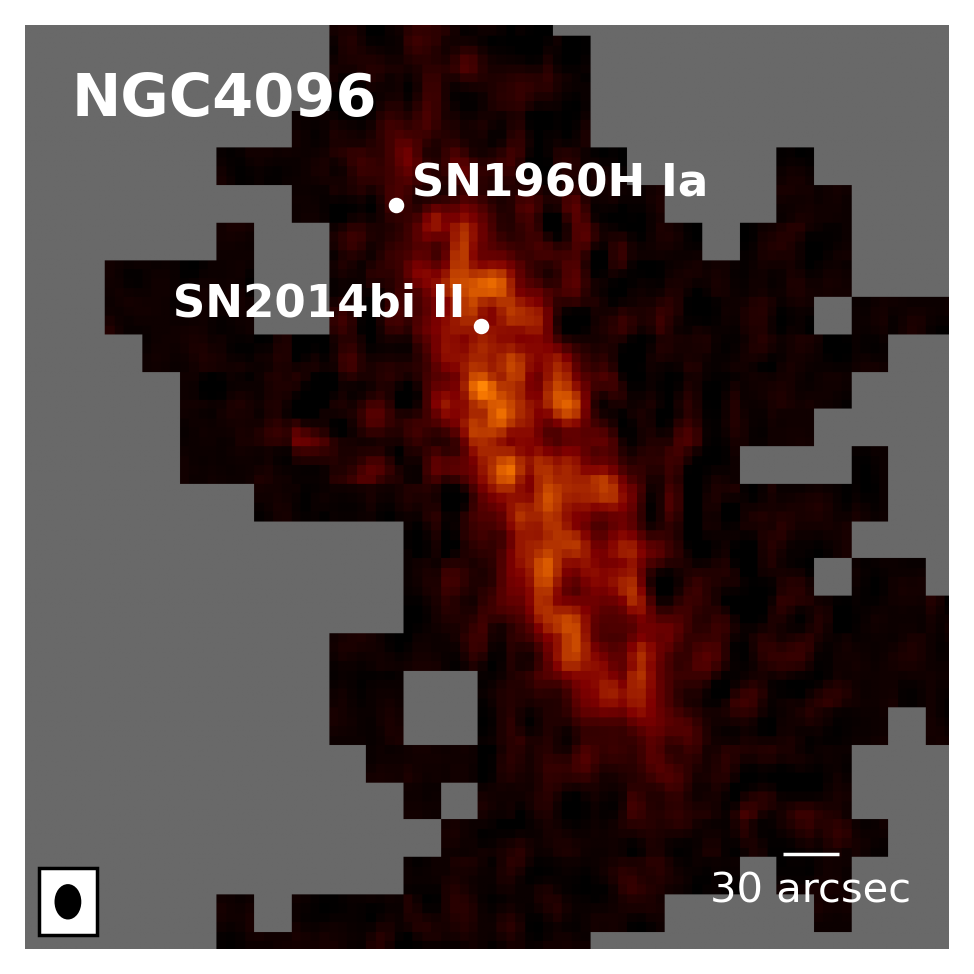}\\
\end{tabular}
\caption{Continued.}
\label{fig:mapIa}
\end{figure*}

\addtocounter{figure}{-1}
\begin{figure*}
\centering
\begin{tabular}{llll}
\includegraphics[width=0.25\textwidth]{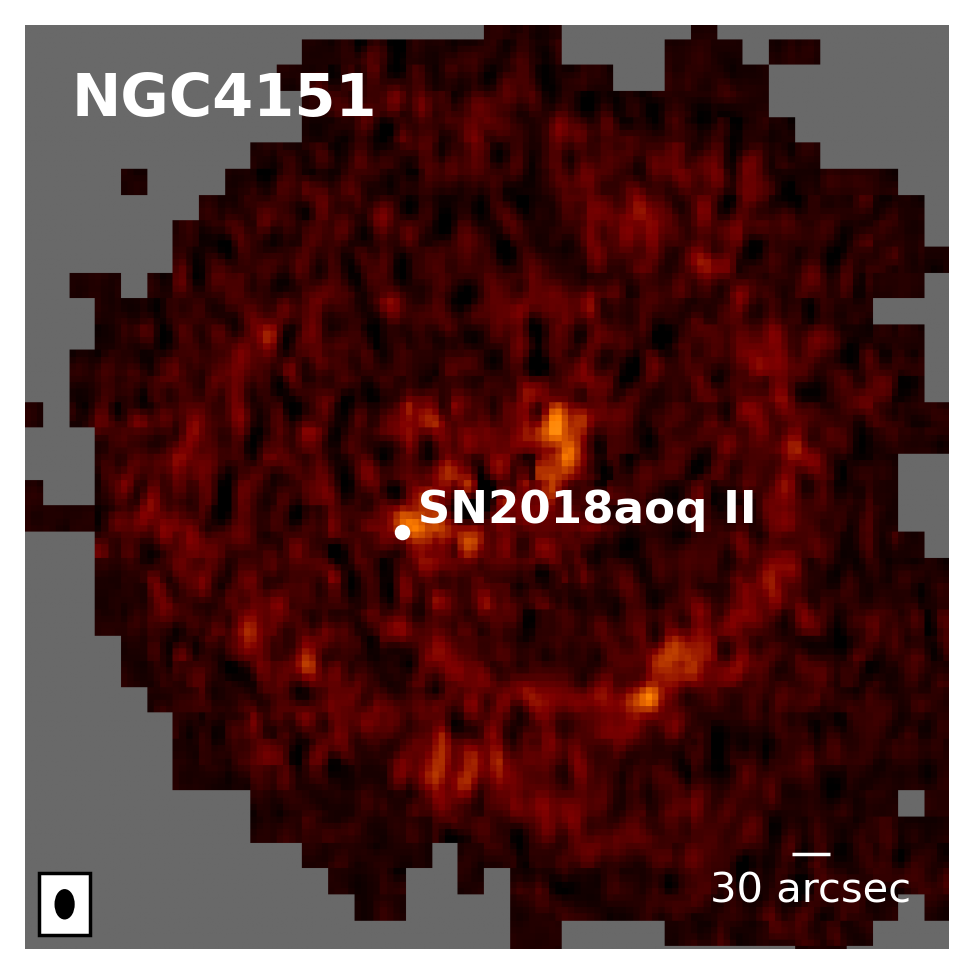}
\includegraphics[width=0.25\textwidth]{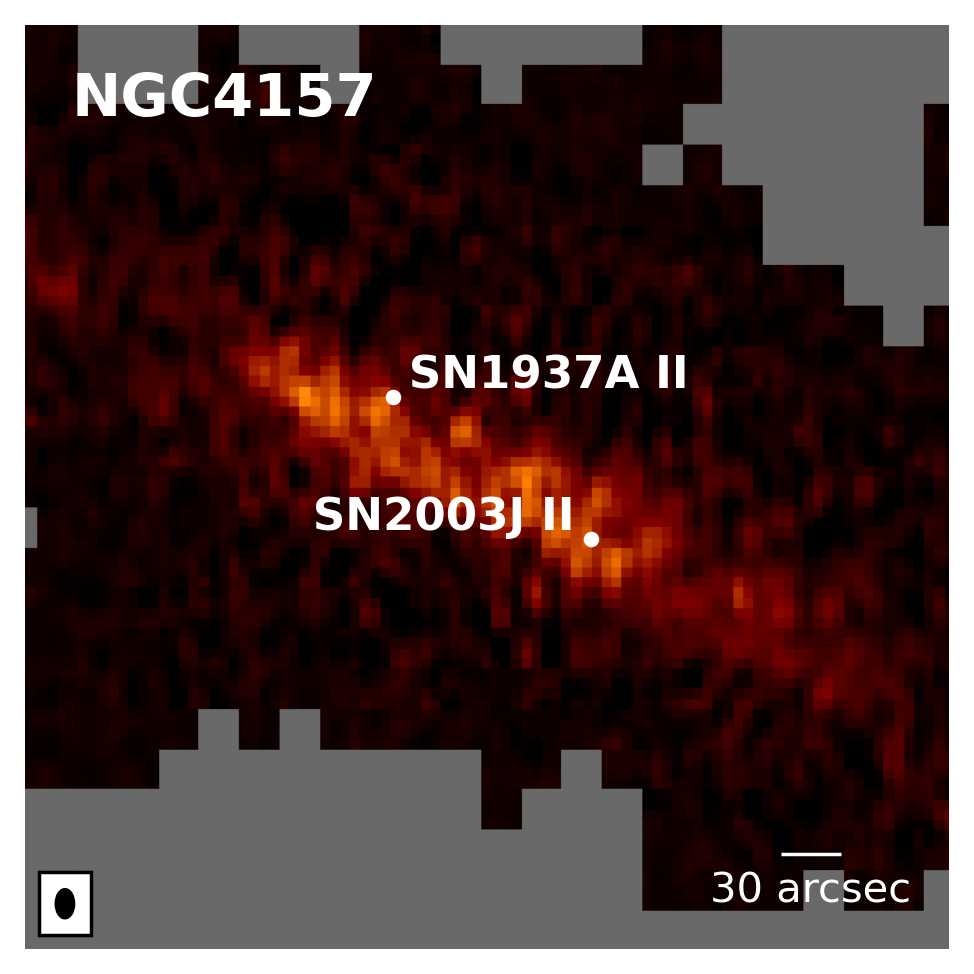}
\includegraphics[width=0.25\textwidth]{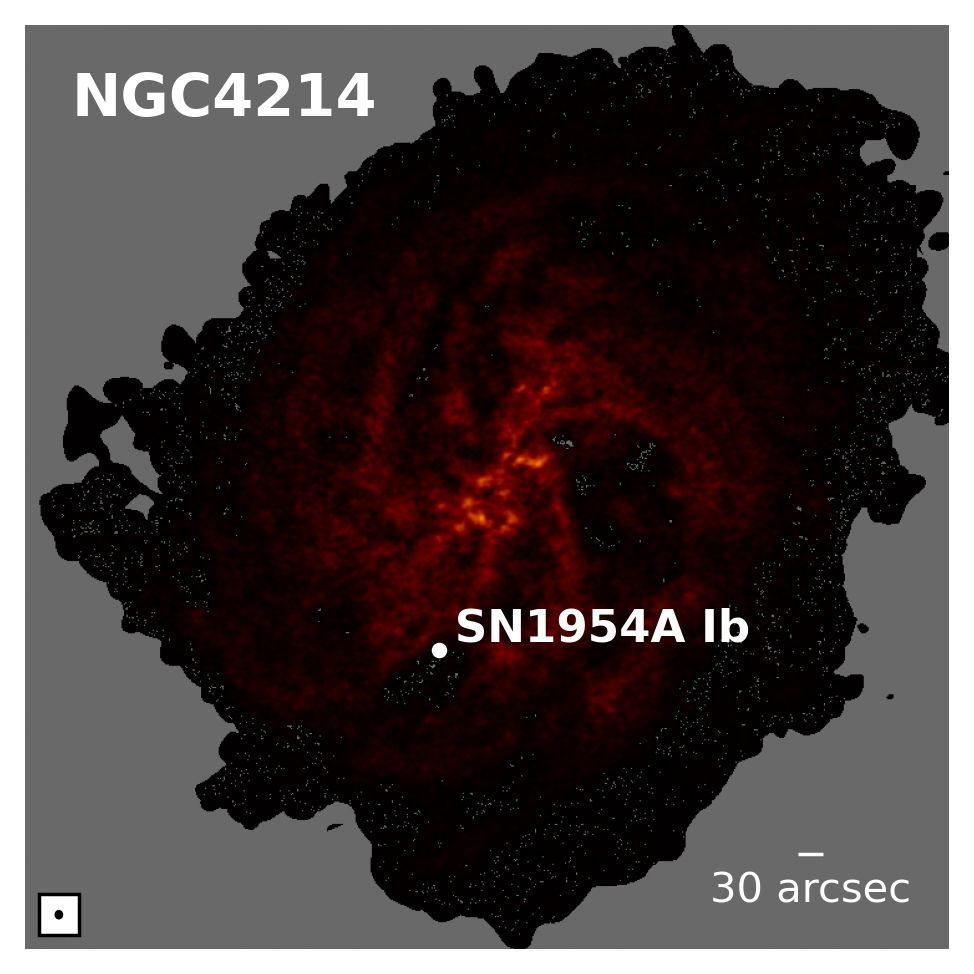}
\includegraphics[width=0.25\textwidth]{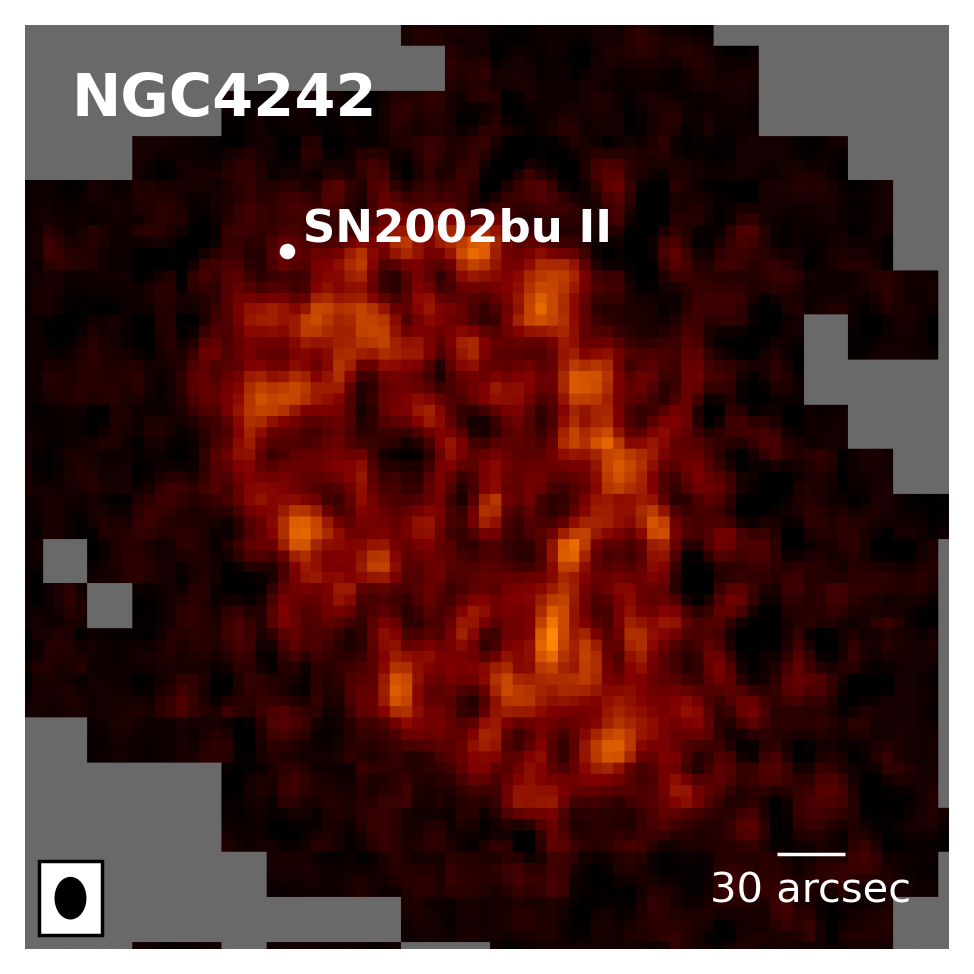}\\
\includegraphics[width=0.25\textwidth]{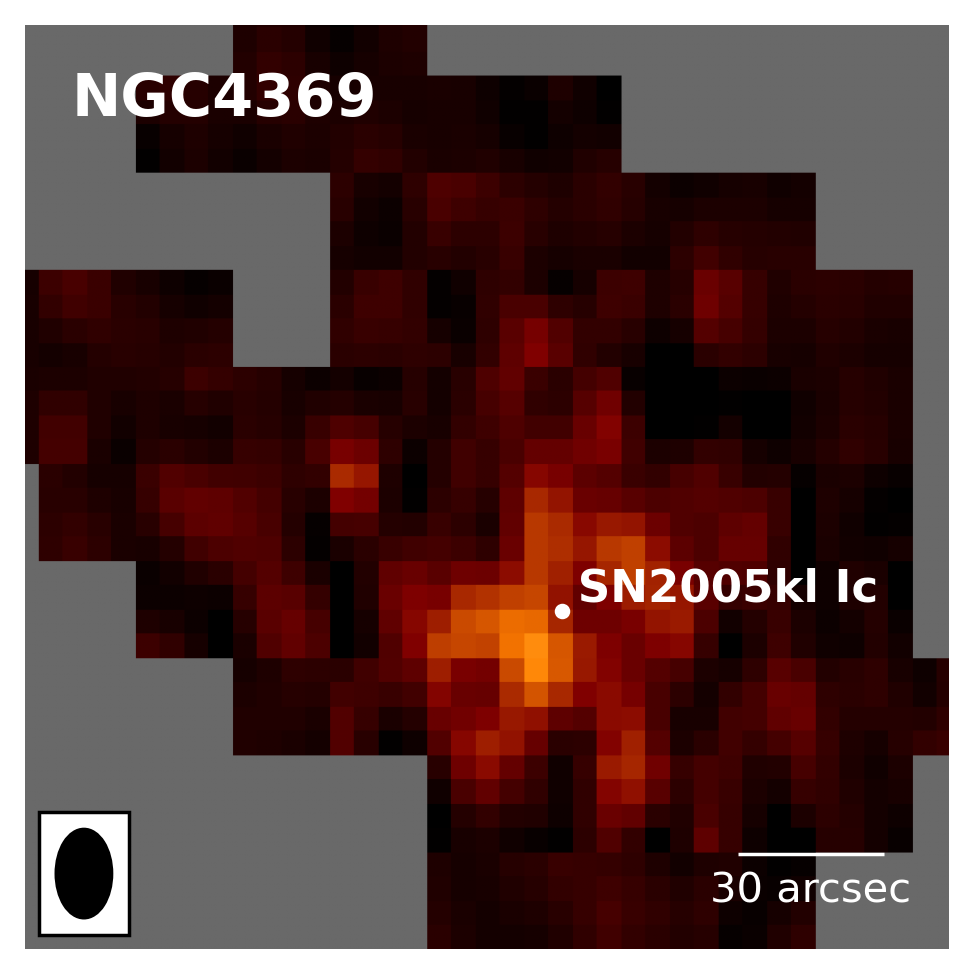}
\includegraphics[width=0.25\textwidth]{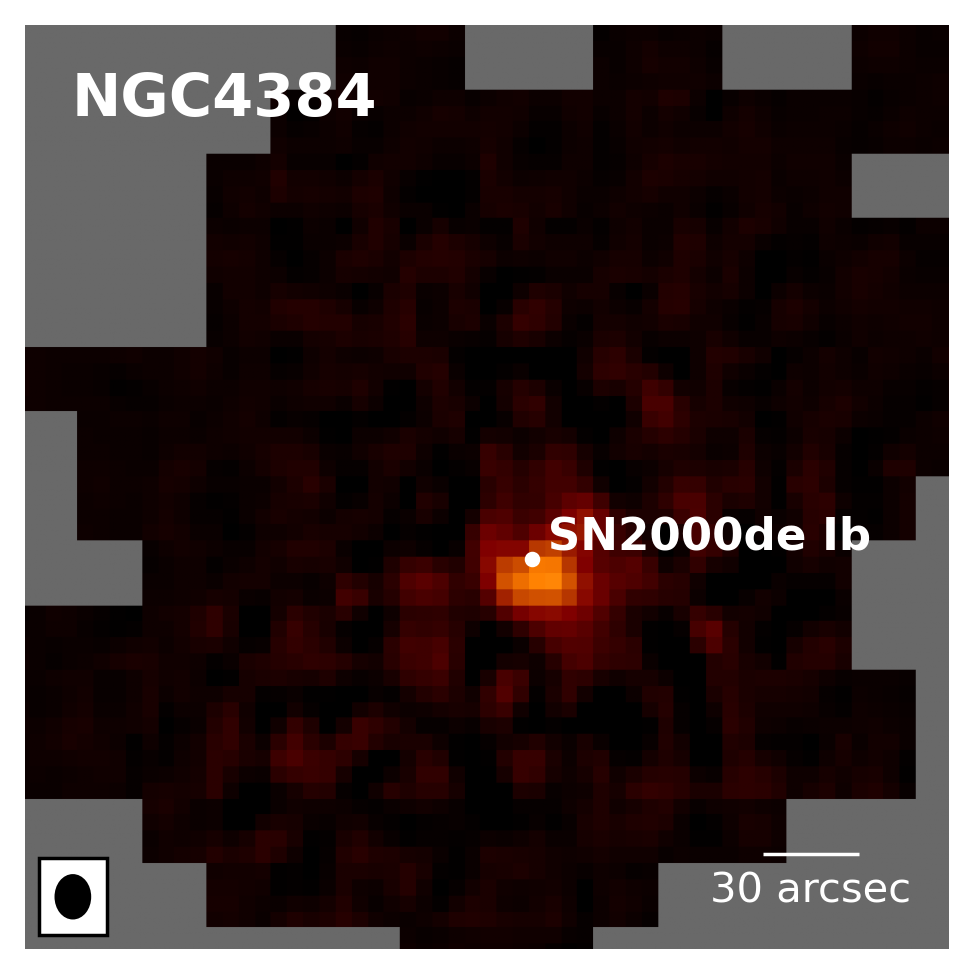}
\includegraphics[width=0.25\textwidth]{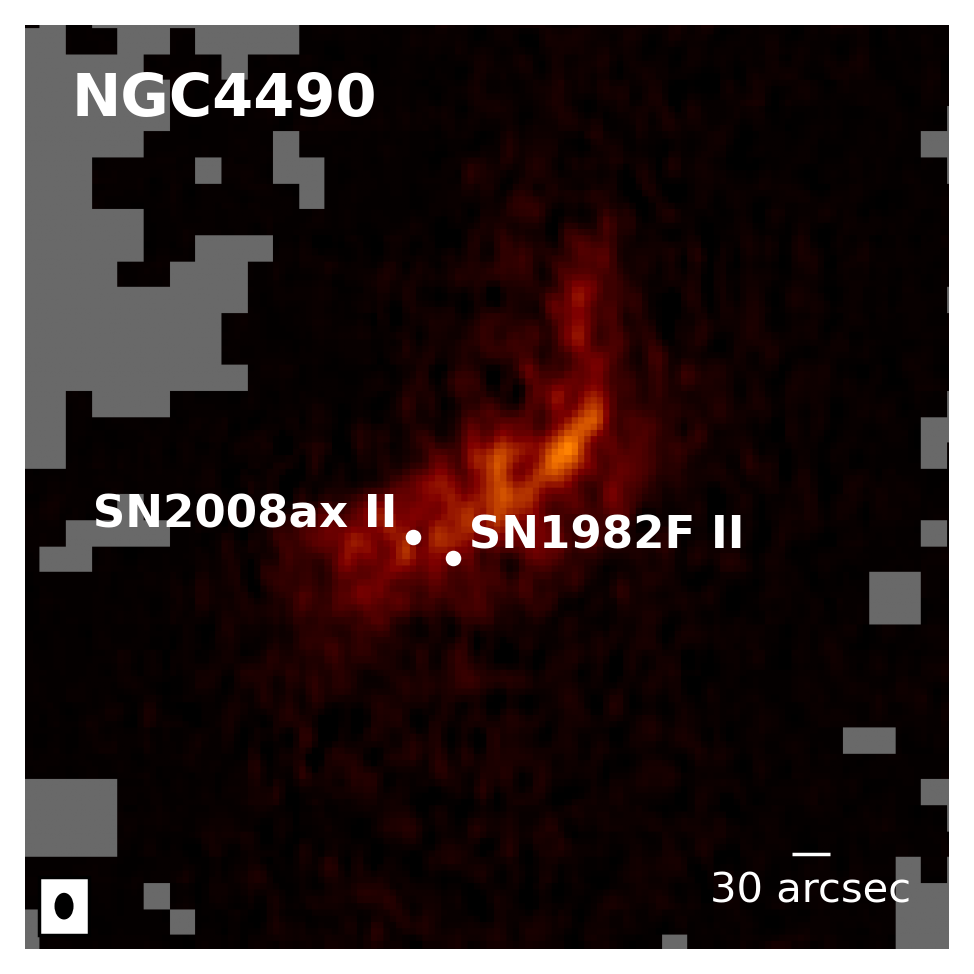}
\includegraphics[width=0.25\textwidth]{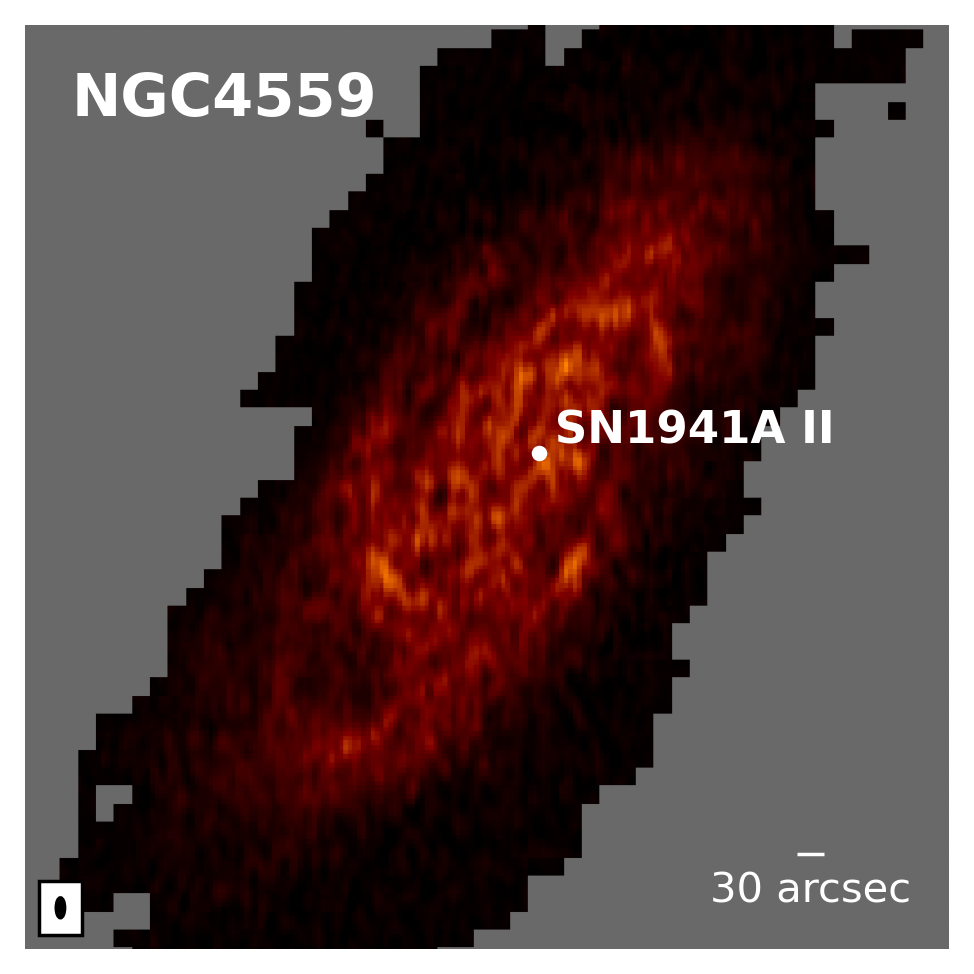}\\
\includegraphics[width=0.25\textwidth]{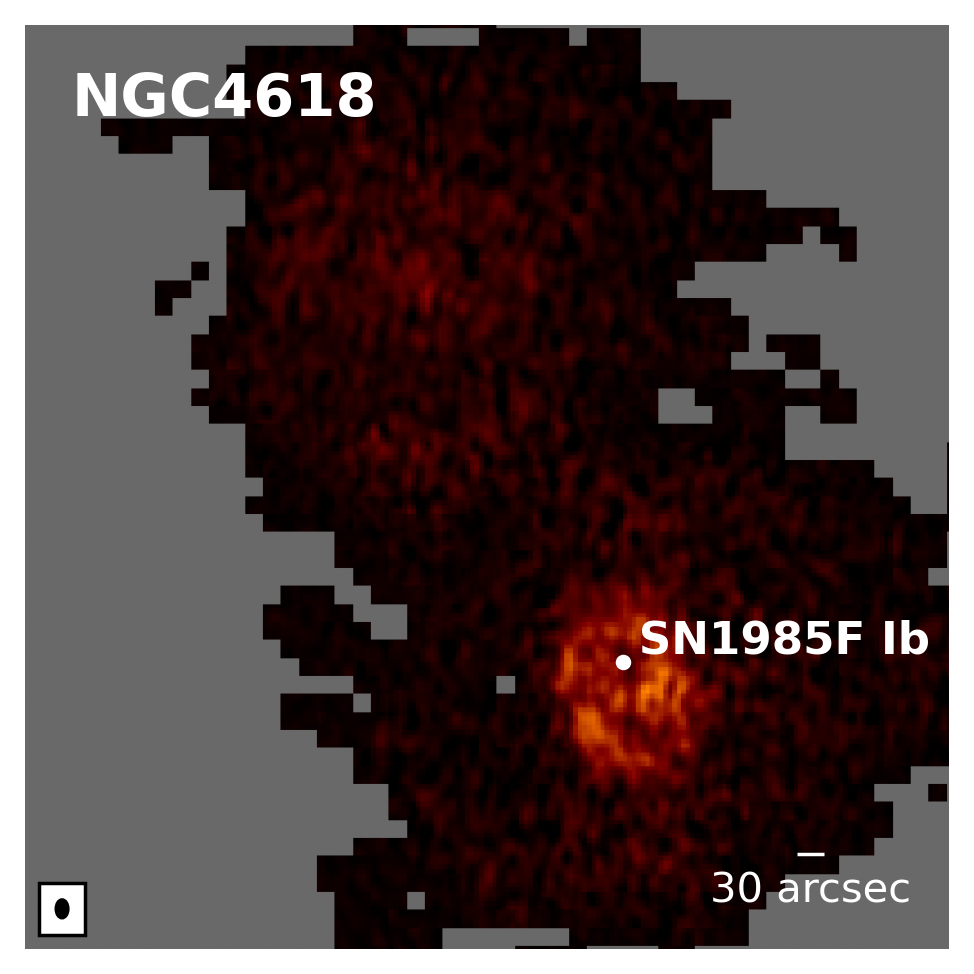}
\includegraphics[width=0.25\textwidth]{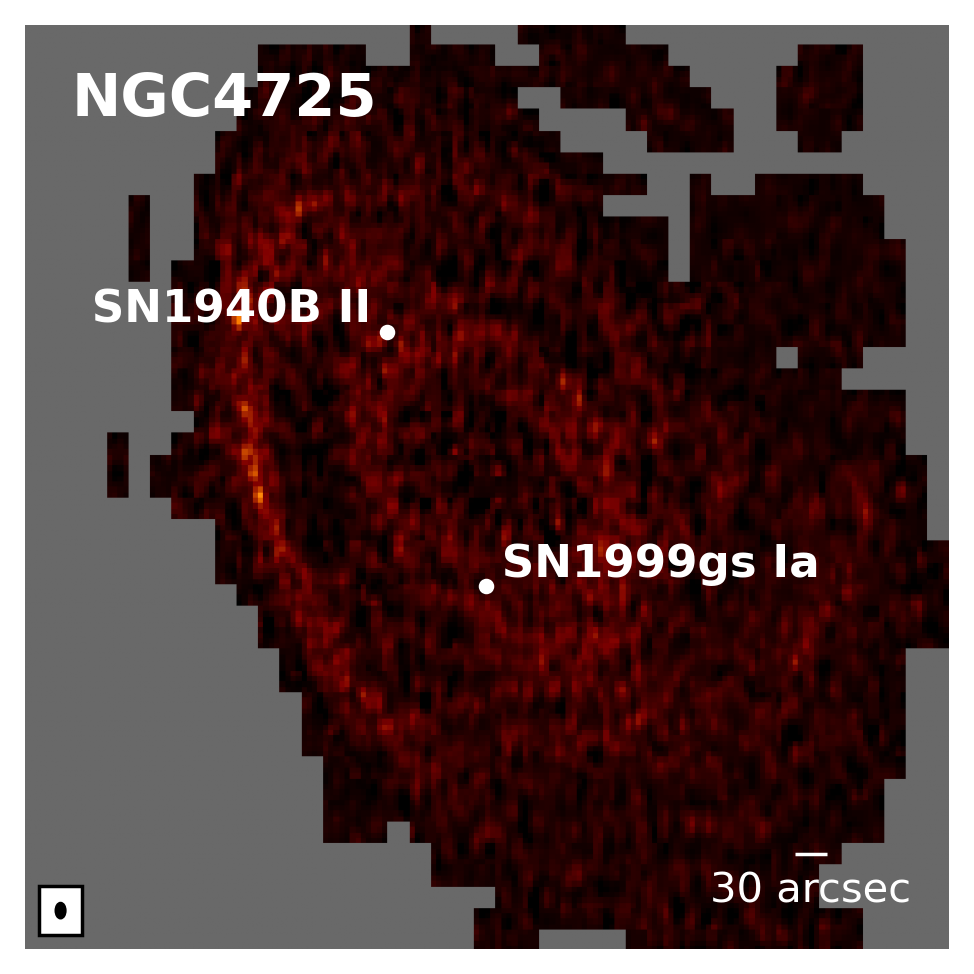}
\includegraphics[width=0.25\textwidth]{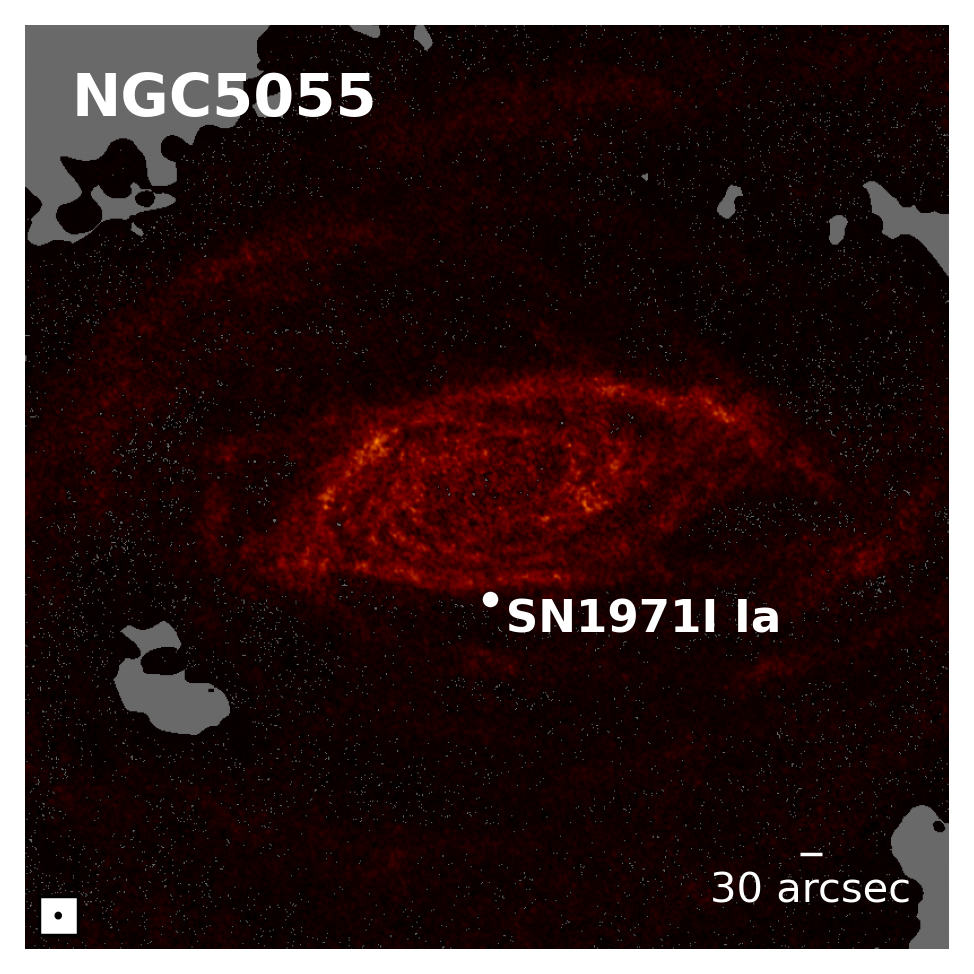}
\includegraphics[width=0.25\textwidth]{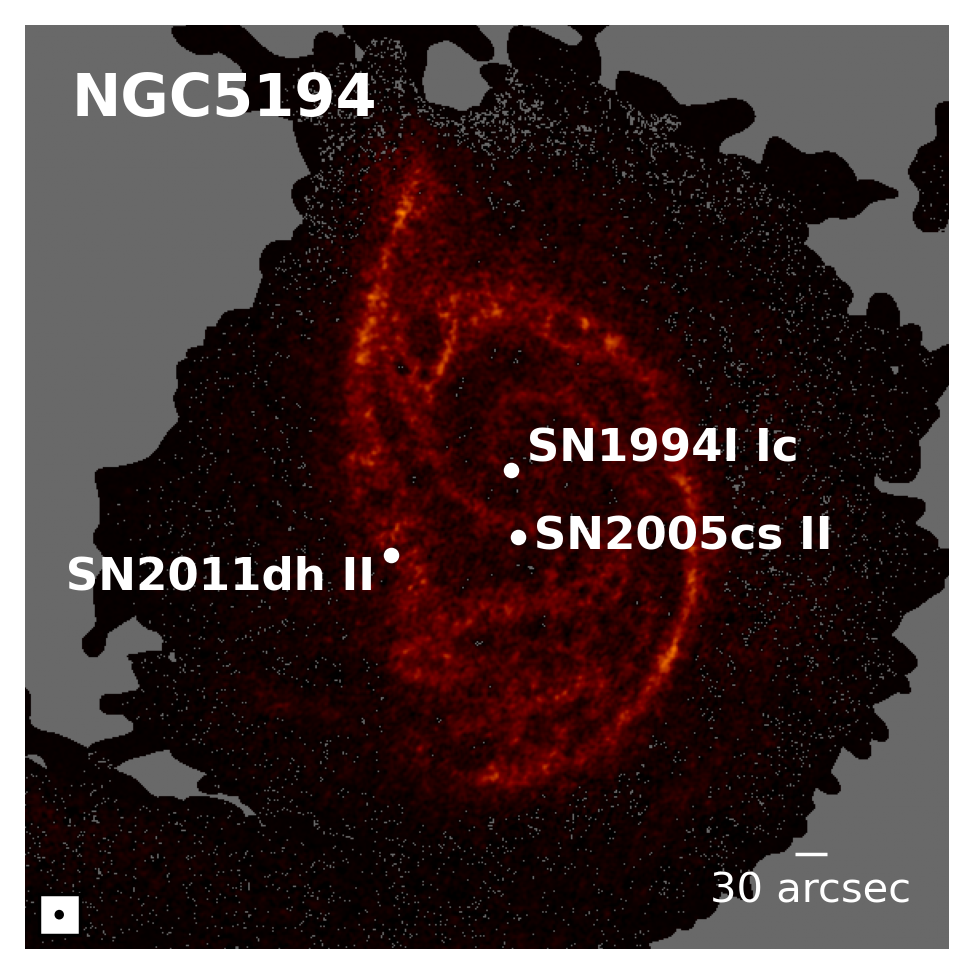}\\
\includegraphics[width=0.25\textwidth]{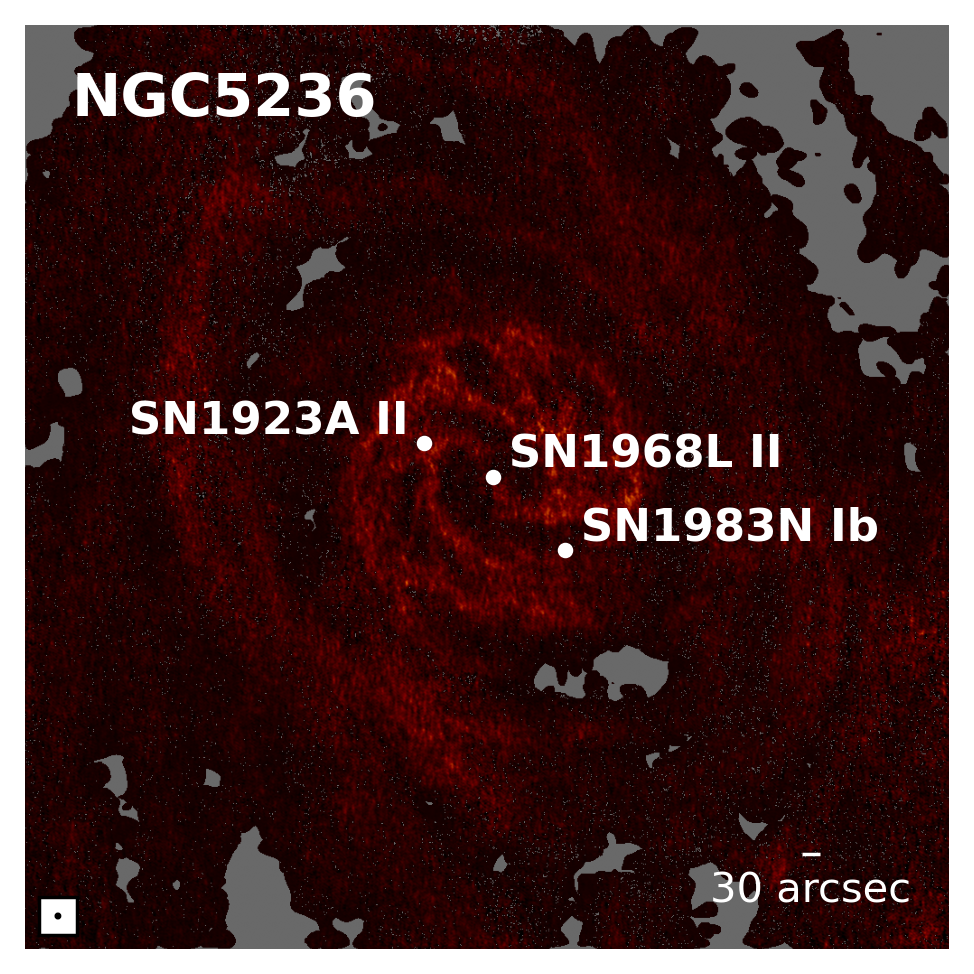}
\includegraphics[width=0.25\textwidth]{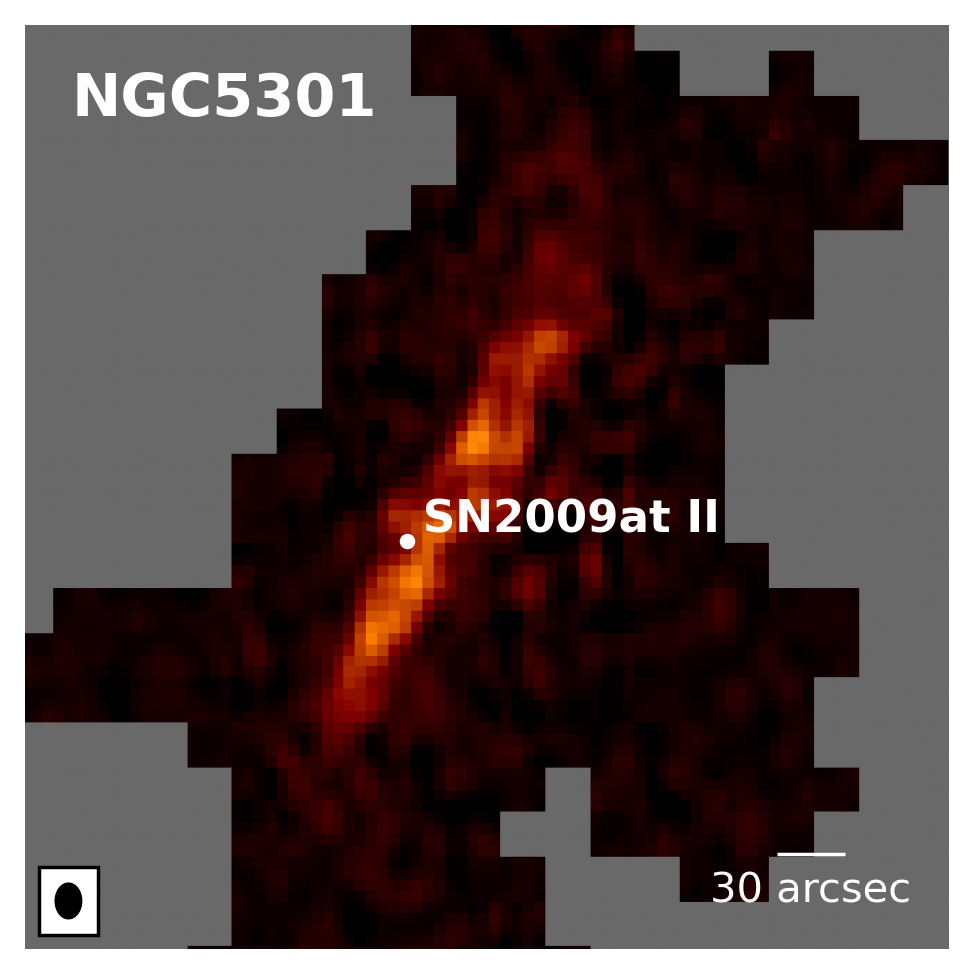}
\includegraphics[width=0.25\textwidth]{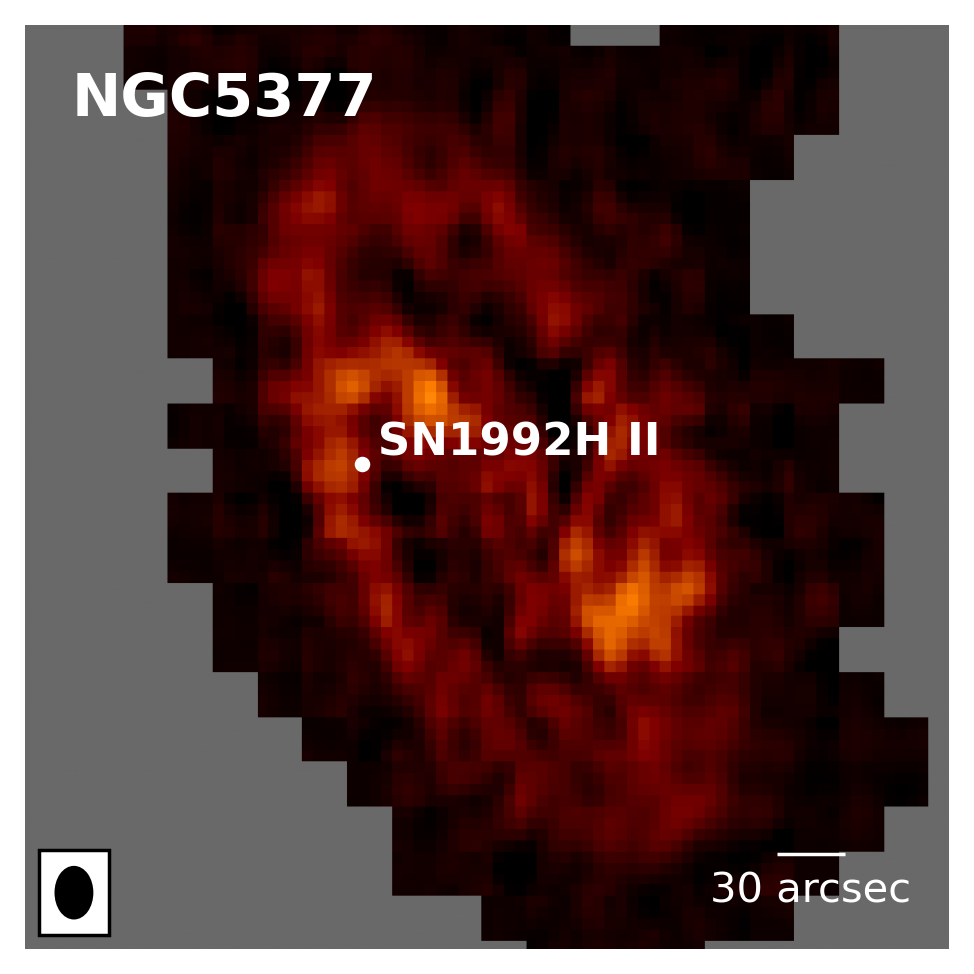}
\includegraphics[width=0.25\textwidth]{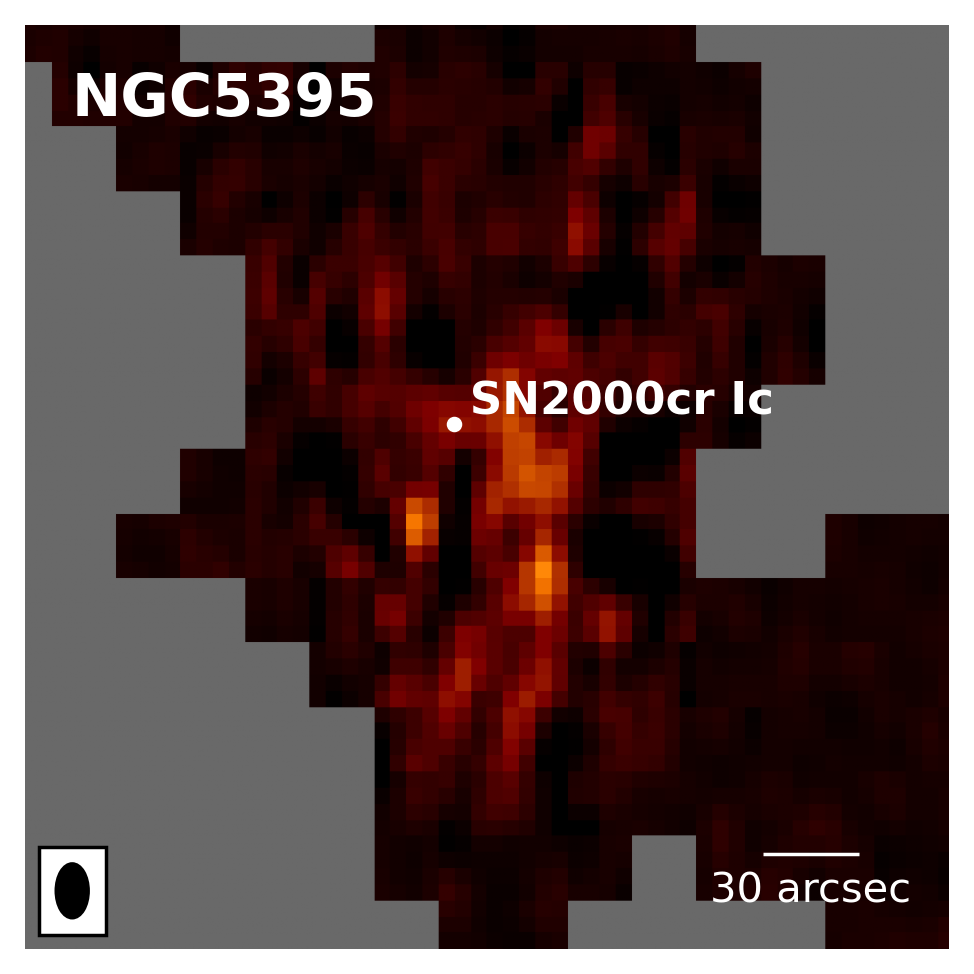}\\
\includegraphics[width=0.25\textwidth]{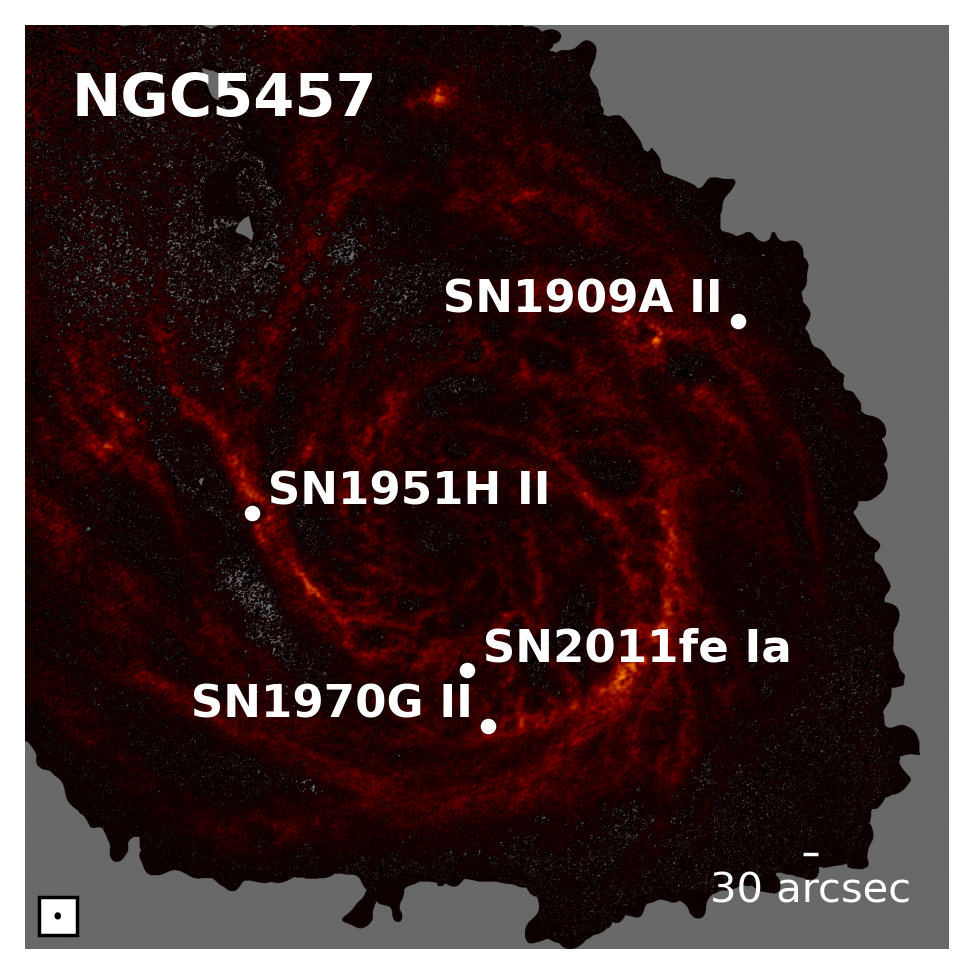}
\includegraphics[width=0.25\textwidth]{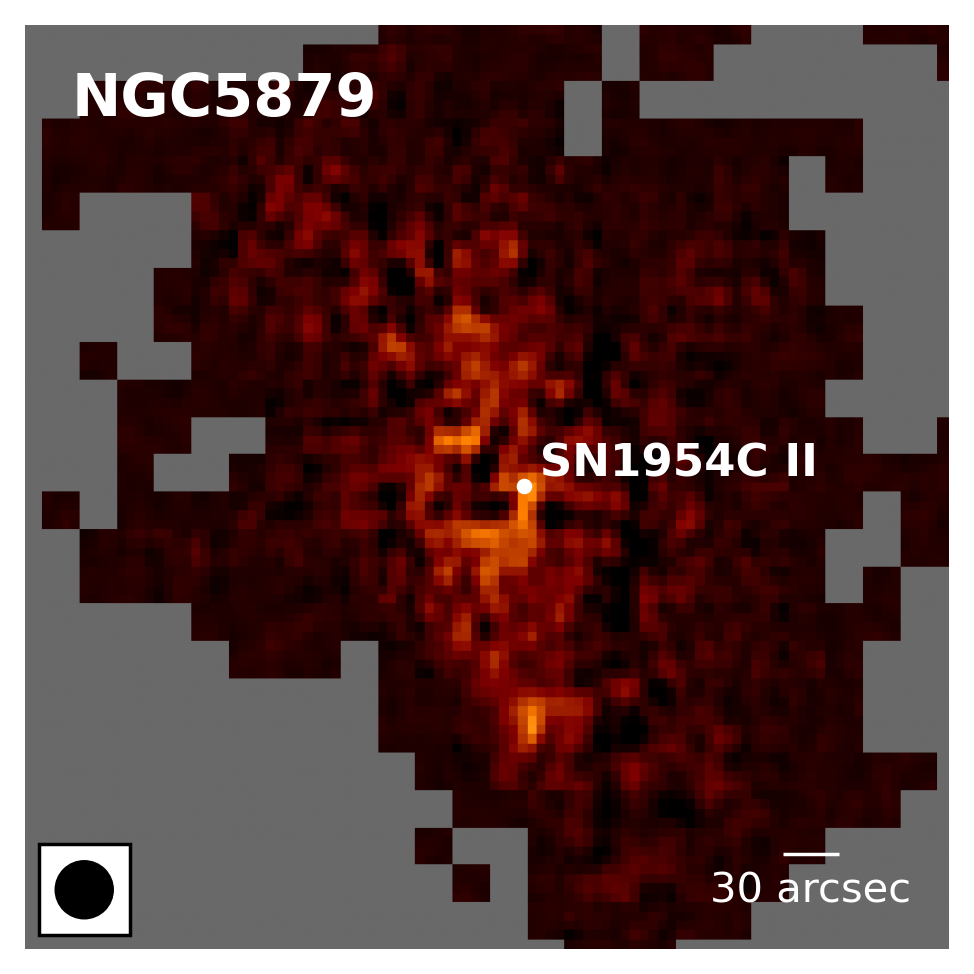}
\includegraphics[width=0.25\textwidth]{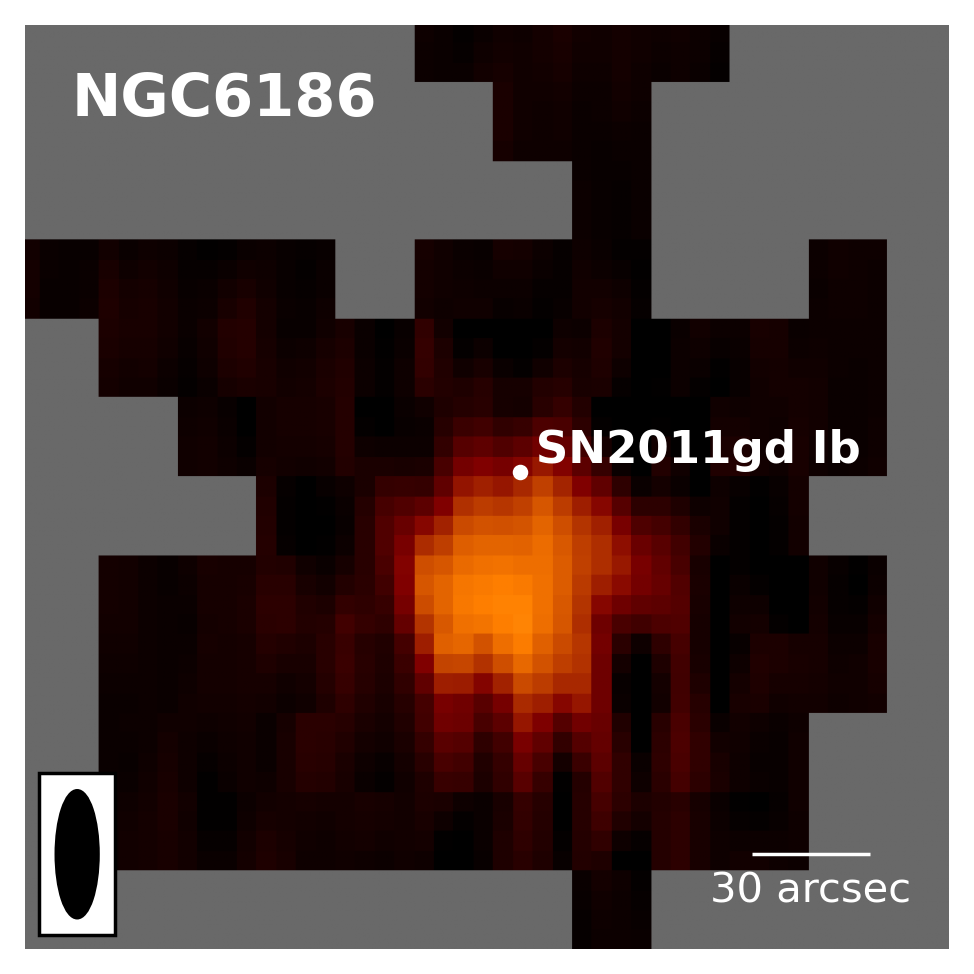}
\includegraphics[width=0.25\textwidth]{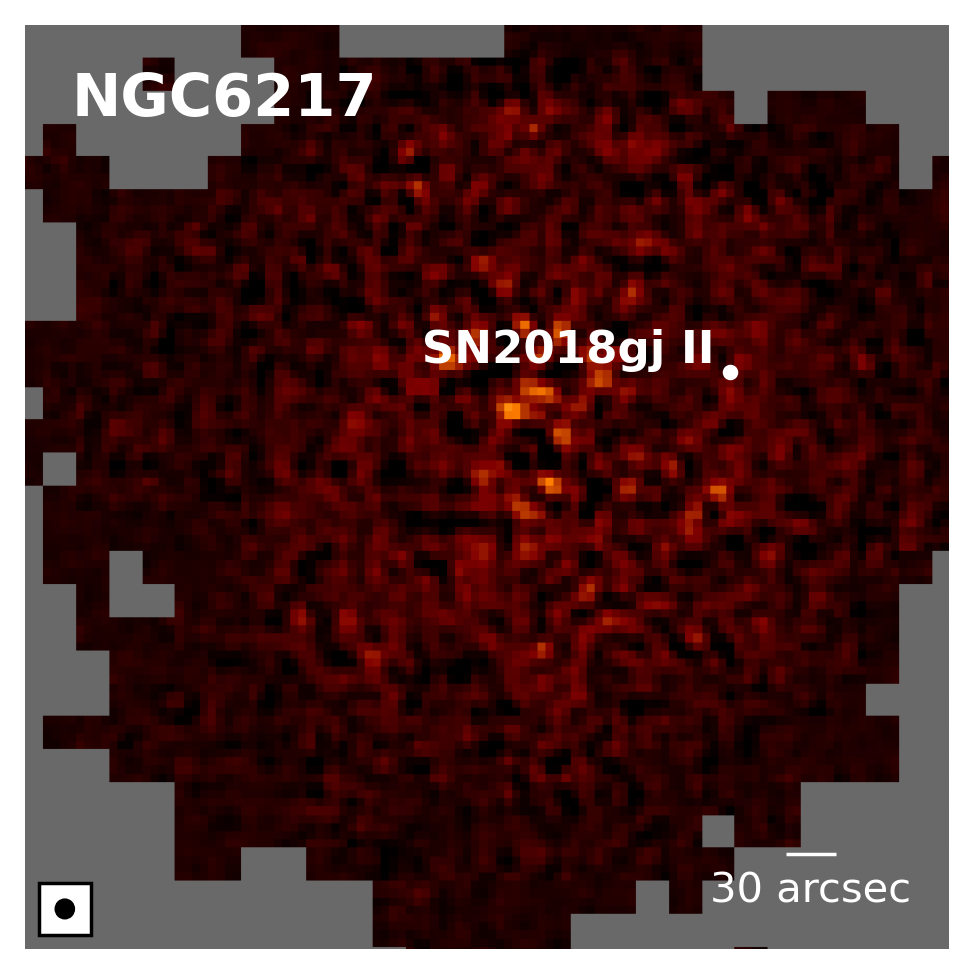}\\
\end{tabular}
\caption{Continued.}
\label{fig:mapIa}
\end{figure*}

\addtocounter{figure}{-1}
\begin{figure*}
\centering
\begin{tabular}{llll}
\includegraphics[width=0.25\textwidth]{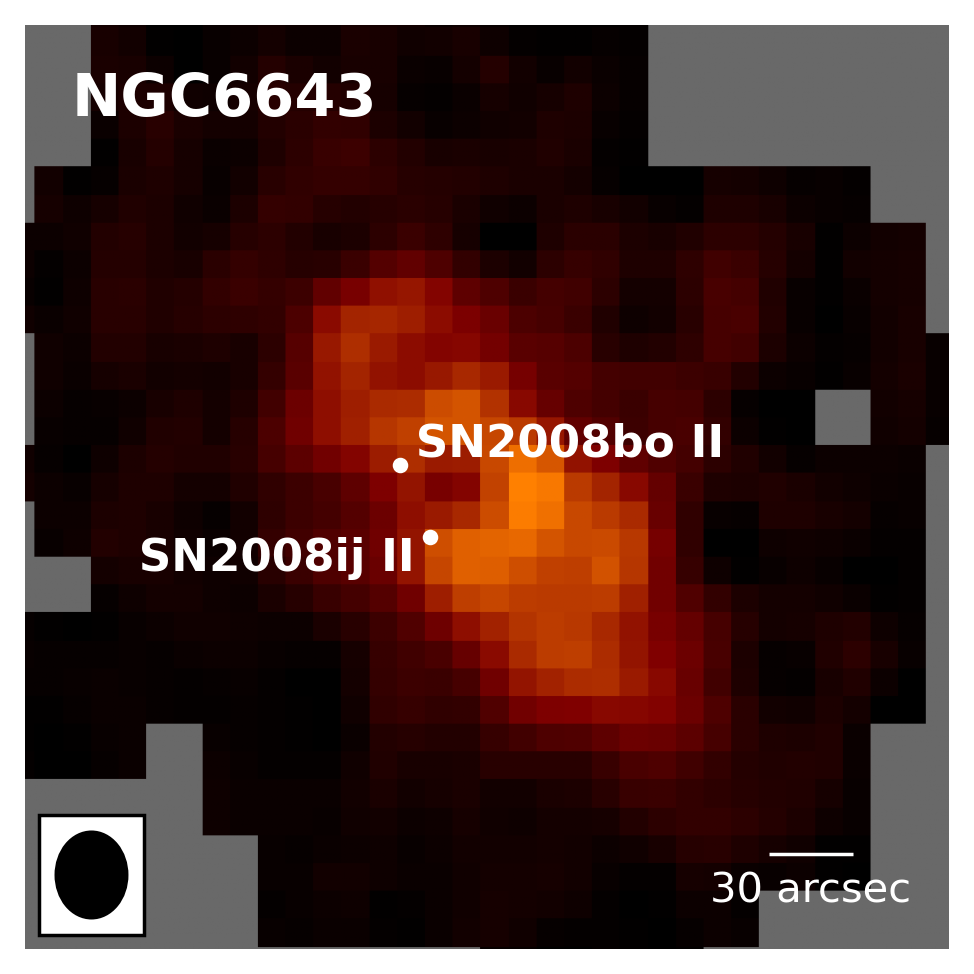}
\includegraphics[width=0.25\textwidth]{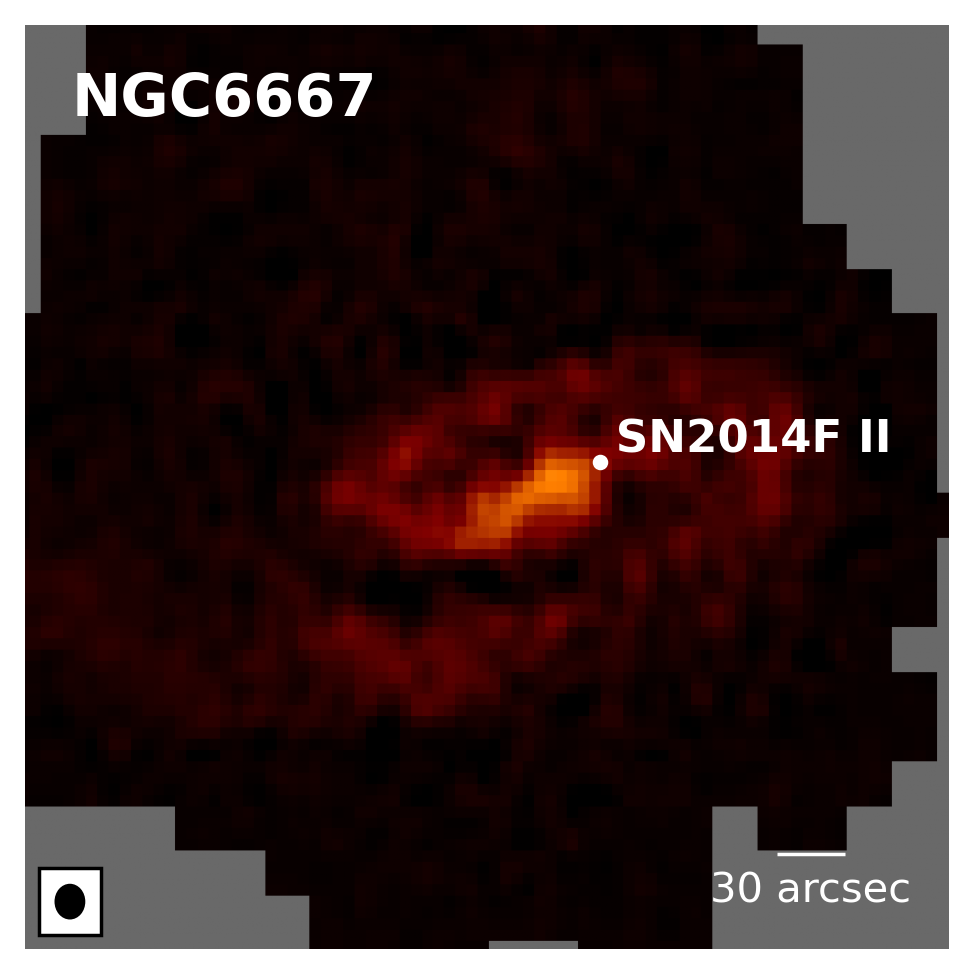}
\includegraphics[width=0.25\textwidth]{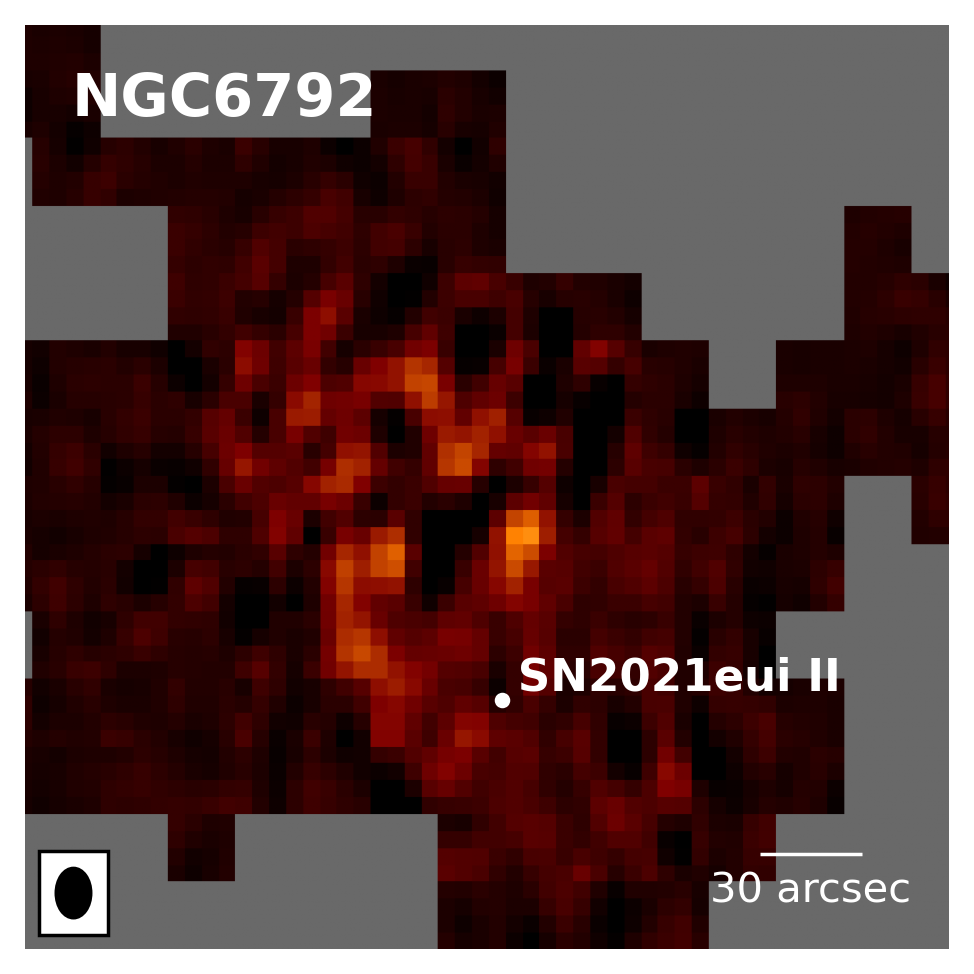}
\includegraphics[width=0.25\textwidth]{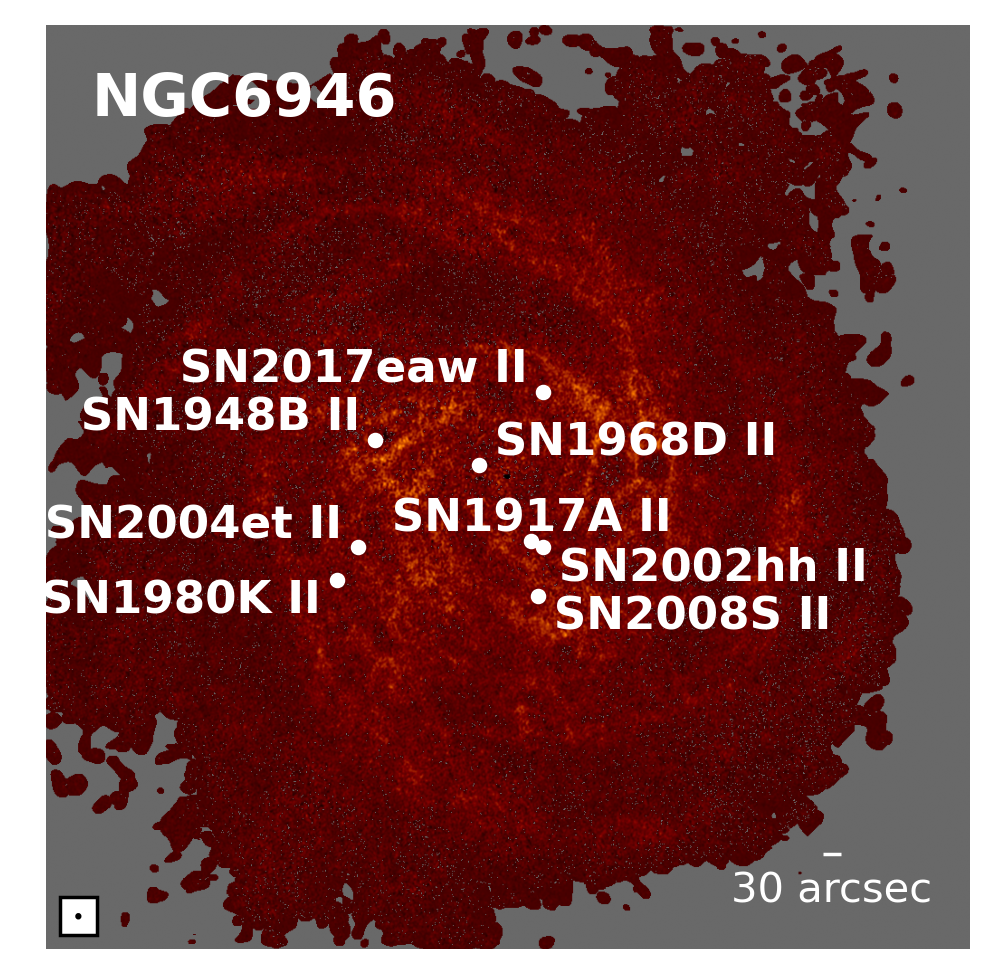}\\
\includegraphics[width=0.25\textwidth]{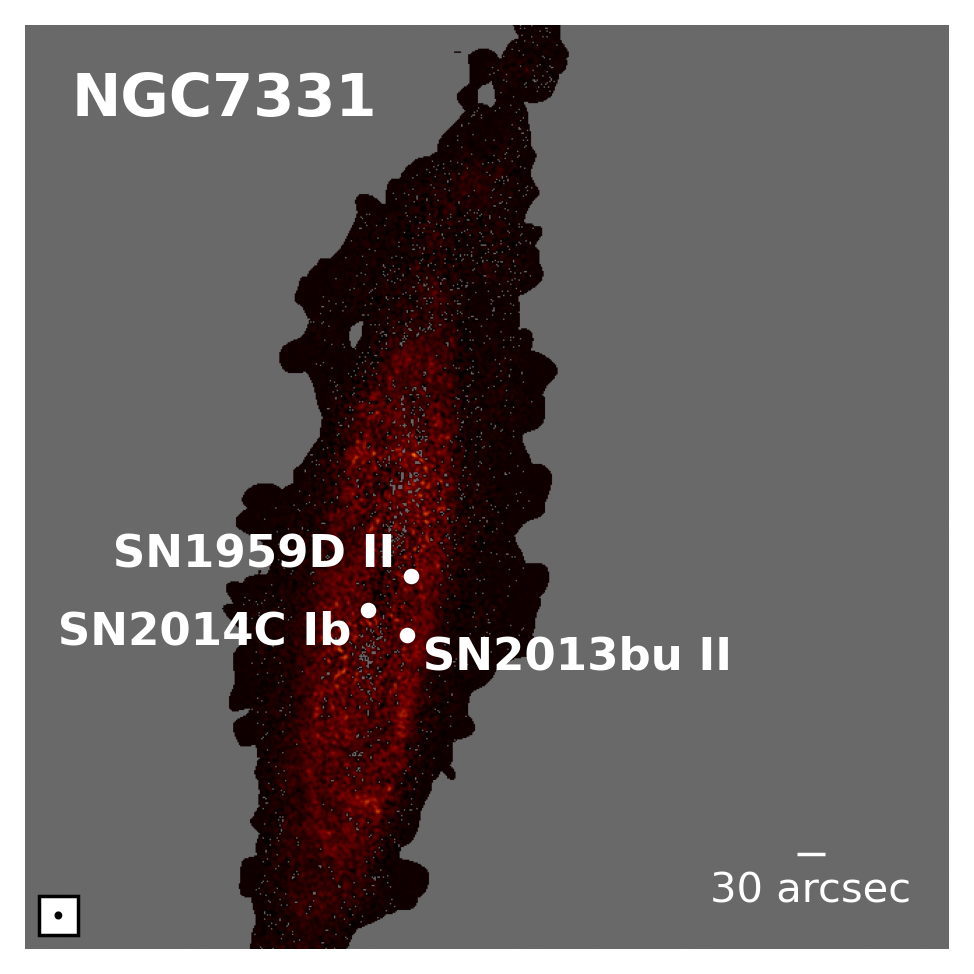}
\includegraphics[width=0.25\textwidth]{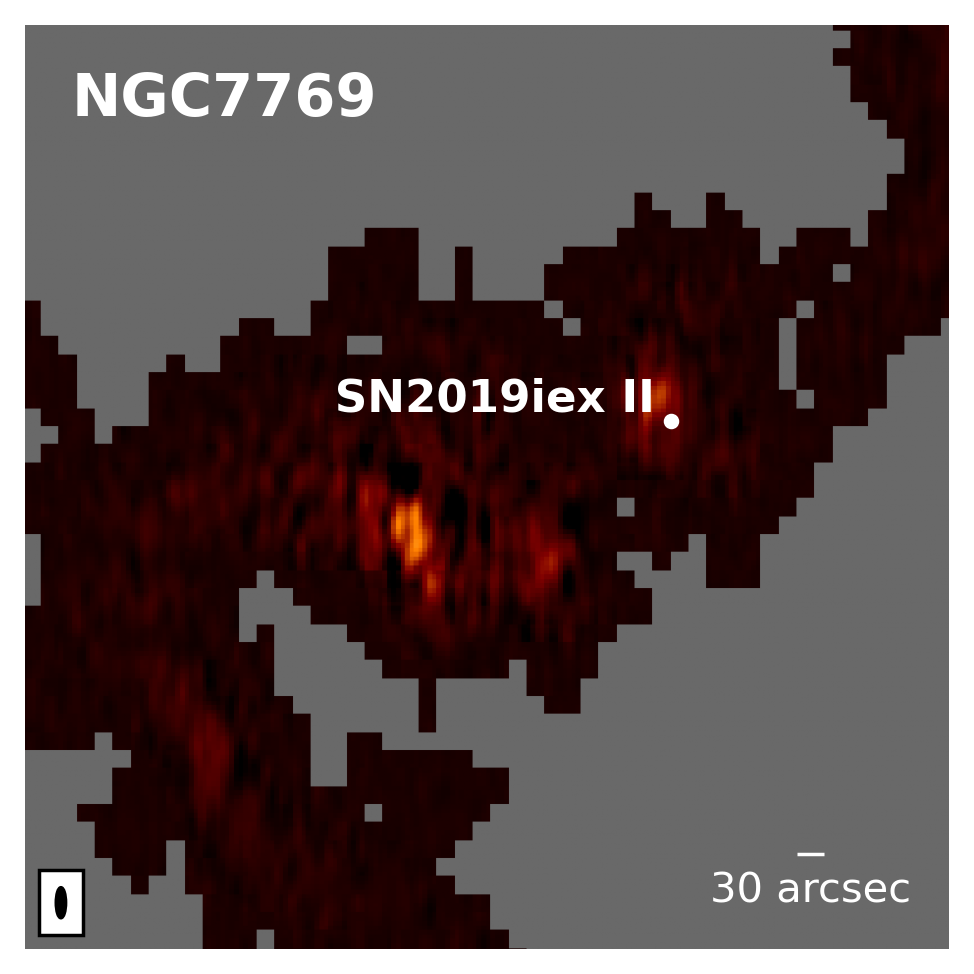}
\includegraphics[width=0.25\textwidth]{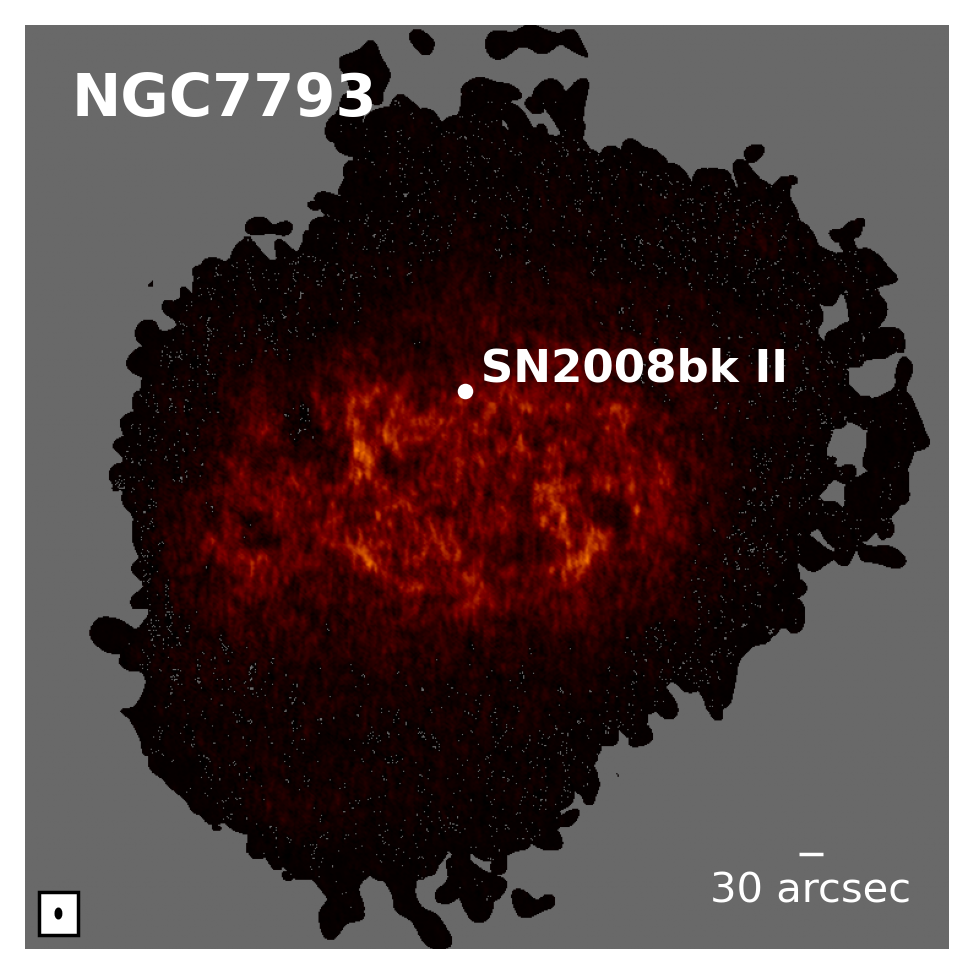}
\includegraphics[width=0.25\textwidth]{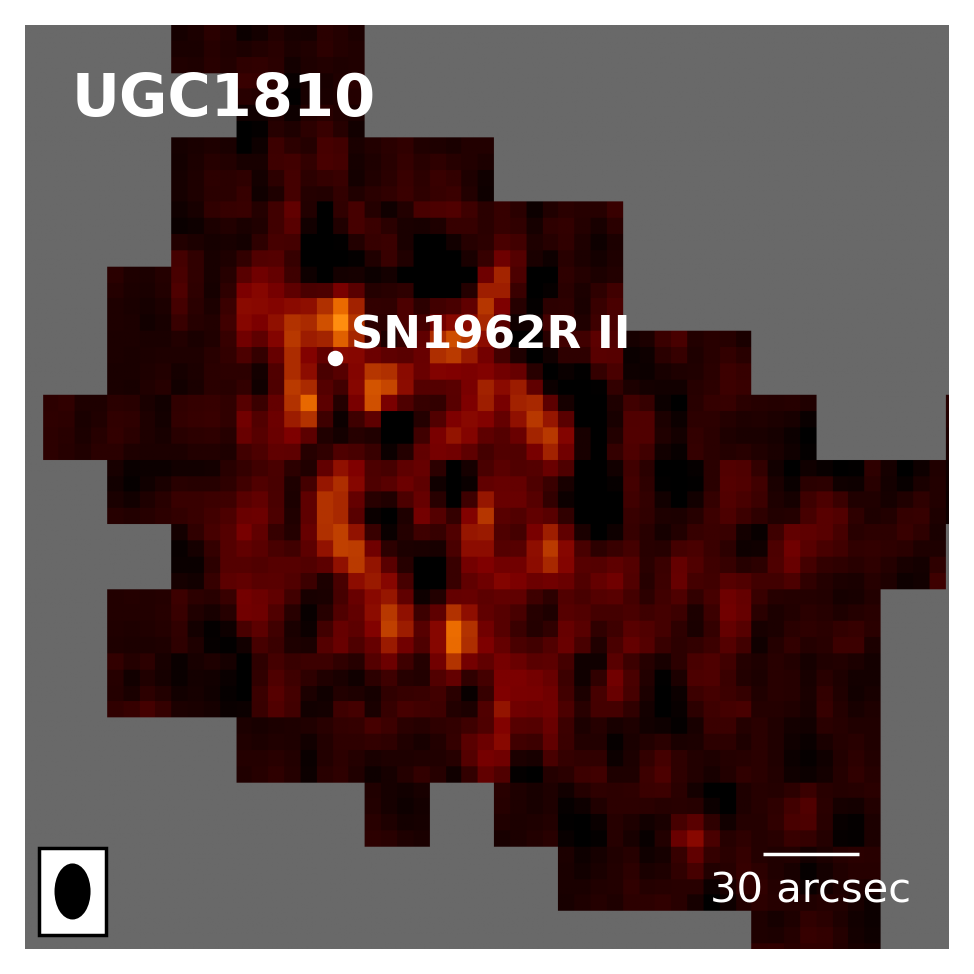}\\
\includegraphics[width=0.25\textwidth]{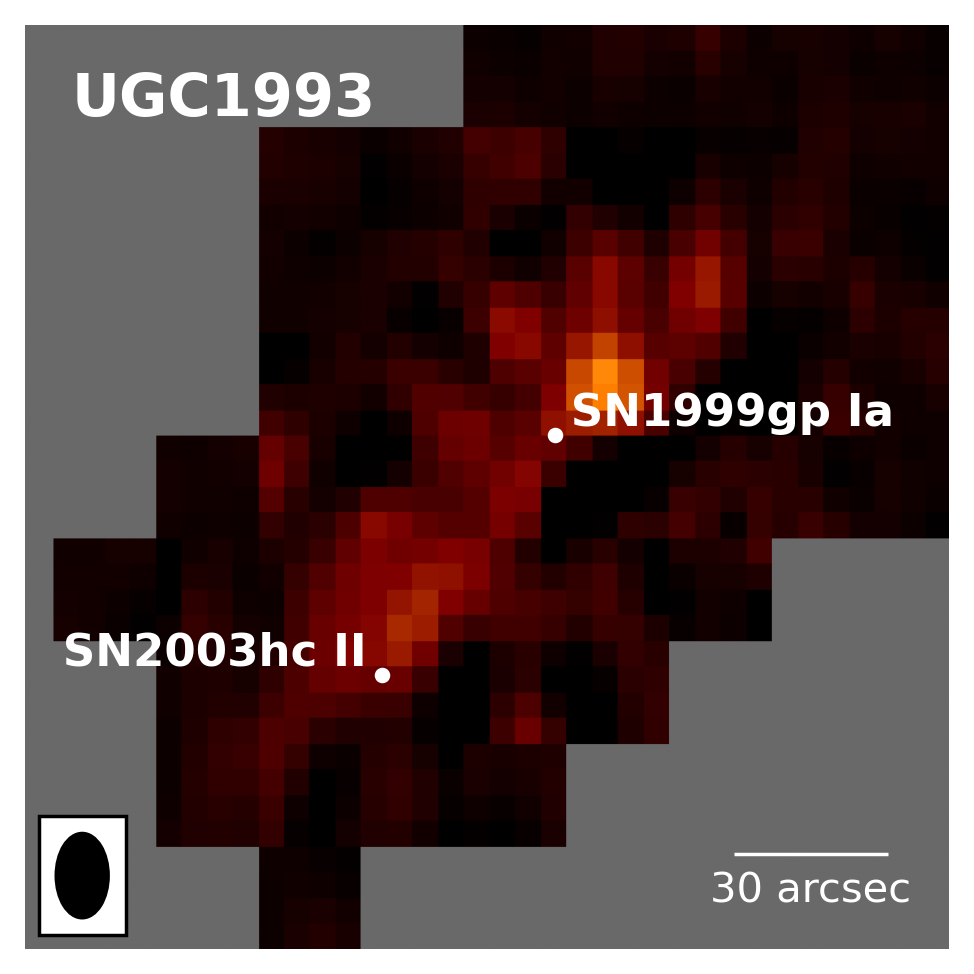}
\includegraphics[width=0.25\textwidth]{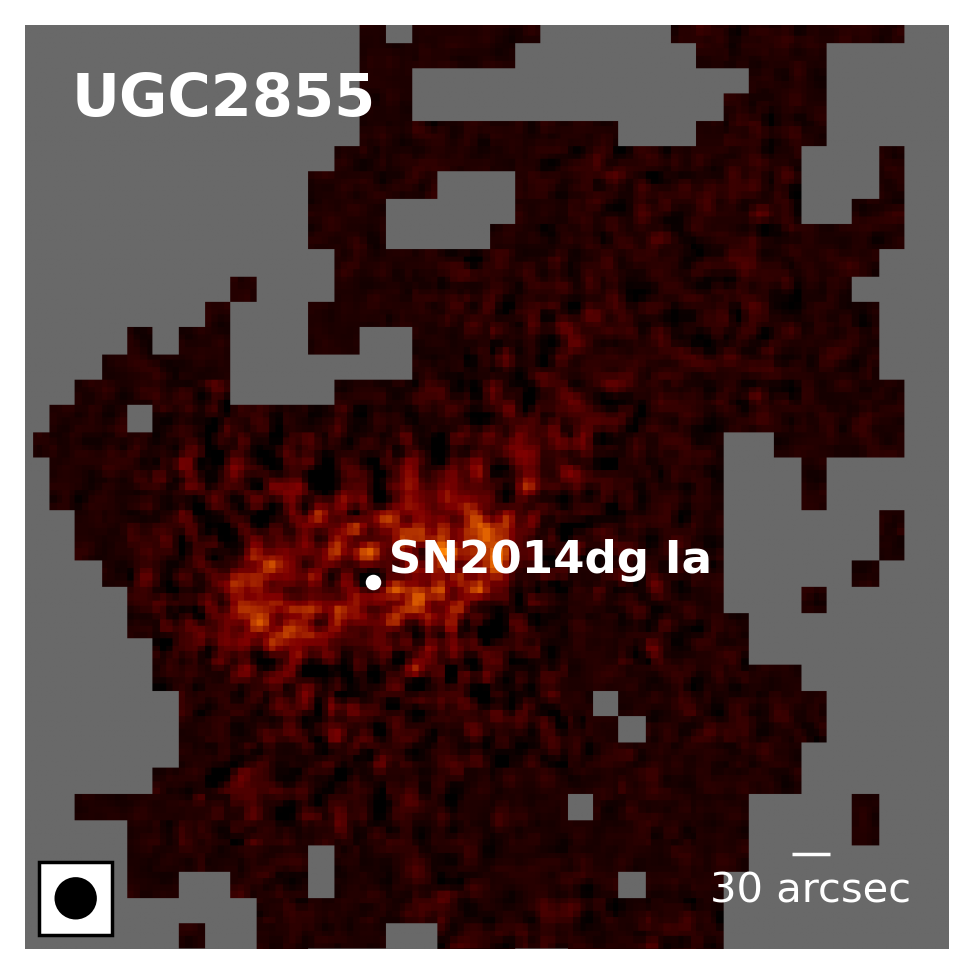}
\includegraphics[width=0.25\textwidth]{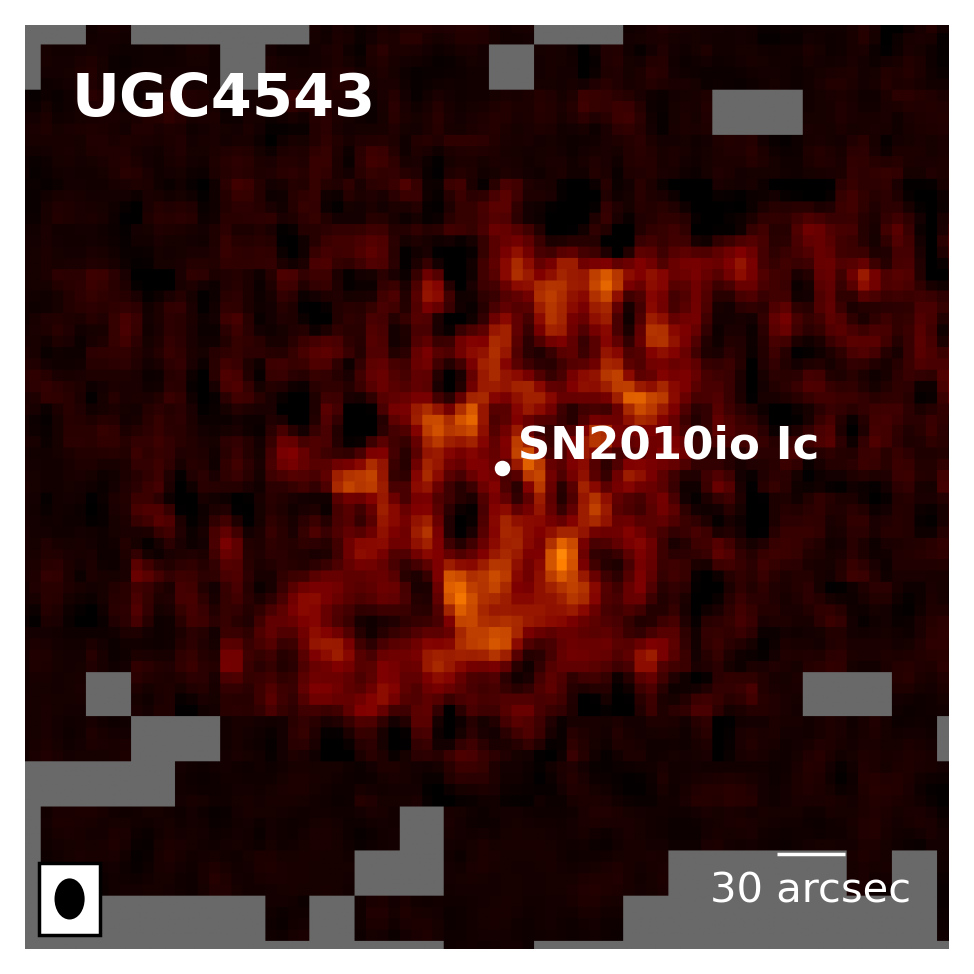}
\includegraphics[width=0.25\textwidth]{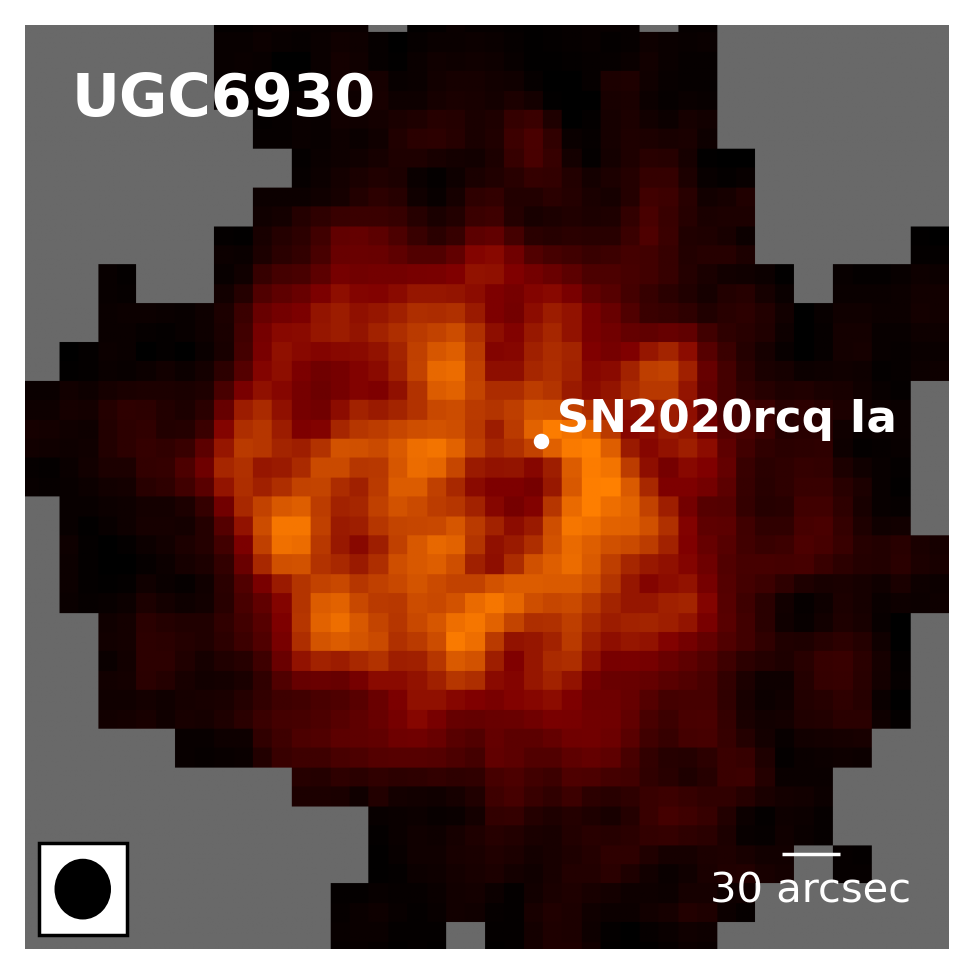}\\
\includegraphics[width=0.25\textwidth]{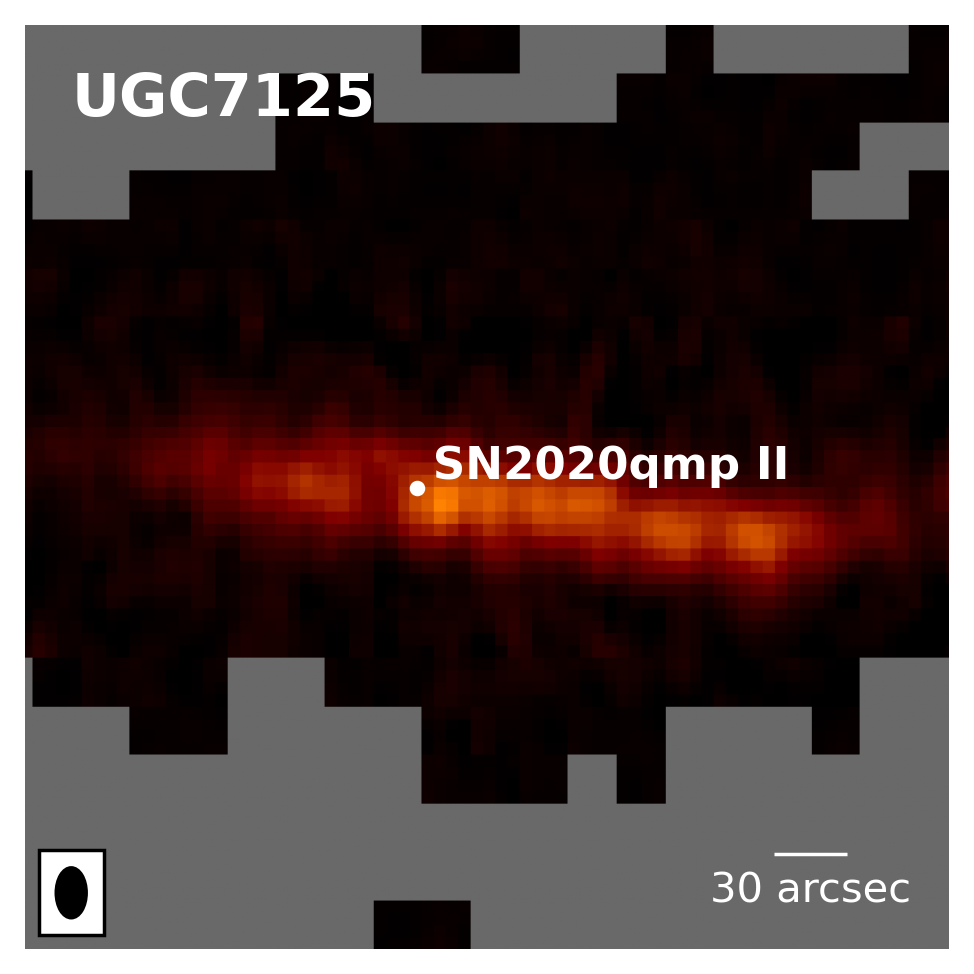}
\includegraphics[width=0.25\textwidth]{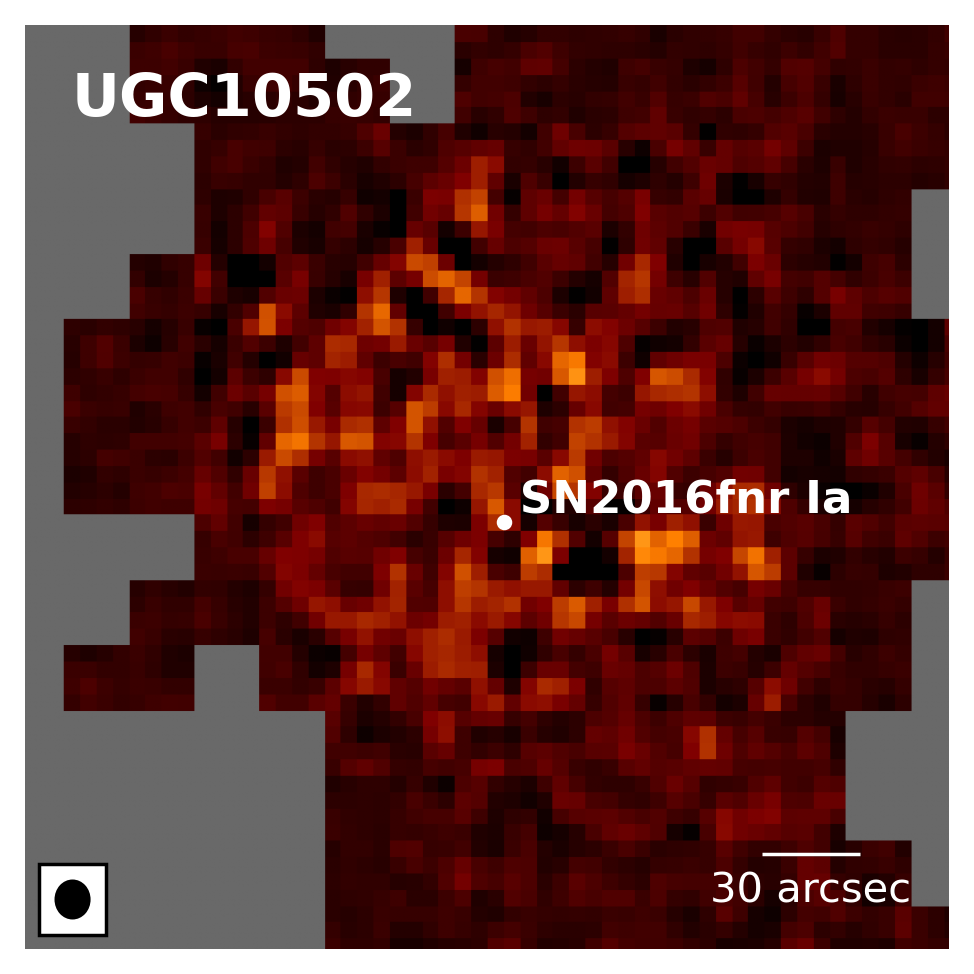}
\includegraphics[width=0.25\textwidth]{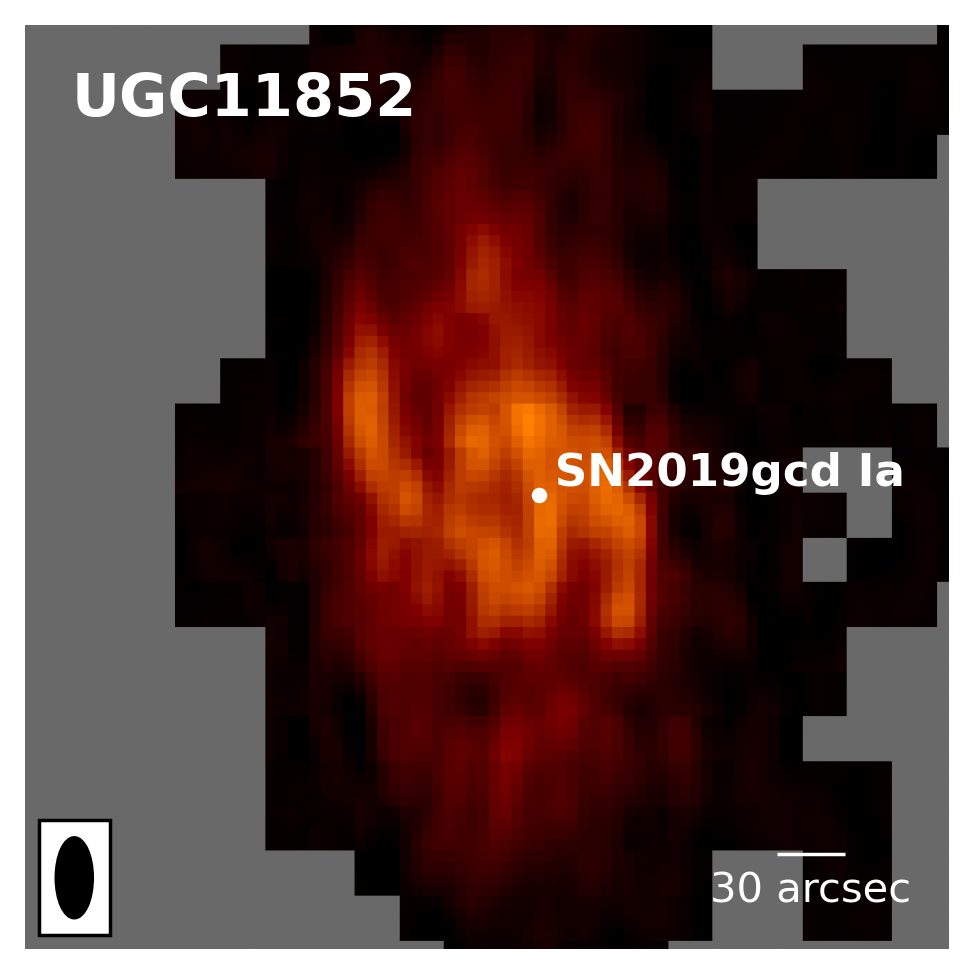}
\includegraphics[width=0.25\textwidth]{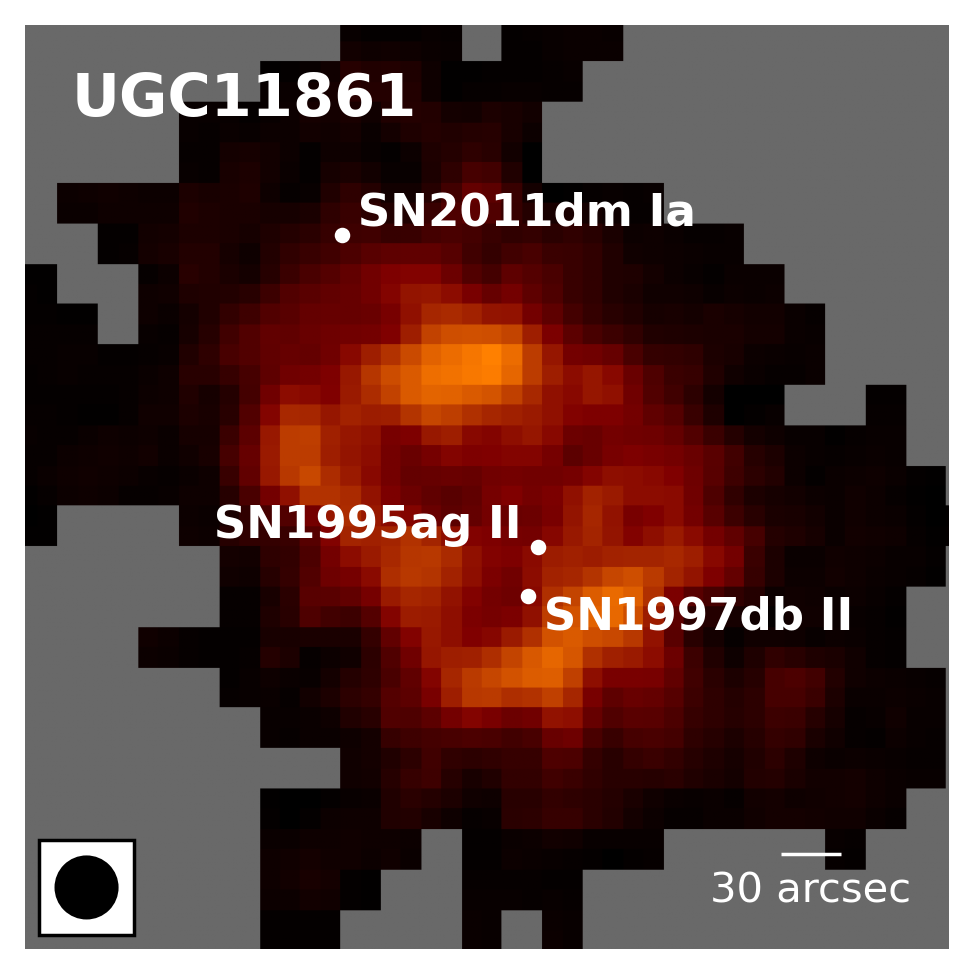}\\
\includegraphics[width=0.25\textwidth]{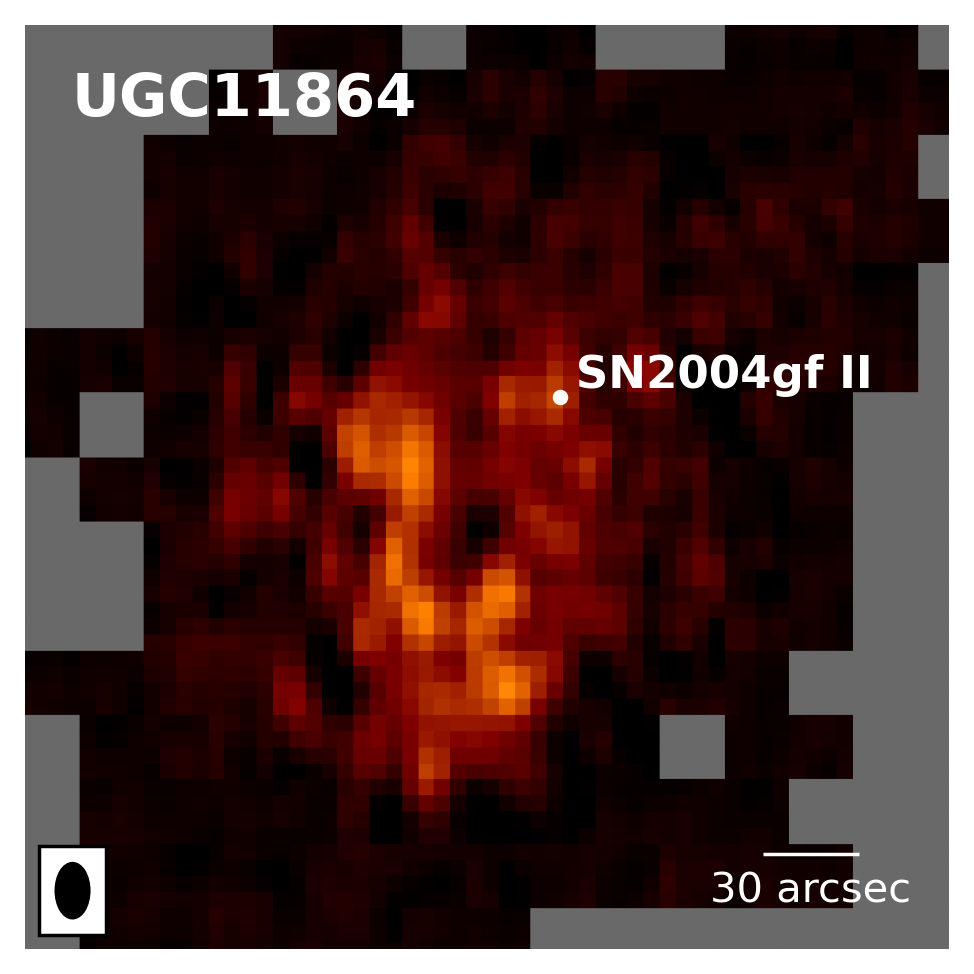}
\includegraphics[width=0.25\textwidth]{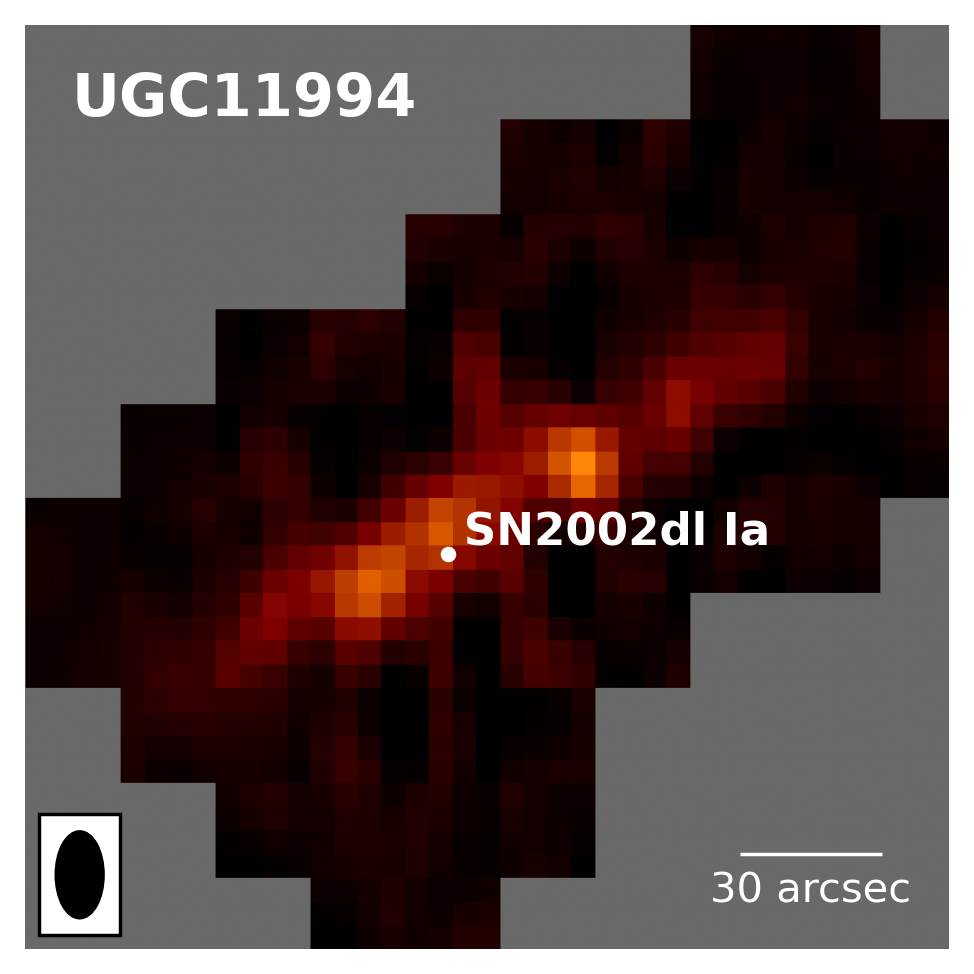}\\
\end{tabular}
\caption{Continued.}
\label{fig:mapIa}
\end{figure*}

\begin{table*}[!h]
\tiny
\centering
\caption{Properties of SNe analysed here: redshift, the telescope with which the data was taken, resolution and $f_p$ and $f_f$ statistics.}
\label{tab:table}
\begin{tabular}{ccccccccc}
\hline \hline
Supernova &Type &Host galaxy &Redshift & Telescope  &Resolution (\arcsec)& Resolution (kpc)& $f_p$ & $f_f$\\ \hline 
&&&&\bf{Ia}&&& \\
\hline
SN1937C & Ia & IC4182 & 0.001071 & WSRT & 30.0 x 30.0 & 0.67 x 0.67 & 0.827 & 0.387 \\ 
SN1957A & Ia & NGC2841 & 0.0021 & VLA & 6.1 x 5.8 & 0.41 x 0.39 & 0.933 & 0.707 \\ 
SN1960H & Ia & NGC4096 & 0.0019 & WSRT & 17.1 x 12.5 & 0.98 x 0.71 & 0.832 & 0.34 \\ 
SN1963J & Ia & NGC3913 & 0.00318 & WSRT & 12.5 x 9.8 & 0.24 x 0.19 & 0.785 & 0.333 \\ 
SN1966J & Ia & NGC3198 & 0.00227 & VLA & 6.5 x 5.2 & 0.44 x 0.35 & 0.902 & 0.654 \\ 
SN1971I & Ia & NGC5055 & 0.00131 & VLA & 5.8 x 5.3 & 0.25 x 0.23 & 0.478 & 0.091 \\ 
SN1979B & Ia & NGC3913 & 0.00318 & WSRT & 12.5 x 9.8 & 0.24 x 0.19 & 0.893 & 0.586 \\ 
SN1989B & Ia & NGC3627 & 0.002425 & VLA & 5.8 x 5.5 & 0.26 x 0.25 & 0.962 & 0.738 \\ 
SN1995E & Ia & NGC2441 & 0.01158 & WSRT & 10.8 x 10.3 & 2.0 x 1.91 & 0.786 & 0.3 \\ 
SN1997bq & Ia & NGC3147 & 0.0094 & WSRT & 10.7 x 10.2 & 1.99 x 1.9 & 0.915 & 0.63 \\ 
SN1998eb & Ia & NGC1961 & 0.01312 & WSRT & 11.2 x 10.6 & 1.73 x 1.64 & 0.989 & 0.91 \\ 
SN1999by & IaPec & NGC2841 & 0.002128 & VLA & 6.1 x 5.8 & 0.41 x 0.39 & 0.773 & 0.323 \\ 
SN1999gp & Ia & UGC1993 & 0.026725 & WSRT & 16.4 x 10.1 & 8.5 x 5.24 & 0.977 & 0.852 \\ 
SN1999gs & Ia & NGC4725 & 0.004 & WSRT & 13.2 x 8.0 & 0.78 x 0.47 & 0.661 & 0.239 \\ 
SN2002bo & Ia & NGC3190 & 0.00424 & WSRT & 26.3 x 9.8 & 3.07 x 1.14 & 0.829 & 0.394 \\ 
SN2002cv & Ia & NGC3190 & 0.00424 & WSRT & 26.3 x 9.8 & 3.07 x 1.14 & 0.989 & 0.908 \\ 
SN2002dl & IaPec & UGC11994 & 0.016258 & WSRT & 18.1 x 9.87 & 5.52 x 3.01 & 0.984 & 0.861 \\ 
SN2004bd & Ia & NGC3786 & 0.00893 & WSRT & 31.1 x 15.9 & 6.51 x 3.33 & 0.981 & 0.666 \\ 
SN2008fv & Ia & NGC3147 & 0.0094 & WSRT & 10.7 x 10.2 & 1.99 x 1.9 & 0.998 & 0.978 \\ 
SN2011dm & Ia & UGC11861 & 0.0049 & WSRT & 30.0 x 30.0 & 2.07 x 2.07 & 0.653 & 0.191 \\ 
SN2011fe & Ia & NGC5457 & 0.000804 & VLA & 7.5 x 6.1 & 0.25 x 0.2 & 0.455 & 0.087 \\ 
SN2014dg & Ia & UGC2855 & 0.004 & WSRT & 30.0 x 30.0 & 2.23 x 2.23 & 0.002 & 0.0 \\ 
ASASSN-15so & Ia & NGC3583 & 0.007125 & WSRT & 13.2 x 10.3 & 2.0 x 1.56 & 0.971 & 0.813 \\ 
SN2016fnr & Ia & UGC10502 & 0.014367 & WSRT & 11.2 x 10.0 & 3.32 x 2.96 & 0.951 & 0.717 \\ 
SN2019np & Ia & NGC3254 & 0.004556 & WSRT & 20.4 x 10.1 & 3.71 x 1.83 & 0.986 & 0.907 \\ 
SN2019gcd & IaPec & UGC11852 & 0.019513 & WSRT & 35.7 x 16.1 & 13.95 x 6.29 & 0.995 & 0.975 \\ 
SN2020rcq & Ia & UGC6930 & 0.002628 & WSRT & 29.6 x 27.4 & 2.42 x 2.24 & 0.927 & 0.767 \\ 
SN2020scc & Ia & NGC2782 & 0.008503 & WSRT & 24.8 x 15.5 & 1.38 x 0.86 & 0.264 & 0.004 \\ 
SN2021hpr & Ia & NGC3147 & 0.0094 & WSRT & 10.7 x 10.2 & 1.99 x 1.9 & 0.838 & 0.439 \\
\hline
&&&&\bf{II}&&& \\ \hline
SN1909A & II & NGC5457 & 0.000804 & VLA & 7.5 x 6.1 & 0.25 x 0.2 & 0.504 & 0.116 \\ 
SN1917A & II & NGC6946 & 0.00015 & VLA & 4.9 x 4.5 & 0.18 x 0.17 & 0.798 & 0.432 \\ 
SN1921B & II & NGC3184 & 0.00197 & VLA & 5.3 x 5.1 & 0.3 x 0.28 & 0.672 & 0.249 \\ 
SN1923A & II & NGC5236 & 0.00308 & VLA & 5.8 x 5.6 & 0.14 x 0.13 & 0.965 & 0.835 \\ 
SN1937A & II & NGC4157 & 0.002608 & WSRT & 13.9 x 8.6 & 0.91 x 0.56 & 0.983 & 0.823 \\ 
SN1937F & II & NGC3184 & 0.00197 & VLA & 5.3 x 5.1 & 0.3 x 0.28 & 0.586 & 0.162 \\ 
SN1940B & II & NGC4725 & 0.004 & WSRT & 13.2 x 8.0 & 0.78 x 0.47 & 0.904 & 0.663 \\ 
SN1948B & II & NGC6946 & 0.00015 & VLA & 4.9 x 4.5 & 0.18 x 0.17 & 0.964 & 0.837 \\ 
SN1941A & II & NGC4559 & 0.002719 & WSRT & 22.1 x 9.3 & 0.74 x 0.31 & 0.962 & 0.834 \\ 
SN1954C & II & NGC5879 & 0.00254 & WSRT & 30.0 x 30.0 & 2.28 x 2.28 & 0.999 & 0.994 \\ 
SN1951H & II & NGC5457 & 0.000804 & VLA & 7.5 x 6.1 & 0.25 x 0.2 & 0.388 & 0.055 \\ 
SN1959D & II & NGC7331 & 0.002722 & VLA & 4.9 x 4.6 & 0.35 x 0.33 & 0.781 & 0.282 \\ 
SN1961F & II & NGC3003 & 0.00493 & WSRT & 20.6 x 9.8 & 1.47 x 0.7 & 0.938 & 0.624 \\ 
SN1961V & IIPec? & NGC1058 & 0.0017 & WSRT & 15.8 x 9.2 & 0.8 x 0.47 & 0.729 & 0.326 \\ 
SN1962R & II & UGC1810 & 0.02508 & WSRT & 16.3 x 10.1 & 8.68 x 5.38 & 0.949 & 0.744 \\ 
SN1963N & II & NGC536 & 0.017322 & WSRT & 17.7 x 9.9 & 4.88 x 2.73 & 0.966 & 0.795 \\ 
SN1968D & II & NGC6946 & 0.00015 & VLA & 4.9 x 4.5 & 0.18 x 0.17 & 0.798 & 0.433 \\ 
SN1968L & II & NGC5236 & 0.00308 & VLA & 5.8 x 5.6 & 0.14 x 0.13 & 0.019 & 0.0 \\ 
SN1968V & II & NGC2276 & 0.008062 & WSRT & 30.0 x 30.0 & 2.66 x 2.66 & 0.972 & 0.798 \\ 
SN1969B & II & NGC3556 & 0.002328 & WSRT & 13.6 x 11.9 & 0.54 x 0.48 & 0.984 & 0.885 \\ 
SN1969L & II & NGC1058 & 0.0017 & WSRT & 15.8 x 9.2 & 0.8 x 0.47 & 0.735 & 0.335 \\ 
SN1970G & II & NGC5457 & 0.000804 & VLA & 7.5 x 6.1 & 0.25 x 0.2 & 0.98 & 0.885 \\ 
SN1973R & II & NGC3627 & 0.002425 & VLA & 5.8 x 5.5 & 0.26 x 0.25 & 0.981 & 0.848 \\ 
SN1980K & II & NGC6946 & 0.00015 & VLA & 4.9 x 4.5 & 0.18 x 0.17 & 0.709 & 0.298 \\ 
SN1982F & II & NGC4490 & 0.00196 & WSRT & 17.9 x 12.0 & 0.82 x 0.55 & 0.989 & 0.768 \\ 
SN1992H & IIP & NGC5377 & 0.005954 & WSRT & 22.4 x 15.8 & 2.74 x 1.93 & 0.938 & 0.735 \\ 
SN1992az & II & NGC818 & 0.014156 & WSRT & 16.3 x 10.0 & 4.48 x 2.75 & 0.105 & 0.0 \\ 
SN1993J & IIb & NGC3031 & -0.00014 & VLA & 7.6 x 7.4 & 0.13 x 0.13 & 0.996 & 0.894 \\ 
SN1993X & II & NGC2276 & 0.008062 & WSRT & 30.0 x 30.0 & 2.66 x 2.66 & 0.971 & 0.793 \\ 
SN1994ak & II & NGC2782 & 0.008503 & WSRT & 24.8 x 15.5 & 1.38 x 0.86 & 0.945 & 0.73 \\ 
SN1995ag & II & UGC11861 & 0.0049 & WSRT & 30.0 x 30.0 & 2.07 x 2.07 & 0.854 & 0.55 \\ 
SN1996cb & IIb & NGC3510 & 0.00238 & WSRT & 27.3 x 12.5 & 1.84 x 0.84 & 0.993 & 0.885 \\ 
SN1997db & II & UGC11861 & 0.0049 & WSRT & 30.0 x 30.0 & 2.07 x 2.07 & 0.85 & 0.54 \\ 
SN1997bs & II & NGC3627 & 0.002425 & VLA & 5.8 x 5.5 & 0.26 x 0.25 & 0.814 & 0.263 \\ 
SN1999bw & IIn & NGC3198 & 0.00227 & VLA & 6.5 x 5.2 & 0.44 x 0.35 & 0.973 & 0.878 \\ 
SN1999gb & II & NGC2532 & 0.01751 & WSRT & 16.8 x 10.7 & 2.84 x 1.81 & 0.943 & 0.64 \\ 
SN1999gi & II & NGC3184 & 0.00197 & VLA & 5.3 x 5.1 & 0.3 x 0.28 & 0.983 & 0.918 \\ 
SN2001ac & IIn & NGC3504 & 0.005104 & WSRT & 35.2 x 15.7 & 1.62 x 0.72 & 0.824 & 0.330 \\ 
SN2001ee & II & NGC2347 & 0.014836 & WSRT & 11.6 x 10.3 & 4.85 x 4.31 & 0.982 & 0.826 \\ 
SN2002bu & IIn & NGC4242 & 0.00176 & WSRT & 17.5 x 12.7 & 0.45 x 0.33 & 0.845 & 0.484 \\ 
SN2002hh & II & NGC6946 & 0.00015 & VLA & 4.9 x 4.5 & 0.18 x 0.17 & 0.934 & 0.736 \\ 
SN2002kg & IIn & NGC2403 & 0.00043 & VLA & 6.0 x 5.2 & 0.09 x 0.08 & 0.646 & 0.23 \\ 
SN2003J & II & NGC4157 & 0.002608 & WSRT & 13.9 x 8.6 & 0.91 x 0.56 & 0.997 & 0.968 \\ 
SN2003gd & IIP & NGC628 & 0.002108 & VLA & 6.7 x 5.6 & 0.28 x 0.24 & 0.952 & 0.759 \\ 
\hline
\hline
\end{tabular}
\end{table*}

\addtocounter{table}{-1}
\begin{table*}[!h]
\tiny
\centering
\caption{Continued.}
\begin{tabular}{ccccccccc}
\hline \hline
Supernova &Type &Host galaxy &Redshift & Telescope  &Resolution (\arcsec)& Resolution (kpc)& $f_p$ & $f_f$\\ \hline 
&&&&\bf{II}&&& \\ \hline
SN2003hc & II & UGC1993 & 0.026725 & WSRT & 16.4 x 10.1 & 8.5 x 5.24 & 0.909 & 0.573 \\ 
SN2003ie & II & NGC4051 & 0.0023 & WSRT & 29.5 x 29.0 & 1.25 x 1.23 & 0.905 & 0.651 \\ 
SN2004dj & IIn & NGC2403 & 0.00043 & VLA & 6.0 x 5.2 & 0.09 x 0.08 & 0.855 & 0.542 \\ 
SN2004et & II & NGC6946 & 0.00015 & VLA & 4.9 x 4.5 & 0.18 x 0.17 & 0.866 & 0.565 \\ 
SN2004gf & II & UGC11864 & 0.014417 & WSRT & 16.8 x 10.1 & 4.99 x 3.0 & 0.985 & 0.894 \\ 
SN2005cs & II & NGC5194 & 0.00156 & VLA & 5.8 x 5.6 & 0.24 x 0.23 & 0.875 & 0.426 \\ 
SN2005dl & II & NGC2276 & 0.008062 & WSRT & 30.0 x 30.0 & 2.66 x 2.66 & 0.914 & 0.53 \\ 
SN2008S & IInPec/LBV & NGC6946 & 0.00015 & VLA & 4.9 x 4.5 & 0.18 x 0.17 & 0.344 & 0.028 \\ 
SN2008ax & IIb & NGC4490 & 0.00196 & WSRT & 17.9 x 12.0 & 0.82 x 0.55 & 0.991 & 0.813 \\ 
SN2008bk & IIP & NGC7793 & 0.000757 & VLA & 10.4 x 5.4 & 0.18 x 0.09 & 0.946 & 0.749 \\ 
SN2008bo & IIb & NGC6643 & 0.004967 & WSRT & 30.7 x 25.3 & 3.04 x 2.5 & 0.953 & 0.767 \\ 
SN2008ij & II & NGC6643 & 0.004967 & WSRT & 30.7 x 25.3 & 3.04 x 2.5 & 0.984 & 0.912 \\ 
SN2009at & II & NGC5301 & 0.005026 & WSRT & 15.0 x 10.9 & 1.35 x 0.98 & 0.964 & 0.661 \\ 
SN2009hd & II & NGC3627 & 0.002425 & VLA & 5.8 x 5.5 & 0.26 x 0.25 & 0.842 & 0.318 \\ 
SN2011dh & IIb & NGC5194 & 0.00156 & VLA & 5.8 x 5.6 & 0.24 x 0.23 & 0.193 & 0.0 \\ 
SN2012aw & IIP & NGC3351 & 0.00256 & VLA & 6.3 x 5.2 & 0.29 x 0.24 & 0.65 & 0.261 \\ 
SN2013ej & II-P/L & NGC628 & 0.002108 & VLA & 6.7 x 5.6 & 0.28 x 0.24 & 0.858 & 0.485 \\ 
SN2013bu & II & NGC7331 & 0.002722 & VLA & 4.9 x 4.6 & 0.35 x 0.33 & 0.824 & 0.37 \\ 
SN2013cc & II & NGC1961 & 0.01312 & WSRT & 11.2 x 10.6 & 1.73 x 1.64 & 0.986 & 0.889 \\ 
SN2014F & II & NGC6667 & 0.008613 & WSRT & 14.4 x 12.2 & 3.1 x 2.63 & 0.994 & 0.912 \\ 
SN2014G & IIn & NGC3448 & 0.00457 & WSRT & 12.9 x 10.8 & 1.42 x 1.19 & 0.991 & 0.89 \\ 
SN2014bi & II & NGC4096 & 0.0019 & WSRT & 17.1 x 12.5 & 0.98 x 0.71 & 0.95 & 0.693 \\ 
SN2015bh & IIn & NGC2770 & 0.00645 & WSRT & 23.9 x 12.6 & 2.78 x 1.46 & 0.964 & 0.69 \\ 
SN2016bkv & II & NGC3184 & 0.00197 & VLA & 5.3 x 5.1 & 0.3 x 0.28 & 0.774 & 0.395 \\ 
SN2016gfy & II & NGC2276 & 0.008062 & WSRT & 30.0 x 30.0 & 2.66 x 2.66 & 0.836 & 0.314 \\ 
SN2016gil & II & NGC2532 & 0.01751 & WSRT & 16.8 x 10.7 & 2.84 x 1.81 & 0.946 & 0.654 \\ 
SN2017eaw & IIP & NGC6946 & 0.00015 & VLA & 4.9 x 4.5 & 0.18 x 0.17 & 0.976 & 0.88 \\ 
SN2018gj & IIb & NGC6217 & 0.004556 & WSRT & 10.7 x 10.4 & 0.81 x 0.79 & 0.82 & 0.405 \\ 
SN2018aoq & II & NGC4151 & 0.003262 & WSRT & 20.8 x 12.9 & 0.39 x 0.24 & 0.979 & 0.898 \\ 
SN2019iex & II & NGC7769 & 0.0139 & WSRT & 33.4 x 10.4 & 15.04 x 4.68 & 0.983 & 0.782 \\ 
SN2020qmp & II & UGC7125 & 0.0034 & WSRT & 20.5 x 12.3 & 1.54 x 0.92 & 0.993 & 0.926 \\ 
SN2021eui & II & NGC6792 & 0.015471 & WSRT & 14.6 x 10.2 & 5.29 x 3.7 & 0.486 & 0.034 \\ 
SN2021gmj & II & NGC3310 & 0.0033 & WSRT & 17.5 x 14.1 & 1.47 x 1.18 & 0.979 & 0.8 \\ 
\hline
&&&&\bf{Ib}&&& \\ \hline
SN1954A & Ib & NGC4214 & 0.000977 & VLA & 7.4 x 6.3 & 0.1 x 0.09 & 0.565 & 0.112 \\ 
SN1972R & Ib & NGC2841 & 0.0021 & VLA & 6.1 x 5.8 & 0.41 x 0.39 & 0.968 & 0.84 \\ 
SN1983N & Ib & NGC5236 & 0.00308 & VLA & 5.8 x 5.6 & 0.14 x 0.13 & 0.908 & 0.664 \\ 
SN1985F & Ib & NGC4618 & 0.0018 & WSRT & 19.8 x 12.9 & 0.69 x 0.45 & 0.903 & 0.593 \\ 
SN1999eh & Ib & NGC2770 & 0.00645 & WSRT & 23.9 x 12.6 & 2.78 x 1.46 & 0.994 & 0.932 \\ 
SN2000de & Ib & NGC4384 & 0.00833 & WSRT & 13.0 x 10.3 & 1.16 x 0.92 & 0.999 & 0.981 \\ 
SN2001is & Ib & NGC1961 & 0.013 & WSRT & 11.2 x 10.6 & 1.73 x 1.64 & 0.856 & 0.434 \\ 
SN2006gi & Ib & NGC3147 & 0.0094 & WSRT & 10.7 x 10.2 & 1.99 x 1.9 & 0.331 & 0.0 \\ 
SN2007uy & IbPec & NGC2770 & 0.0065 & WSRT & 23.9 x 12.6 & 2.78 x 1.46 & 0.971 & 0.739 \\ 
SN2008D & Ib & NGC2770 & 0.006521 & WSRT & 23.9 x 12.6 & 2.78 x 1.46 & 0.982 & 0.825 \\ 
SN2011gd & Ib & NGC6186 & 0.0098 & WSRT & 32.2 x 10.6 & 6.4 x 2.11 & 0.918 & 0.462 \\ 
SN2014C & Ib & NGC7331 & 0.002722 & VLA & 4.9 x 4.6 & 0.35 x 0.33 & 0.883 & 0.522 \\ 
\hline
&&&&\bf{Ib/c}&&& \\ \hline
SN1988ac & Ib/c & NGC3995 & 0.010854 & WSRT & 27.2 x 14.7 & 4.07 x 2.2 & 0.963 & 0.641 \\ 
SN2010br & Ib/c & NGC4051 & 0.0023 & WSRT & 29.5 x 29.0 & 1.25 x 1.23 & 0.885 & 0.594 \\ 
\hline
&&&&\bf{Ic}&&& \\ \hline
SN1983I & Ic & NGC4051 & 0.0023 & WSRT & 29.5 x 29.0 & 1.25 x 1.23 & 0.93 & 0.734 \\ 
SN1991N & Ic & NGC3310 & 0.0033 & WSRT & 17.5 x 14.1 & 1.47 x 1.18 & 0.927 & 0.554 \\ 
SN1994I & Ic & NGC5194 & 0.00156 & VLA & 5.8 x 5.6 & 0.24 x 0.23 & 0.597 & 0.083 \\ 
SN1997ei & Ic & NGC3963 & 0.01063 & WSRT & 12.2 x 9.6 & 1.03 x 0.81 & 0.633 & 0.114 \\ 
SN1999bu & Ic & NGC3786 & 0.008933 & WSRT & 32.2 x 15.9 & 6.74 x 3.33 & 0.979 & 0.649 \\ 
SN2002ap & IcBL & NGC628 & 0.002108 & VLA & 6.7 x 5.6 & 0.28 x 0.24 & 0.905 & 0.604 \\ 
SN2000cr & Ic & NGC5395 & 0.012 & WSRT & 16.7 x 9.9 & 3.15 x 1.87 & 0.98 & 0.848 \\ 
SN2002hn & Ic & NGC2532 & 0.01751 & WSRT & 16.8 x 10.7 & 2.84 x 1.81 & 0.761 & 0.171 \\ 
SN2005kl & Ic & NGC4369 & 0.003486 & WSRT & 18.2 x 11.4 & 2.8 x 1.76 & 0.975 & 0.829 \\ 
SN2007gr & Ic & NGC1058 & 0.0017 & WSRT & 15.8 x 9.2 & 0.8 x 0.47 & 0.898 & 0.657 \\ 
SN2010io & Ic & UGC4543 & 0.0065 & WSRT & 16.8 x 11.9 & 2.42 x 1.71 & 0.915 & 0.514 \\ 
SN2012fh & Ic & NGC3344 & 0.001935 & WSRT & 24.4 x 10.2 & 1.15 x 0.48 & 0.94 & 0.79 \\ 
SN2021do & Ic & NGC3147 & 0.0094 & WSRT & 10.7 x 10.2 & 1.99 x 1.9 & 0.207 & 0.0 \\ 
\hline
\end{tabular}
\end{table*}


\section{Anderson-Darling Test}
\label{app:B}

In Tables \ref{tab:ADsim} and \ref{tab:AD} we present the results of comparison between the observed data and the simulations, as well as between different types of SNe, obtained using the Anderson-Darling test. As seen from the tables, the results are qualitatively the same as the ones obtained by the KS test.

\begin{table}
\caption{Anderson-Darling $p$-values for $f_f$ between each SN type and the simulations. The numbers come from tabulated values capped at $0.1\%$. As indicated in the table, this means that some $p$-values could be lower.}
\tiny
\center
\begin{tabular}{cccc}
\hline \hline
$f_f$ \vline & Ia & II & Ib/c \\ \hline
Random & $\leq 0.001$  & $ \leq0.001$  & $ \leq0.001$   \\ 
HI & 0.37  & 0.0023   & 0.10  \\
Near-IR & 0.11  &  $\leq 0.001$  & 0.07   \\ 
\hline
\end{tabular}
\label{tab:ADsim}
\end{table}

\begin{table}
\caption{Anderson-Darling test $p$-values for $f_f$ between different SN types.}
\tiny
\center
\begin{tabular}{ccc}
\hline \hline
$f_f$ \vline  & II & Ib/c \\ \hline
Ia    & 0.68  & 0.77  \\ 
II   &   & 0.52 \\
\hline
\label{tab:AD}
\end{tabular}
\end{table}

\end{appendix}

\end{document}